\documentclass[preprint]{aastex631}
\shorttitle{TRAO observations of Orion A and Ophiuchus }
\shortauthors{Yun et al.}

\usepackage{mathtools}
\begin{document}

\title{TIMES I: a Systematic Observation in Multiple Molecular Lines Toward the Orion A and Ophiuchus Clouds}

\author[0000-0001-6842-1555]{Hyeong-Sik Yun}
\affiliation{School of Space Research, Global Campus, Kyung Hee University, 1732 Deogyeong-daero, Giheung-gu, Yongin-si, Gyeonggi-do, 17104, Republic of Korea\\(jeongeun.lee@khu.ac.kr, hs-yun@khu.ac.kr)}
\author[0000-0003-3119-2087]{Jeong-Eun Lee}
\affiliation{School of Space Research, Global Campus, Kyung Hee University, 1732 Deogyeong-daero, Giheung-gu, Yongin-si, Gyeonggi-do, 17104, Republic of Korea\\(jeongeun.lee@khu.ac.kr, hs-yun@khu.ac.kr)}
\author{Yunhee Choi}
\affiliation{Korea Astronomy and Space Science Institute, 776, Daedeok-daero, Yuseong-gu, Daejeon, 34055, Republic of Korea}
\author{Neal J. Evans II}
\affiliation{Department of Astronomy, The University of Texas at Austin, 2515 Speedway, Austin, TX 78712, USA}
\affiliation{Korea Astronomy and Space Science Institute, 776, Daedeok-daero, Yuseong-gu, Daejeon, 34055, Republic of Korea}
\affiliation{Humanitas College, Global Campus, Kyung Hee University, 1732 Deogyeong-daero, Giheung-gu, Yongin-si, Gyeonggi-do, 17104, Republic of Korea\\(jeongeun.lee@khu.ac.kr, hs-yun@khu.ac.kr)}
\author{Stella S. R. Offner}
\affiliation{Department of Astronomy, The University of Texas at Austin, 2515 Speedway, Austin, TX 78712, USA}
\author{Mark H. Heyer}
\affiliation{Department of Astronomy, University of Massachusetts Amherst, 710 N. Pleasant Street, Amherst, MA 01003, USA}
\author{Brandt A. L. Gaches}
\affiliation{Physikalisches Institut, Universit\"{a}t zu K\"{o}ln, Z\"{u}lpicher Stra$\beta$e 77, 50937, K\"{o}ln, Germany}
\affiliation{Center of Planetary Systems Habitability, The University of Texas at Austin, 2515 Speedway, Austin, TX 78712, USA}
\author{Yong-Hee Lee}
\affiliation{School of Space Research, Global Campus, Kyung Hee University, 1732 Deogyeong-daero, Giheung-gu, Yongin-si, Gyeonggi-do, 17104, Republic of Korea\\(jeongeun.lee@khu.ac.kr, hs-yun@khu.ac.kr)}
\author{Giseon Baek}
\affiliation{School of Space Research, Global Campus, Kyung Hee University, 1732 Deogyeong-daero, Giheung-gu, Yongin-si, Gyeonggi-do, 17104, Republic of Korea\\(jeongeun.lee@khu.ac.kr, hs-yun@khu.ac.kr)}
\author{Minho Choi}
\affiliation{Korea Astronomy and Space Science Institute, 776, Daedeok-daero, Yuseong-gu, Daejeon, 34055, Republic of Korea}
\author{Hyunwoo Kang}
\affiliation{Korea Astronomy and Space Science Institute, 776, Daedeok-daero, Yuseong-gu, Daejeon, 34055, Republic of Korea}
\author{Seokho Lee}
\affiliation{National Astronomical Observatory of Japan, 2-21-1 Osawa, Mitaka, Tokyo 181-8588, Japan}
\author{Ken'ichi Tatematsu}
\affiliation{Nobeyama Radio Observatory, National Astronomical Observatory of Japan, National Institutes of Natural Sciences, 462-2 Nobeyama, Minamimaki, Minamisaku, Nagano 384-1305, Japan}
\affiliation{Department of Astronomical Science, SOKENDAI (The Graduate University for Advanced Studies), 2-21-1 Osawa, Mitaka, Tokyo 181-8588, Japan}
\author{Yao-Lun Yang}
\affiliation{Department of Astronomy, University of Virginia, Charlottesville, VA 22904-4235, USA} 
\affiliation{RIKEN Cluster for Pioneering Research, Wako-shi, Saitama, 351-0106, Japan}
\author{How-Huan Chen}
\affiliation{Department of Astronomy, The University of Texas at Austin, 2515 Speedway, Austin, TX 78712, USA}
\author{Youngung Lee}
\affiliation{Korea Astronomy and Space Science Institute, 776, Daedeok-daero, Yuseong-gu, Daejeon, 34055, Republic of Korea}
\author{Jae Hoon Jung}
\affiliation{Korea Astronomy and Space Science Institute, 776, Daedeok-daero, Yuseong-gu, Daejeon, 34055, Republic of Korea}
\author{Changhoon Lee}
\affiliation{Korea Astronomy and Space Science Institute, 776, Daedeok-daero, Yuseong-gu, Daejeon, 34055, Republic of Korea}
\author{Jungyeon Cho}
\affiliation{Department of Astronomy and Space Science, Chungnam National University, 99, Daehak-ro Yuseong-gu, Daejeon, 34134, Republic of Korea}

\begin{abstract}
We have used the Taeduk Radio Astronomy Observatory to observe the Orion A and Ophiuchus clouds in the $J=$1$-$0\ lines of $^{13}$CO, C$^{18}$O, HCN, HCO$^+$, and N$_2$H$^+$ and the $J=$2$-$1\ line of CS. The fully sampled maps with uniform noise levels are used to create moment maps. The variations of the line intensity and velocity dispersion with total column density, derived from dust emission maps, are presented and compared to previous work. The CS line traces dust column density over more than one order of magnitude, and the N$_2$H$^+$ line best traces the highest column density regime ($\log(N_\mathrm{H_2}$) $>$ 22.8). Line luminosities, integrated over the cloud, are compared to those seen in other galaxies. The HCO$^+$-to-HCN luminosity ratio in the Orion A cloud is similar to that of starburst galaxies, while that in the Ophiuchus cloud is in between those of active galactic nuclei and starburst galaxies.
\end{abstract}


\section{Introduction} \label{Sec_intro}
Gas motion in molecular clouds (MCs) is generally turbulent \citep{Lar81,Elm04,Hey04}. Supersonic turbulence on a large scale acts as an internal pressure against the global gravitational collapse and also produces dense clumps in small scales via shocks \citep{Eva99,Pad99,Pad01b,Mac04}. As turbulence dissipates,  star formation becomes easier in the clumps \citep{Pad01b,Ber07}. Thus, turbulence plays a critical role in the evolution of clouds and star-forming regions, and understanding the properties of turbulence is key to understand its role in star formation, especially for the earliest phase \citep{Ber07}. However, the relation between the turbulence and  star formation is still controversial. 

To obtain the properties of turbulence in MCs, $J$=1$-$0 transitions of $^{12}$CO and $^{13}$CO have been used \citep{Hey04,Pad06,Bru09,Pad09,Koc17} because these transitions can easily trace molecular gas in the interstellar medium (ISM). However, these lines become optically thick toward the high column density regions and cannot trace the turbulent motions in a dense environment. The properties of turbulence derived from these optically thick lines might be ineffective for assessing the relation between turbulence and star formation since stars are generally formed in a dense environment. Recent surveys, such as the CARMA Large Area Star Formation Survey \citep[CLASSy; ][]{Sto14} and the Green Bank Ammonia Survey \citep[GAS; ][]{Fri17,Mon18}, used optically thin lines and found that the turbulence affect the formation of the kinematical and morphological structures of dense gas \citep{Sto14,Sto16,Kir17,Che19}. However, the CLASSy survey observed small areas limited to clump-scale (about 1~pc $\times$ 1~pc). The small map size would limit the spatial size of turbulent motion that we can investigate, and the supersonic turbulent motion in large scales would not be probed. The GAS survey, using the NH$_3$ lines, is only focused on cold and dense gas.

To investigate the gas motions in various densities and spatial scales, it is necessary to map the entire MC in different molecular lines which can trace various density environments. \citet{Gac15} simulated a star-forming MC using a hydrodynamic simulation with post-processed three-dimensional photodissociation astrochemistry. They categorized molecular transitions into three groups ($\textit{diffuse}$, $\textit{intermediate}$, and $\textit{dense}$ tracers) which trace different density environments. Therefore, the spectral maps in different molecular transitions would represent the turbulent motions in different density environments if we chose the transitions from these three groups \citep{Goo98}. This systematic study of MCs can also provide detailed initial properties and constraints for the simulation of turbulent star-forming clouds.

To compare the properties of turbulence in different star-forming environments, we should observe MCs that have different star-forming environments. The Orion A cloud can be divided into three regions: the Integral Shaped Filament (ISF), L1641, and L1647 from the north to the south \citep{Lyn62,Mei16}. Among these regions, the ISF encompasses active massive star-forming regions \citep{Ike07,Meg12,Fur16}, and the other regions include low-mass star-forming regions \citep{All08,Nak12,Meg12,Fur16}. In the Ophiuchus cloud, low-mass stars are actively forming in L1688 \citep{Mot98,Wil08,Zha09,Dun15}. In this region, many dense cores (Oph-A through L) have been identified using DCO$^+$ \citep{Lor90}, N$_2$H$^+$ \citep{Pan17} and millimeter continuum observations \citep{Joh04,Pat15}. Some of these cores and their sub-structures were identified as ``droplets", which are the pressure confined coherent cores \citep{Che19,Che20}. Also, a filamentary structure that stretches from L1688 to the north-east, L1709, contains one starless core and one protostellar core \citep{Lor90,Dun15,Pat15}. The star formation in L1709 is less efficient than that in L1688 \citep{Pat15}. Because of their various star-forming environments and proximity \citep[389$-$443 and 137~pc for the Orion A and Ophiuchus clouds, respectively;][]{Ort17,Kou18}, these clouds are ideal targets to compare the properties of the turbulence in the different star-forming environments.

The Orion A cloud has been mapped in various lines, including $J=$1$-$0 of $^{13}$CO \citep{Bal87,Tat93,Nag98,Rip13,Shi14,Kon18}, C$^{18}$O \citep{Shi11,Kon18}, and N$_2$H$^+$ \citep{Tat08,Nak19}. The Ophiuchus cloud has been mapped in the $J=$1$-$0 line of $^{13}$CO \citep{Lor89a,Rid06}, HCN \citep{Shi17}, HCO$^+$ \citep{Shi17}, and N$_2$H$^+$ \citep{Pan17}. However, most of the observations focused on the northern part of Orion A \citep[the ISF and L1641-N;][]{Tat93,Shi14,Kon18} or L1688 in Ophiuchus \citep{Pan17,Shi17}, which are the most active star-forming regions in each cloud. In addition, there are maps of the entire MCs, but these were done with larger beams and/or fewer transitions \citep{Bal87,Lor89a,Nag98}.

We performed a systematic observation toward the Orion A and Ophiuchus clouds in multiple molecular lines using the Taeduk Radio Astronomy Observatory \citep[TRAO;][]{Roh99,Jeo19} 13.7-m telescope. All spectral maps were obtained by the TRAO Key Science Program (KSP), ``mapping Turbulent properties In star-forming MolEcular clouds down to the Sonic scale" (TIMES; PI: Jeong-Eun Lee). Our program aims to obtain spectral maps of the entire Orion A and Ophiuchus clouds in multiple molecular lines in order to investigate the properties of turbulence in MCs that have different star-forming environments. This is the first observational study in multiple molecular lines that can trace various density environments \citep{Gac15} toward the entire area of the target clouds with a consistent observational scheme, high velocity resolution, and high sensitivity. We especially designed our observations to achieve uniform noise levels throughout the maps in order to calculate turbulence statistics. 

This first paper presents our observations and simple analyses. Further analysis of the turbulence in the clouds will be presented in the second paper. We describe the details of the observation program in Section \ref{Sec_obs}. In Section \ref{Sec_anal}, the method to produce moment 0, 1, and 2 maps with a high signal-to-noise ratio is described. We assess the uniformity of the data quality and morphological/kinematical features of the observed clouds in Section \ref{Sec_rst}, where we also compare the integrated intensities of the observed lines with column densities derived from the dust continuum. Section \ref{Sec_discs} discusses the physical properties of the line emitting gas in both clouds, and Section \ref{Sec_sum} presents the Summary.

\section{Observations} \label{Sec_obs}
\subsection{The TRAO 13.7~m Telescope} \label{Sec_TRAO}
We obtained six molecular line maps toward each cloud using the 13.7-m radio telescope at TRAO in Daejeon, South Korea. The SEQUOIA-TRAO receiver, which has 16-pixels arranged in a 4$\times$4 array, can obtain two molecular lines at 85$-$100 GHz or 100$-$115~GHz, simultaneously. TRAO also provides the On-The-Fly (OTF) observing mode, so that the combination of the simultaneous observation of two lines and the OTF mode with the multi-beam receiver makes the TRAO telescope an excellent facility to map multiple molecular transitions toward a large area efficiently. 

The backend is an FFT2G spectrometer that can accept the 32 IF outputs from SEQUOIA-TRAO. The FFT2G spectrometer has a bandwidth of 62.5~MHz with 4096 channels. Thus, its spectral resolution is about 15~kHz,  corresponding to a velocity resolution of about 0.04~km~s$^{-1}$ at 110~GHz. The main beam of the TRAO telescope has an almost circular pattern with a beam size of about $57\arcsec$ and $49\arcsec$ at 86 and 110~GHz, respectively \citep{Jeo19}.  

\subsection{Mapping the Orion A and Ophiuchus Clouds}
The Orion A and Ophiuchus clouds were divided into multiple 20$\arcmin$ $\times$ 20$\arcmin$ areas (submaps). The OTF mapping was performed toward each submap along with the right ascension and declination directions. Each OTF data possibly contains a \textit{scanning noise} \citep{Eme88}. The \textit{scanning noise} is manifested by noise features along the scanning direction which originate from the variation of weather conditions during the scanning process. We minimize these noise features by combining the OTF maps in R.A. and Dec. directions and achieve a uniform noise distribution on the covered area. The observed submaps were combined to build the spectral maps for the entire clouds. During the observation, the pointing uncertainty is less than 10$\arcsec$. The system noise temperature ranges from 250 to 400~K at 110~GHz and varies depending on the weather condition and elevation of the clouds.

Both clouds were mapped in six molecular lines that can trace the diffuse to dense gas in MCs: $^{13}$CO $J$=1$-$0, C$^{18}$O $J$=1$-$0, HCN $J$=1$-$0, HCO$^+$ $J$=1$-$0, N$_2$H$^+$ $J$=1$-$0, and CS $J$=2$-$1. Two of these lines were observed together; each of the $^{13}$CO/C$^{18}$O, HCN/HCO$^+$, and N$_2$H$^+$/CS line pairs was simultaneously observed. Table \ref{tbl_lines} shows the line frequency, velocity resolution ($\Delta V$), critical density ($n_\mathrm{crit}$), and main-beam efficiency for each observed line. Note that the main-beam efficiencies are obtained by interpolation of the efficiencies measured by \citet{Jeo19}. All spectral maps were observed from 2016 January to 2019 April.

\begin{figure}
\epsscale{0.7}
\plotone{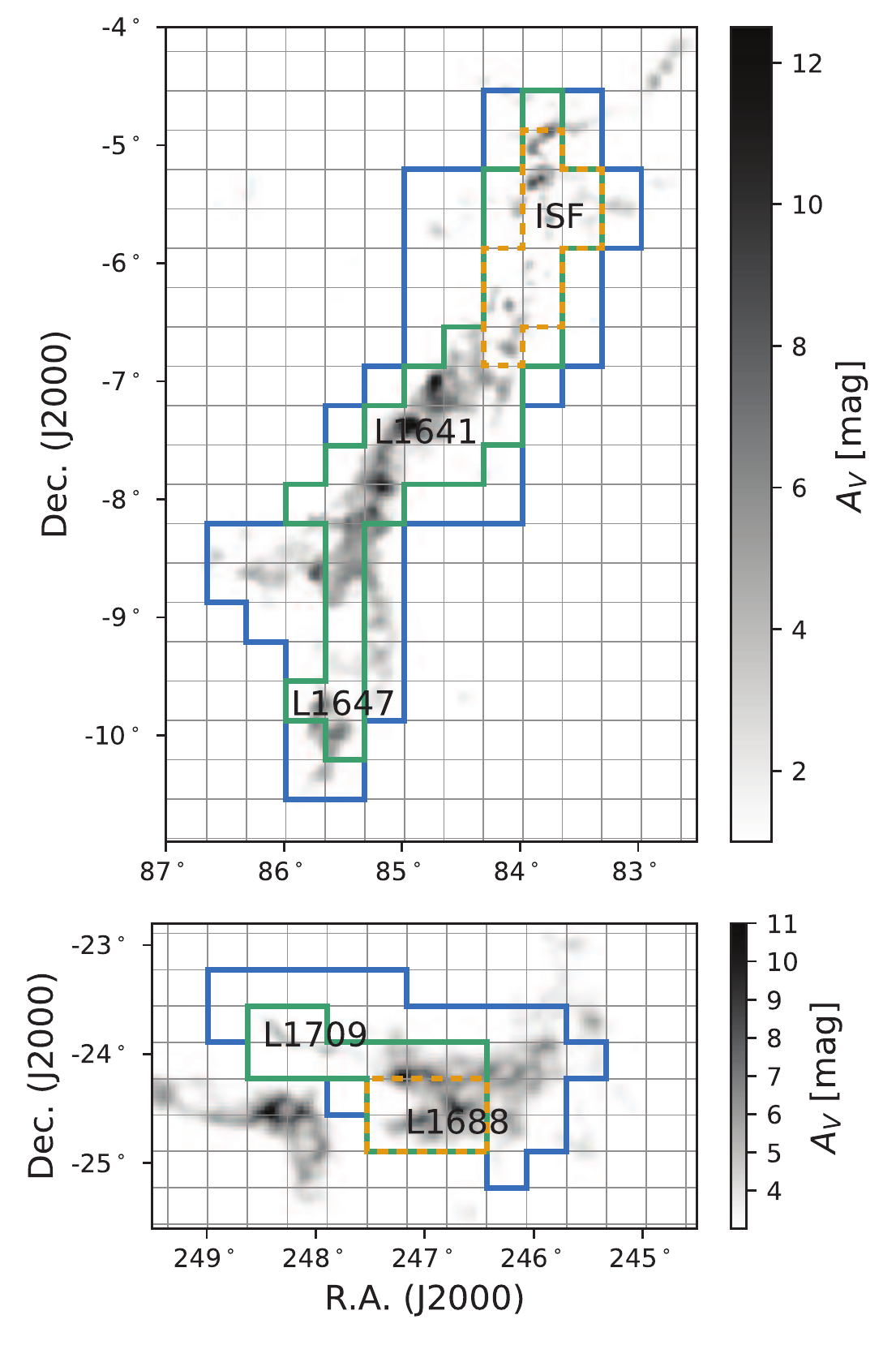}
\caption{Visual extinction map of the Orion A (top) and Ophiuchus (bottom) clouds provided by \citet{Dob05} overlaid with outlines showing 20$\arcmin$ $\times$ 20$\arcmin$ submaps (the gray vertical and horizontal lines). The mapped area in $^{13}$CO $J$=1$-$0, C$^{18}$O $J$=1$-$0, HCN $J$=1$-$0, and HCO$^+$ $J$=1$-$0 is presented in the blue solid line. The green solid line shows the area where N$_2$H$^+$ $J$=1$-$0 and CS $J$=2$-$1 are mapped. For the active star-forming regions, such as ISF, L1641-N cluster (in the Orion A cloud) and L1688 (in the Ophiuchus cloud) that outlined with the orange dotted lines, the N$_2$H$^+$ and CS lines are mapped more deeply. \label{fig_cloud_2MASS}}
\end{figure}

The observed areas toward the Orion A and Ophiuchus clouds are presented in Figure \ref{fig_cloud_2MASS}. The $^{13}$CO/C$^{18}$O and HCN/HCO$^+$ lines were mapped within the same area through the entire clouds. We obtained the visual extinction ($A_V$) maps provided by \citet{Dob05} and selected initially the submaps, which have $A_V$ higher than a certain value. We adopted $A_V$ of 2.0 and 4.0 for the Orion A and Ophiuchus clouds, respectively, which reasonably outline the structures of the clouds. The OTF observation was initially performed toward the selected submaps, and subsequently, we extended the observation if the $^{13}$CO line had been significantly detected on the boundary of the observed area. Total mapped areas in the $^{13}$CO/C$^{18}$O and HCN/HCO$^+$ lines are $\sim$ 8.7 deg$^2$ of the Orion A and 3.9 deg$^2$ of Ophiuchus clouds. 
For the N$_2$H$^+$/CS lines, observations were made toward the submaps where the observed C$^{18}$O map exhibits clump-like structures. The mapped areas are $\sim$ 4.0 deg$^2$ and $\sim$ 1.6 deg$^2$ in the Orion A and Ophiuchus clouds, respectively. We observed the N$_2$H$^+$/CS lines deeper than the other lines because of their weak line intensities. Moreover, we chose representative star-forming regions for each cloud: the selected regions are the ISF and L1641-N cluster in the Orion A cloud and the L1688 region in the Ophiuchus cloud. For these regions, we observed the N$_2$H$^+$/CS lines even deeper to obtain high-sensitivity spectral maps. The boundaries of the mapped areas are marked in Figure \ref{fig_cloud_2MASS}. The total observed time to obtain all data was about 1672 hours; 1073 hours for the Orion A and 549 hours for the Ophiuchus clouds.

The obtained maps were processed using the OTFTOOL and GILDAS/CLASS \footnote{\url{http://www.iram.fr/IRAMFR/GILDAS}} programs with a cell size of $20\arcsec$ and $\Delta V$ of about 0.1~km~s$^{-1}$. Baselines were removed using a least $\chi^2$ fitting method with a first-order polynomial. The baseline fitting is performed for the line-free spectra in the velocity ranges which are outside of the velocity windows ($V_\mathrm{win}$). To obtain good baseline, regions in velocity space ($V_\mathrm{space}$) that are almost three times broader than $V_\mathrm{win}$ are used. Some of the observed lines show broad wing structures toward OMC-1 where the energetic out-flowing source Orion KL is located (see Section \ref{Sec_Ori}). For these line spectra, we applied velocity windows broader than $V_\mathrm{win}$ of the other locations in the cloud. The $V_\mathrm{win}$ and $V_\mathrm{space}$ values for each spectral map are listed in Table \ref{tbl_vrng}. 

\section{Moment Maps with a Moment Masking Method} \label{Sec_anal}
We produced the moment 0, 1, and 2 maps for the observed lines which are equivalent to the maps of integrated intensity ($I_\mathrm{tot}$), intensity weighted mean velocity ($V_\mathrm{lsr}$), and velocity dispersion ($\sigma_V$). In this process, the moment masking method \citep{Dam11} is applied. The moment masking method is an efficient way to avoid the noise effect which degrades the signal-to-noise ratio of the moment maps. This method identifies emission-free pixels from the smoothed data and removes noise signals within the spectral cube data. The $I_\mathrm{tot}$ maps of the $^{13}$CO line in the Orion A and Ophiuchus clouds are presented in Figures \ref{fig_13CO_Ori} and \ref{fig_13CO_Oph}, respectively. The $V_\mathrm{lsr}$ and $\sigma_V$ maps for the $^{13}$CO line of the Orion A cloud are presented in Figures \ref{fig_Ori_13CO_mmt1} and \ref{fig_Ori_13CO_mmt2} while those for the Ophiuchus clouds are presented in Figures \ref{fig_Oph_13CO_mmt1} and \ref{fig_Oph_13CO_mmt2}, respectively. The other moment maps for the Orion A cloud are exhibited in Appendix \ref{App_mmtmaps_Ori}, and those for the Ophiuchus cloud are exhibited in Appendix \ref{App_mmtmaps_Oph}. 

We also calculated the uncertainties of the $I_\mathrm{tot}$, $V_\mathrm{lsr}$, and $\sigma_V$ values ($\epsilon_\mathrm{mom0}$, $\epsilon_\mathrm{mom1}$, and $\epsilon_\mathrm{mom2}$, respectively). The uncertainties were derived via the noise propagation. We measured $T_\mathrm{rms}$ from the line-free spectra of the regions outside $V_\mathrm{win}$ and adopted it as an uncertainty of the line intensity. The $\epsilon_\mathrm{mom0}$, $\epsilon_\mathrm{mom1}$, and $\epsilon_\mathrm{mom2}$ values are generally dominated by $T_\mathrm{rms}$ and the number of channels that are included in calculation. The contributions of the velocity uncertainty is minor compared to the other contributions.

For the HCN and N$_2$H$^+$ lines which have multiple hyperfine transitions, we consider a single transition line to derive the $V_\mathrm{lsr}$ and $\sigma_V$ correctly. For the HCN line, we adopted the strongest line at 88.631~GHz. We initially assumed that $V_\mathrm{lsr}$ of the 88.631~GHz line is the same as that of the CS line ($V^\mathrm{CS}_\mathrm{lsr}$). The moment values are derived from the velocity range between $V^\mathrm{CS}_\mathrm{lsr}$ $-$ 4 and $V^\mathrm{CS}_\mathrm{lsr}$ + 3~km~s$^{-1}$, which separate the hyperfine transitions of HCN. The hyperfine transitions of the HCN line are generally blended  in the ISF, therefore $\sigma_V$ in the ISF region would be underestimated. For the N$_2$H$^+$ line, we adopted an isolated transition at 93.176~GHz and derived the moment values from $V^\mathrm{CS}_\mathrm{lsr}$ $-$ 10.87 to $V^\mathrm{CS}_\mathrm{lsr}$ + 4.87~km~s$^{-1}$. Finally, we added 7.87~km~s$^{-1}$, which is the velocity difference between the rest frequency and the isolated hyperfine transition, to the derived $V_\mathrm{lsr}$.

\section{Results} \label{Sec_rst}
\subsection{Homogeneous Data Quality} \label{Sec_qual}
One way to obtain the properties of turbulence from spectral cube data is by statistical analyses, such as the probability distribution functions, two- or three-dimensional power spectra, and a wavelet transform of density or velocity fields \citep{Gil90,Kle00,Oss02,Kow07,Bur09,Koc17}. For some of these statistical analyses, it is quite important to have well characterized uncertainties, so we focus here on the data uniformity.

Figure \ref{fig_rms_PDF} displays the probability distribution functions (PDFs) of $T_\mathrm{rms}$. The $T_\mathrm{rms}$ PDF for each of the $^{13}$CO, C$^{18}$O, HCN, and HCO$^+$ lines have a well-defined single Gaussian-like distribution. The mean and standard deviation of $T_\mathrm{rms}$ are given in Figure \ref{fig_rms_PDF}. For the N$_2$H$^+$ and CS lines, the $T_\mathrm{rms}$ PDFs of the N$_2$H$^+$ and CS lines have a pair of Gaussian-like components because of the high-sensitivity maps toward the selected star-forming regions. The filled histograms on the bottom panel are the PDFs for the high-sensitivity maps. The mean $T_\mathrm{rms}$ values for the N$_2$H$^+$ and CS lines range from 0.10 to 0.11~K, and their standard deviations are less than 0.025~K. For the selected star-forming regions, the means and standard deviations of $T_\mathrm{rms}$ are $\sim$ 0.09 and 0.006~K, respectively. 

These small standard deviation values imply a homogeneous $T_\mathrm{rms}$ in the spectral maps. The $^{13}$CO, C$^{18}$O, HCN, and HCO$^+$ data have similar mean $T_\mathrm{rms}$ values, and each of them has a uniform $T_\mathrm{rms}$ distribution across the observed area. Also, the PDFs of $T_\mathrm{rms}$ for the N$_2$H$^+$ and CS data imply that $T_\mathrm{rms}$ does not significantly vary within each of the selected star-forming regions and the other areas. 

\subsection{The Morphological and Kinematical Features of the Clouds}
\subsubsection{The Orion A Cloud} \label{Sec_Ori}
\begin{figure}
\epsscale{1.0}
\plotone{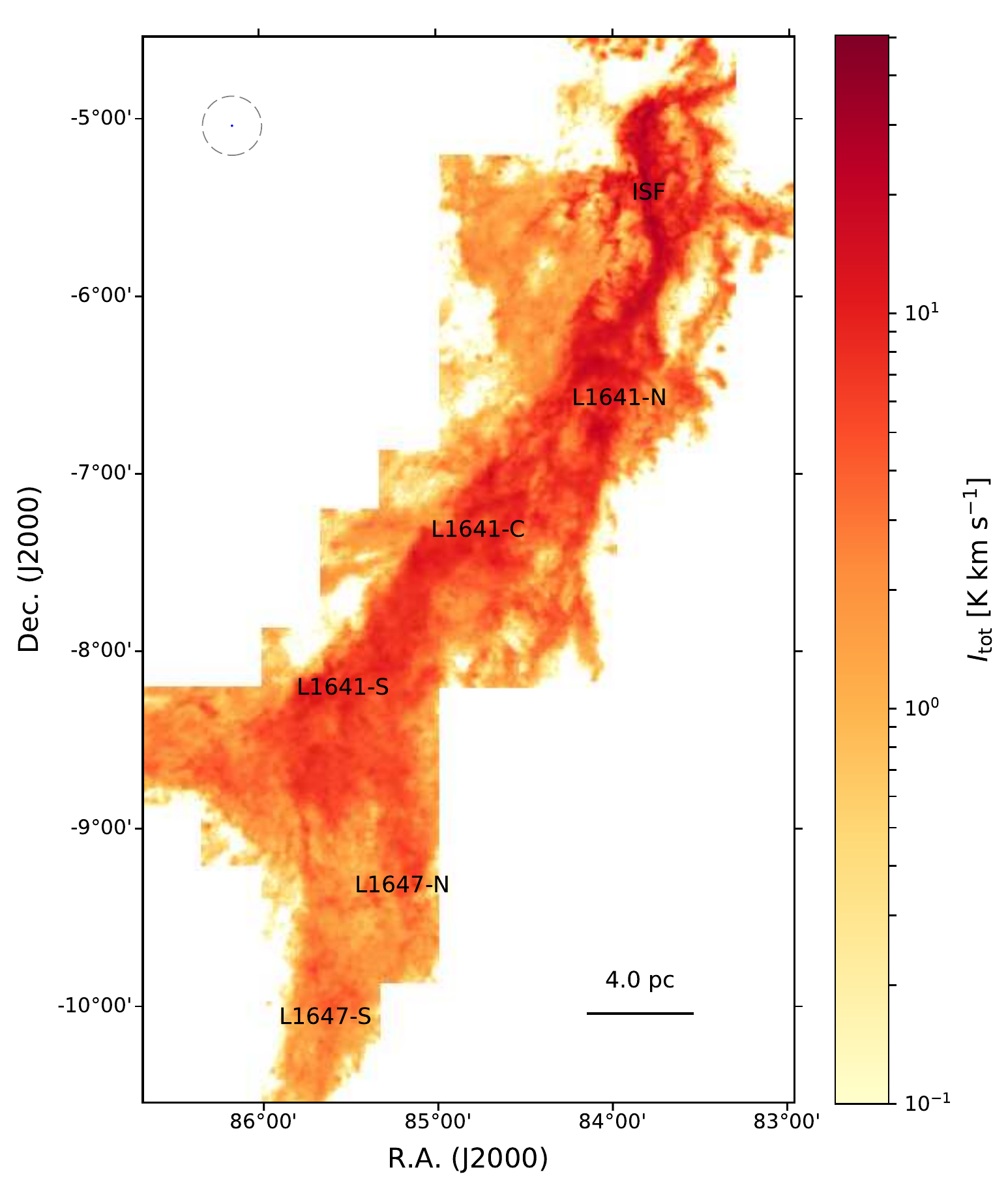}
\caption{The moment 0 (the equivalent of integrated intensity; $I_\mathrm{tot}$) map of the $^{13}$CO line toward the Orion A cloud. The blue circle in the gray dashed circle on the upper-left corner indicates the beam size of the TRAO telescope at 110~GHz. We annotated the $^{13}$CO maps with the names of the sub-regions \citep{Lyn62,Mei16,Gro18}. \label{fig_13CO_Ori}}

\end{figure}

In the Orion A cloud, spatial distributions of the observed lines are generally well correlated. All the lines follow a filamentary structure extending from the north to the south (from ISF to L1647-S). The $I_\mathrm{tot}$ in all lines are generally strong in ISF (Dec. $>$ $-$6.2$\degr$). Otherwise, all the lines in the southern filamentary structure are weaker than those in the ISF. Also the HCN, HCO$^+$, N$_2$H$^+$, and CS lines are not detected in the L1647 region (Dec. $<$ $-$9$\degr$; see Figure \ref{fig_13CO_Ori}, where these various regions are identified).

The observed lines reveal various structures within the Orion A cloud. The C$^{18}$O, HCN, HCO$^+$, N$_2$H$^+$, and CS lines are detected toward the regions where the $I_\mathrm{tot}$ of $^{13}$CO is strong. Among these lines, the C$^{18}$O and N$_2$H$^+$ lines show clumpy structures while the CS line shows relatively extended structures across the Orion A cloud. The HCN and HCO$^+$ lines reveal extended filamentary structures in ISF, while they reveal clumpy structures in the other regions. 

The spatial distributions of the C$^{18}$O, HCN, HCO$^+$, N$_2$H$^+$, and CS lines are well correlated with that of young embedded protostars (Class 0/I young stellar objects and flat-spectrum sources) identified using  \textit{Spitzer} \citep{Meg12} and \textit{Herschel} observations \citep{Fur16}. Especially, the C$^{18}$O line presents a tight correlation between the spatial distribution of $I_\mathrm{tot}$ and the embedded protostars (Figure \ref{fig_Ori_ctlg}). Also, the HCN, HCO$^+$, and CS lines exhibit a similar spatial distribution (Figures \ref{fig_Ori_HCN}. \ref{fig_Ori_HCOp}, and \ref{fig_Ori_CS}). These lines are mainly found in the active star-forming regions (ISF and star-forming clusters) and Herbig-Haro objects (HH1, HH2, and HH43). 

Figures \ref{fig_Ori_13CO_mmt1} and \ref{fig_Ori_13CO_mmt2} present the $V_\mathrm{lsr}$ and $\sigma_V$ maps of the $^{13}$CO line in the Orion A cloud. The $V_\mathrm{lsr}$ map of the Orion A cloud exhibits a global velocity gradient from the north to the south \citep{Hey92,Tat93,Ike07,Shi11,Kon18}. This global velocity gradient seems to be the motion of the overall Orion A cloud and could represent large-scale rotation \citep{Bal87}, expansion \citep{Kut77,Mad86}, or gravitational collapse \citep{Har07}. Statistical analyses without considering the overall motion of MC can cause misunderstanding of turbulence. Therefore, we should take the motion of the Orion A cloud into account when investigating the properties of turbulence in future studies. The $\sigma_V$ ranges from 0.2 to 2.0~km s$^{-1}$ throughout the Orion A cloud except for some regions with the high-$\sigma_V$ values. These regions are located in the eastern part of ISF and the L1647-N region. In these regions, there are multiple cloud components with different $V_\mathrm{lsr}$. We will discuss these high-$\sigma_V$ regions in Appendix \ref{App_highdv}.

Another notable feature is a broad wing structure in the observed lines toward OMC-1 \citep{Kui80,Ryd81,Olo82,Has84}. The $^{13}$CO, HCN, HCO$^+$, and CS lines present the blue and red shifted broad wing structures (see Figure \ref{fig_KL_lines}). For the HCN, HCO$^+$ and CS lines, the broad wing structures result in very high $I_\mathrm{tot}$ values: the $I_\mathrm{tot}$ toward OMC-1 are 241, 71, and 51~K~km s$^{-1}$ in the HCN, HCO$^+$, and CS lines, respectively. The $\sigma_V$ values for the HCO$^+$ and CS lines are also high near OMC-1, while the $\sigma_V$ value for the HCN line is not significantly high because of the limited velocity range that we adopted. 

The C$^{18}$O and N$_2$H$^+$ lines do not present clear wing structures in their spectra toward OMC-1. We compared the integrated intensities of the central peak ($I_\mathrm{center}$) and the broad wing structures ($I_\mathrm{wing}$) to check whether or not the weak broad structures exist in the C$^{18}$O and N$_2$H$^+$ lines. The velocity ranges that $I_\mathrm{center}$ and $I_\mathrm{wing}$ are derived over are summarized in Table \ref{tbl_vrng_wing}. The observed lines generally peak at the velocity of 9~km~s$^{-1}$. We thus derived $I_\mathrm{center}$ over a velocity range from 5 to 13~km~s$^{-1}$ assuming that the central peak extends up to $\pm$4~km~s$^{-1}$ from the line center. $I_\mathrm{wing}$ was calculated over the velocity ranges from -11 to 5 and from 13 to 29~km~s$^{-1}$ assuming that the wings extend up to $\pm$20~km~s$^{-1}$ from the line center. For the HCN and N$_2$H$^+$ lines, the velocity ranges for obtaining $I_\mathrm{center}$ were set to cover all hyperfine components. Also, we set the velocity ranges of the wing structures to consider the broad emission features at the outermost parts of the observed lines. 

Figure \ref{fig_KL_ratio} presents the $I_\mathrm{center}$ and $I_\mathrm{wing}$ for each line. $I_\mathrm{wing}$ values are generally proportional to $I_\mathrm{center}$. The $^{13}$CO, HCN, HCO$^+$, and CS lines have relatively strong $I_\mathrm{wing}$ values that are higher than 10~K~km~s$^{-1}$. $I_\mathrm{wing}$ of C$^{18}$O is barely detected (1.1 $\pm$ 0.3~K~km~s$^{-1}$). Only the N$_2$H$^+$ line is not detected with a value of $-$0.16 $\pm$ 0.17~K~km~s$^{-1}$. Figure \ref{fig_broad_map} presents the distributions of the broad wing emission in $^{13}$CO, C$^{18}$O, HCN, HCO$^+$ and CS. Note that the distribution of wing emission in C$^{18}$O is indistinguishable from the noise.

\subsubsection{The Ophiuchus Cloud} \label{Sec_Oph}
\begin{figure}
\epsscale{1.0}
\plotone{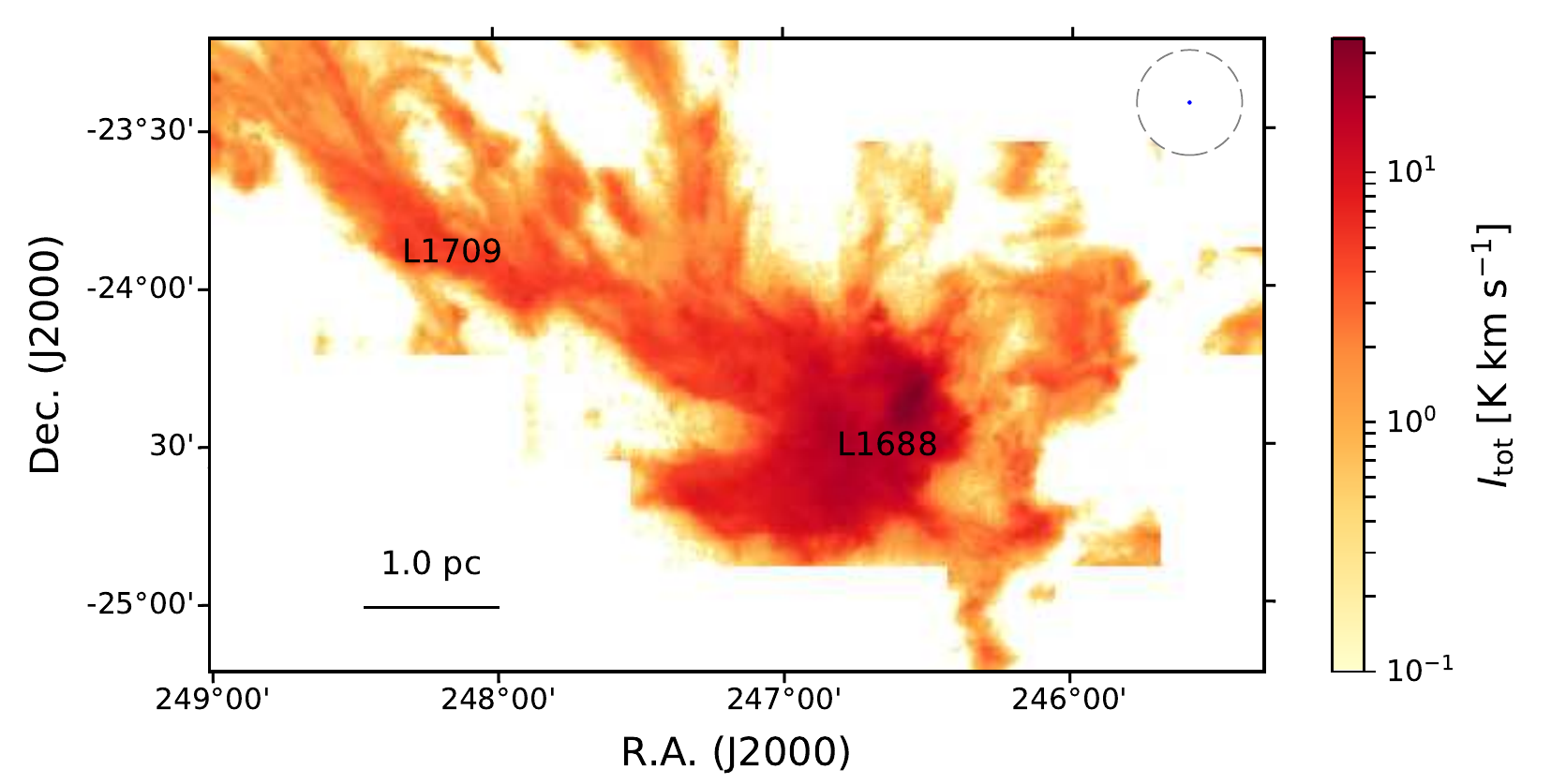}
\caption{Same as Figure \ref{fig_13CO_Ori} but for the Ophiuchus cloud. The beam size is shown in the upper-right corner of the map. The map is annotated with the names of the sub-regions \citep{Lyn62,Lor89a}. \label{fig_13CO_Oph}}
\end{figure}

\begin{figure}
\epsscale{1.0}
\plotone{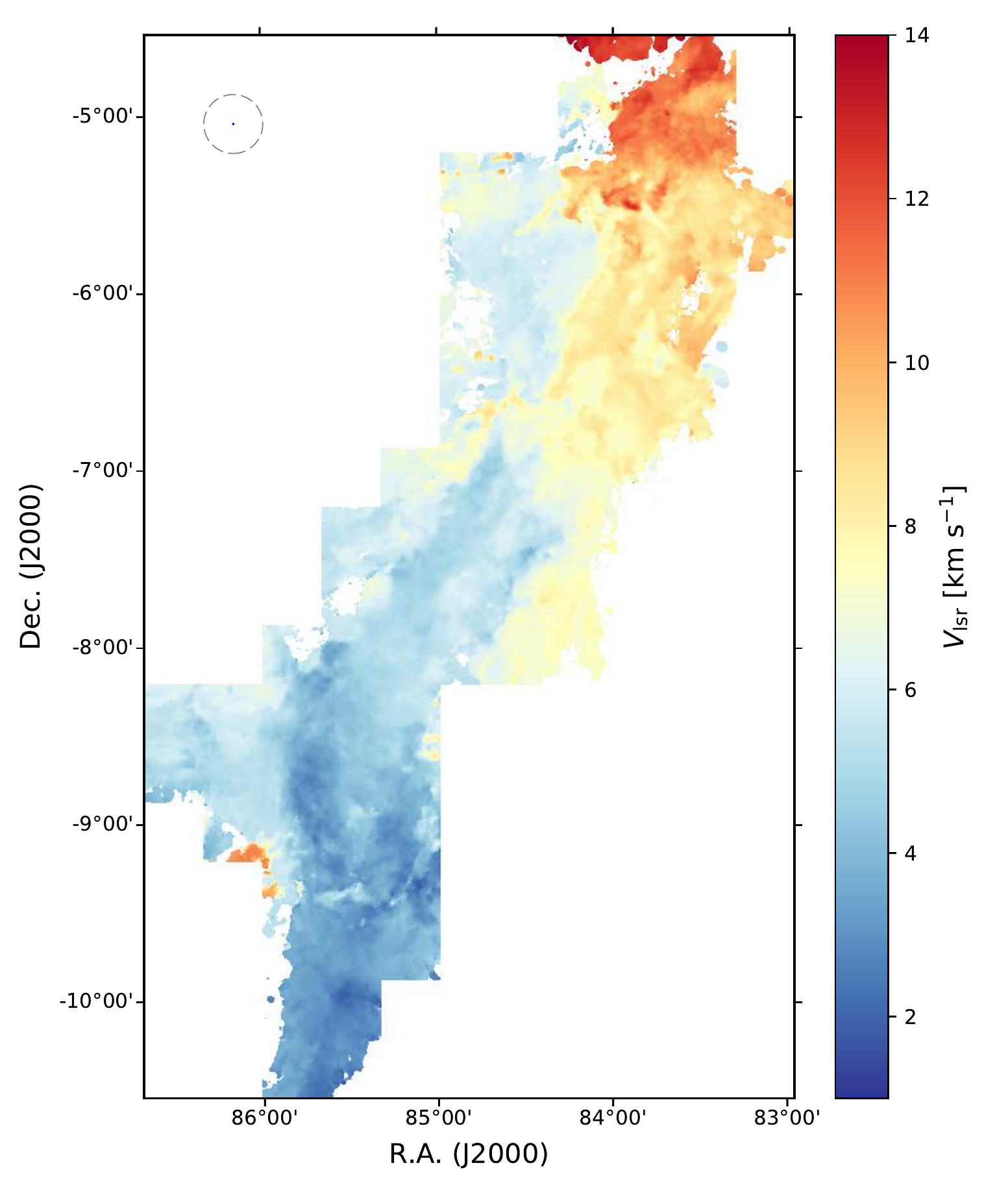}
\caption{The moment 1 (the equivalent of mean line of sight velocity; $V_\mathrm{lsr}$) map for the $^{13}$CO line in the Orion A cloud. \label{fig_Ori_13CO_mmt1}}
\end{figure}

In the Ophiuchus cloud, the observed lines exhibit spatial distribution trends that are similar to those in the Orion A cloud. The $^{13}$CO line traces extended cloud structures from the L1688 to L1709 regions (the regions of the cloud are identified in Figure \ref{fig_13CO_Oph}). The other lines mainly trace small and clumpy structures in the cloud. The line intensities are generally strong in L1688 (R.A. $<$ 247.5), which is the most active star-forming region in this cloud. Also, the spatial distribution of the C$^{18}$O line emission is well correlated with that of the young embedded protostars identified using  \textit{Spitzer} \citep[][ see Figure \ref{fig_Oph_ctlg}]{Dun15}.  

In L1688, the observed lines are generally strong toward the star-forming cores (Oph-A through L; see Figure \ref{fig_Oph_N2Hp}). But, their relative strengths change depending on the lines. The star-forming cores have non-uniform conditions \citep{Pat15,Pun16}. Oph-A and -C are affected by the external heating from the B2V star HD 147889 \citep{Pat15,Pun16}. Oph-B1 and -B2 cores are the coldest among the cores \citep{Pat15}; Oph-B1 is one of the quiescent cores while Oph-B2 is the most turbulent core \citep{Pun16}. Oph-E and -F cores are the most evolved regions in L1688 \citep{Pat15}; Oph-E is strongly pressure-confined while Oph-F is marginally pressure-confined. Also, Oph-F has a similar temperature to that of Oph-A without any external heating. These different environments result in the different relative strength of the observed lines. 

The $I_\mathrm{tot}$ of the $^{13}$CO line in Oph-A is stronger than that in Oph-C while that of C$^{18}$O in Oph-A is similar to that in Oph-C. \citet{Lad80} found that $^{13}$CO $J=$1$-$0 is optically thick with a self-absorption feature in some positions. We thus investigated the $^{13}$CO and C$^{18}$O lines toward the dense cores where the $^{13}$CO line can probably be optically thick. Figure \ref{fig_Oph_lines} presents the $^{13}$CO and C$^{18}$O line spectra toward the 13 cores in L1688 \citep{Pan17} and two DCO$^+$ cores in L1709 \citep{Lor90}. Some of the dense cores present self-absorption features in their $^{13}$CO line spectra. 

Figures \ref{fig_Oph_13CO_mmt1} and \ref{fig_Oph_13CO_mmt2} show the $V_\mathrm{lsr}$ and $\sigma_V$ maps for the $^{13}$CO line in the Ophiuchus cloud. The $V_\mathrm{lsr}$ value varies from $+$1.5 to $+$4.0~km s$^{-1}$ across the Ophiuchus cloud. The $V_\mathrm{lsr}$ map shows that the L1688 and L1709 regions have different $V_\mathrm{lsr}$. The $V_\mathrm{lsr}$ values of L1688 are around 3.5~km~s$^{-1}$ while that in L1709 is around 2.5~km~s$^{-1}$. From this result, \citet{Lor89b} suggested that L1709 may be separate from L1688. Neither L1688 nor L1709 regions have any overall motions in the $^{13}$CO line. We also checked the optically thinner C$^{18}$O line, but there is no overall motion in each region (see Appendix \ref{App_random}). The $\sigma_V$ is about 0.5~km~s$^{-1}$ across the Ophiuchus cloud. 

The moment maps of the $^{13}$CO line imply that the kinematic features of the Ophiuchus cloud are quite different from those of the Orion A cloud. The systematic variation of the $V_\mathrm{lsr}$ in the Ophiuchus cloud is relatively small compared to that in the Orion A cloud. Also, the typical $\sigma_V$ value in the Ophiuchus cloud is smaller than that in the Orion A cloud. The small variation of $V_\mathrm{lsr}$ and small $\sigma_V$ values imply that the Ophiuchus cloud is kinematically quiescent compared to the Orion A cloud \citep{Lor89b}.

The difference between the spatial distributions of the HCN and HCO$^+$ lines is striking (see Figures \ref{fig_Oph_HCN} and \ref{fig_Oph_HCOp}). The HCN line is mainly detected toward Oph-A, -B1, -B2, and -B3 cores while the HCO$^+$ line is predominantly detected toward Oph-A, -C, -E, and -F cores. The peak $I_\mathrm{tot}$ of the HCN and HCO$^+$ lines also appear in different cores. This result  indicates that the HCN and HCO$^+$ lines trace different physical or chemical conditions in the Ophiuchus cloud. 

\subsection{Column Density Maps and $I_\mathrm{tot}$ Variations with Column Density} \label{Sec_cold}
Unbiased mapping toward two MCs in multiple molecular lines provides a good opportunity to assess how the line intensities respond to the physical parameters within clouds. Therefore, we investigated the variation of $I_\mathrm{tot}$ as a function of column density ($N_\mathrm{H_2}$). 

We derived $N_\mathrm{H_2}$ from the observations of dust continuum emission that can trace the amount of gas in a cloud \citep{Goo09}. The $N_\mathrm{H_2}$ maps were derived by fitting a modified blackbody (MBB) into the spectral energy distribution (SED) of the continuum emission from cold dust. Archival \textit{Herschel} PACS (160~$\mu$m) and SPIRE (250, 350, and 500~$\mu$m) continuum observations that were obtained as part of the \textit{Herschel} Gould Belt Survey \citep{And10} are adopted. Note that the PACS 100 and 70~$\mu$m observations are not included in the SED fitting. The 100~$\mu$m continuum observation was not covered by the \textit{Herschel} Gould Belt Survey \citep{And10}. For the 70~$\mu$m band, the MBB fitting with a single temperature cannot fit the observed emission in some cases because of the contamination due to non-equilibrium emission from small dust grains \citep{Roy13}. Even if the detailed dust model is applied to the SED fitting including the 70~$\mu$m data, the derived column density is not significantly different from that derived using the single temperature MBB fitting without the 70~$\mu$m data \citep{Bia13}.

The continuum emission maps were calibrated using \textit{Planck} observations of the same regions via the method in \citet{Che19}. Using the calibrated data, the column density maps were derived in two steps \citep{Fri17,Che19}: (1) dust temperature ($T_\mathrm{d}$) and optical depth ($\tau_{\nu_0}$) maps were found via the SED fitting, and (2) the $N_\mathrm{H_2}$ maps were obtained from the $\tau_{\nu_0}$ map by multiplying by a conversion factor assuming a gas-to-dust mass ratio of 100, 
\begin{equation}
\frac{N_\mathrm{H_2}}{\tau_\mathrm{\nu_0}} = \frac{1}{100\ \kappa_\mathrm{\nu_0}\ \mu_\mathrm{H_2}\ m_\mathrm{H}},
\end{equation} 
where $\kappa_\mathrm{\nu_0}$ is the opacity of 0.1~cm$^2$~g$^{-1}$ at $\nu_0$ of 1000~GHz \citep{Hil83}, $\mu_\mathrm{H_2}$ is a mean molecular weight per H$_2$ molecule of 2.8, and $m_\mathrm{H}$ is the mass of a hydrogen atom. The final map has a beam size of 36$\arcsec$, which is the resolution of the SPIRE 500~$\mu$m data. We thus convolved the $N_\mathrm{H_2}$ map to have a resolution of 50$\arcsec$, which is comparable to the beam sizes of the TRAO maps (see Table \ref{tbl_lines}).

The $I_\mathrm{tot}$ of each observed line generally increases as $N_\mathrm{H_2}$ increases up to a certain $N_\mathrm{H_2}$ and remains relatively constant after that certain $N_\mathrm{H_2}$ (see Figures \ref{fig_line_cold_Ori} and \ref{fig_line_cold_Oph} for the Orion A and Ophiuchus clouds, respectively). We fit the $I_\mathrm{tot}$ variation with a power-law relation to characterize how $I_\mathrm{tot}$ of a molecular line changes as $N_\mathrm{H_2}$ increases. In this process, we divide the Orion A cloud into two sub-regions, the ISF region (Dec. $>$ $-6\degr$) and the rest of the cloud (Dec. $<$ $-6\degr$) because ISF is more likely to be affected by the PDR \citep{Shi14}. Hereafter, ISF, the other regions in the Orion A cloud, and the Ophiuchus cloud are referred as the ISF, L1641, and Ophiuchus regions.

For each line in each of the ISF, L1641, and Ophiuchus regions, the mean values of the $I_\mathrm{tot}$, peak temperature ($T_\mathrm{peak}$), $T_\mathrm{d}$, and $\sigma_V$ for a given $N_\mathrm{H_2}$ are investigated. The line spectra were separated into regularly spaced $\log(N_\mathrm{H_2})$-bins with a size of $\Delta \log(N_\mathrm{H_2})$ $=$ 0.1. We derived the  mean values of $I_\mathrm{tot}$ and $\sigma_V$ for each bin, weighting the values by the inverse of the square of the uncertainty. For the $T_\mathrm{peak}$ and $T_\mathrm{d}$, the arithmetic mean values are adopted. Note that the variation of $T_\mathrm{d}$ seems to be relatively constant because it generally varies from 12 to 20~K in most areas except near the heating sources (such as OMC-1 and NGC 1977 in the Orion A cloud and HD147889 in the Ophiuchus cloud). 

We investigated the power-law indices which can explain the variation of the mean $I_\mathrm{tot}$ values ($\bar{I}_\mathrm{tot}$). We fitted the data with a power law: $\log I_\mathrm{fit} = \log b + {\alpha}\log N_\mathrm{H_2}$. The $\chi^2$ minimization technique was used to obtain the fit parameters. The $\chi^2$ value is defined as follows, 
\begin{equation}
\chi^2 = \sum (\frac{\bar{I}_\mathrm{tot}-I_\mathrm{fit}}{\bar{\epsilon}_\mathrm{mom0}})^2, 
\end{equation} 
where the sum is over the bin numbers, $I_\mathrm{fit}$ is an expected $I_\mathrm{tot}$ value from the power-law fit, and $\bar{\epsilon}_\mathrm{mom0}$ is the uncertainty of $\bar{I}_\mathrm{tot}$. For C$^{18}$O, HCN, HCO$^+$, N$_2$H$^+$, and CS, there are several data points with very weak and relatively constant $I_\mathrm{tot}$ at a low-$N_\mathrm{H_2}$ regime ($N_\mathrm{H_2}$ $<$ 2$\times$10$^{21}$~cm$^{-2}$). These data points are the weighted means of a few pixels. Their line spectra are generally dominated by the noise although the moment masking method identified them as emission lines. Therefore, we excluded these data points from the power-law fitting. The $N_\mathrm{H_2}$ ranges for power-law fits were visually defined. The $N_\mathrm{H_2}$ ranges for the fitting and the best-fit power-law indices ($\alpha$) are listed in Table \ref{tbl_fit}.

Figure \ref{fig_line_cold_CO} shows the $I_\mathrm{tot}$ variation of the $^{13}$CO and C$^{18}$O lines. For these lines, we divided the $N_\mathrm{H_2}$ range into three regimes: (1) the low-$N_\mathrm{H_2}$ regime where $I_\mathrm{tot}$ steeply increase ($\alpha$ $>$ 2.0), (2) the intermediate-$N_\mathrm{H_2}$ regime where $I_\mathrm{tot}$ is proportional to $N_\mathrm{H_2}$ ($\alpha$ $\sim$ 1.0), and (3) the high-$N_\mathrm{H_2}$ regime where $I_\mathrm{tot}$ becomes relatively constant ($\alpha$ $<$ 1.0). The detailed $N_\mathrm{H_2}$ ranges are different depending on the lines and regions, however, their $I_\mathrm{tot}$ variations are similar. 

Figure \ref{fig_line_cold_HC} shows the $I_\mathrm{tot}$ variations in HCN, HCO$^+$, and CS lines. The variations of $I_\mathrm{tot}$ are quite different depending on the lines and regions and cannot be explained with a single trend. Their $I_\mathrm{tot}$ variations should be interpreted individually. The $I_\mathrm{tot}$ variation of the N$_2$H$^+$ line is presented in Figure \ref{fig_line_cold_NH}. The N$_2$H$^+$ line is only detected at the high-$N_\mathrm{H_2}$ regions where $N_\mathrm{H_2}$ is higher than 10$^{22}$~cm$^{-2}$. We will discuss these results in Sections \ref{Sec_vari} and \ref{Sec_dv}.

\section{Discussion} \label{Sec_discs}
\subsection{Broad Wing Structures in the Orion KL Spectra} 
As noted above, the Orion-KL region shows wide line wings in $^{13}$CO, HCN, HCO$^+$, and CS, marginally in C$^{18}$O, and not in N$_2$H$^+$. \citet{Wan77} suggested that $^{12}$CO J=1-0 is optically thin in the wings. The $^{13}$CO and C$^{18}$O lines thus would be optically thinner at the line wings. In this case, the intensity ratio between the $^{13}$CO and C$^{18}$O lines is probably close to their abundance ratio \citep{Wan77}. The ratio between $I_\mathrm{wing}$ of the $^{13}$CO and C$^{18}$O lines is 10.5 $\pm$ 2.8. This value is close to the abundance ratio determine by \citet{Shi14} ($X_\mathrm{^{13}CO}/X_\mathrm{C^{18}O}$ = 12.14) within a 1-$\sigma$ range.

The ratios between $I_\mathrm{wing}$ and $I_\mathrm{center}$ for the HCN, HCO$^+$ and CS lines are slightly different. The origins of the broad wing structures were discussed in previous works \citep{Kui80,Ryd81,Olo82,Has84}. The bipolar outflows aligned with the line of sight are suggested as the origin of the high velocity wings \citep{Olo82}. In this case, the high velocity emission is spatially confined close to Orion KL. For the HCO$^+$ line, the contribution of the shocked gas is also suggested \citep{Ryd81,Joh84,Olo82}. In addition, the abundance enhancement of HCN and HCO$^+$ in the high velocity components is also reported \citep{Ryd81,Joh84}.  \citet{Has84} analyzed the CS line and suggested that the existence of a large gas disk around Orion KL nebula. They also mentioned that the high velocity emission corresponding to the bipolar outflows was not observed in CS. These results may be related to the differences in ratios between $I_\mathrm{wing}$ and $I_\mathrm{center}$.

\subsection{Variation of $I_\mathrm{tot}$ as a Function of $N_\mathrm{H_2}$} \label{Sec_vari}
If {\it all} the following assumptions are true (the lines are optically thin, stimulated radiative processes can be ignored, excitation temperatures are much lower than the kinetic temperature ($T_\mathrm{K}$), and the abundance of the species is constant), the line intensity is proportional to $n^2$ because every collision leads to a photon. If the line of sight depth is constant, the line intensity is also proportional to $N_\mathrm{H_2}^2$, as is generally true for optical or infrared emission lines. All these conditions are rarely met for millimeter-wave molecular emission lines. The fact that the excitation temperature is limited below by the radiation temperature and above by the kinetic temperature constrains the regime of densities over which every collision leads to a photon. Optical depth increasing with $N_\mathrm{H_2}$ also leads intensity to increase more slowly with $N_\mathrm{H_2}$. For lines that reach optical depth near unity and excitation temperatures near the kinetic temperature, the intensity should plateau with $T_\mathrm{peak} \approx T_\mathrm{K}$. The dependence of $I_\mathrm{tot}$ on $N_\mathrm{H_2}$ is a complicated function of density, temperature, abundance, and velocity field. 

Examination of Figures \ref{fig_line_cold_CO} and \ref{fig_line_cold_HC} and the fits in Table \ref{tbl_fit} show broad agreement with these predictions. At low $N_\mathrm{H_2}$, the fits to intensity versus $\log (N_\mathrm{H_2})$ are super-linear, but they become closer to linear at higher $N_\mathrm{H_2}$ before reaching plateaus at levels that depend on $T_\mathrm{K}$. The middle plots of Figures \ref{fig_line_cold_CO} and \ref{fig_line_cold_HC} show $T_\mathrm{peak}$ and $T_\mathrm{d}$ from Herschel. For densities above a few times 10$^4$ cm$^{-3}$, $T_\mathrm{K} \approx T_\mathrm{d}$. For $^{13}$CO, $T_\mathrm{peak}$ indeed levels off near $T_\mathrm{d}$, as predicted for optically thick, thermalized lines. None of the other transitions reach this point, so the fits to their slopes indicate that they lie primarily in the intermediate zone between growing like $N_\mathrm{H_2}^2$ and the plateau. N$_2$H$^+$ is the exception, with slope near the predicted value of 2 over a substantial range of $N_\mathrm{H_2}$ in the ISF, where $T_\mathrm{K}$ is relatively high.

\citet{kau17} analyzed the $^{13}$CO, C$^{18}$O, HCN, and N$_2$H$^+$ lines in the northern part of ISF and presented the normalized line-to-mass ratio as a function of the visual extinction ($A_\mathrm{V}$) derived from Herschel column density. Figure 2 in \citet{kau17} showed that there is a certain $A_\mathrm{V}$ regime where the line-to-mass ratios remain relatively constant which means that the $I_\mathrm{tot}$ is proportional to $N_\mathrm{H_2}$. Table \ref{tbl_kauff} shows the $A_\mathrm{V}$ regimes where the line-to-mass ratio remains constant and corresponding $N_\mathrm{H_2}$ values for each line for the lines studied by  \citet{kau17} and the corresponding values for our study.

For the $^{13}$CO, C$^{18}$O, and N$_2$H$^+$ lines, the $N_\mathrm{H_2}$ regime where $\alpha$ of about 1.0 is generally consistent with that from \citet{kau17}. The $N_\mathrm{H_2}$ regimes for $\alpha \sim$ 1.0 that are derived in this study are larger than those of \citet{kau17}. For the HCN line, the $N_\mathrm{H_2}$ regime is quite different from what \citet{kau17} presented. \citet{kau17} only adopted a small region north of -5:10:00 Dec. (J2000) to avoid the effect of radiation which is emitted from the Orion Nebula. However, the whole region north of -6:00:00 Dec. (J2000) including OMC-1 is included in this study. The different results for column density regimes might be caused by the difference in the areas that are included in each analysis. 

\subsubsection{$^{13}$CO $J=$1$-$0} \label{Sec_13CO}
The $^{13}$CO line has $\alpha > 2$  where $N_\mathrm{H_2}< 10^{22}$~cm$^{-2}$. The uncertainty of $\alpha$ in these regimes is larger than that in the other regimes. Thus, detailed interpretation of $I_\mathrm{tot}$ variation using $\alpha$ could be uncertain.
The $\alpha$ values in intermediate-$N_\mathrm{H_2}$ regimes are close to 1.0 and become shallower in the high-$N_\mathrm{H_2}$ regime ($\alpha$ $\sim$ 0.3). This result indicates that the $^{13}$CO lines in both clouds are optically thick toward high-$N_\mathrm{H_2}$ regions and cannot trace all the gas along the line of sight. In the Ophiuchus cloud,  self-absorption features appear in the $^{13}$CO line spectra toward the star-forming cores. These results are consistent with previous studies showing that $^{13}$CO line can be optically thick in the Orion A \citep{Shi14} and Ophiuchus \citep{Lad80} clouds. 

In this analysis, we neglected the effect of abundance variation within MCs. In fact, the $I_\mathrm{tot}$ variation of the $^{13}$CO line also can be affected by the chemical difference between the sub-regions. The ISF region is also affected by the photon-dominated regions (PDRs) heated by the Trapezium stars \citep{Shi14}. In cold dense regions, $^{13}$CO can freeze onto dust grains.  

\subsubsection{C$^{18}$O $J=$1$-$0} \label{Sec_C18O}
The observed C$^{18}$O lines preferentially trace  regions of higher column density than are traced by the $^{13}$CO line. This is understood as the effect of lower abundance \citep{Wil94} which translates into lower optical depth and less radiative trapping. When $N_\mathrm{H_2}$ is smaller than 3$\times$10$^{22}$~cm$^{-2}$, only a few weak emission lines were detected. If we observe both clouds more deeply, the $I_\mathrm{tot}$ variation in very low $N_\mathrm{H_2}$ regimes would be accessible.

In the intermediate-$N_\mathrm{H_2}$ regime, where $N_\mathrm{H_2}$ is between 1.6$\times$10$^{22}$ and 5.0$\times$10$^{22}$~cm$^{-2}$, the $\alpha$ values are about one. If $N_\mathrm{H_2}$ exceeds 5.0$\times$10$^{22}$~cm$^{-2}$, the slopes become shallower. In the ISF and Ophiuchus regions, $\alpha$ becomes comparable to zero. For the ISF region, \citet{Shi14} found that C$^{18}$O $J=$1$-$0 is optically thin, and the abundance of C$^{18}$O would be affected by the selective photodissociation in PDR chemistry. Abundance variations caused by photodissociation or freeze-out \citep{Cas99} can also affect the line emission. In the Ophiuchus region, $I_\mathrm{tot}$ sharply increases beyond $N_\mathrm{H_2}$ of 10$^{23}$~cm$^{-2}$. This increase results from a selection bias; the highest $N_\mathrm{H_2}$ bin originates only from the Oph-A core that has the highest $N_\mathrm{H_2}$ and $T_\mathrm{d}$ within the Ophiuchus region.

\subsubsection{HCN $J=$1$-$0 and HCO$^+$ $J=$1$-$0} \label{Sec_HCN_HCOp}
The $I_\mathrm{tot}$ for the HCN and HCO$^+$ lines shows different variation depending on the lines and regions. Also, the $\alpha$ for a given line and region changes in different ways depending on the $N_\mathrm{H_2}$ regimes. Only the $I_\mathrm{tot}$ variation for the HCO$^+$ line in the ISF region is similar to what the $^{13}$CO and C$^{18}$O lines exhibited. This difference seen in the HCO$^+$ and HCN lines may be due to different star formation activities in different regions. HCO$^+$ and HCN have been known as good tracers of star-formation activities because they become abundant in the gas affected by shocks and high energy UV photons.

\subsubsection{CS $J=$1$-$0} \label{Sec_CS}
The $I_\mathrm{tot}$ variation of the CS line shows that $\alpha$ decreases as $N_\mathrm{H_2}$ increases, similar to what was seen for $^{13}$CO and C$^{18}$O lines. However, the $\alpha$ values are generally greater than or similar to one. Only the Ophiuchus cloud has $\alpha$ significantly smaller than one ($\alpha$ $=$ 0.63) when $\log(N_\mathrm{H_2})$ exceeds 22.6.

Table \ref{tbl_fit} shows that the $I_\mathrm{tot}$ variation of CS can be explained with $\alpha$ of about 1.0 across the large $N_\mathrm{H_2}$ regime. In the ISF and L1641 regions, the $N_\mathrm{H_2}$ regimes with $\alpha$ $\sim$ 1.0 are extended over one order of magnitude. This result indicates that the $I_\mathrm{tot}$ of CS is proportional to $N_\mathrm{H_2}$ over the broad range of $N_\mathrm{H_2}$ in the Orion A cloud. \citet{Pet17} also mentioned that CS $J=$2$-$1 is one of the more useful column density tracers in the Orion B cloud.  

\subsubsection{N$_2$H$^+$ $J=$1$-$0} \label{Sec_N2Hp}
The N$_2$H$^+$ line increases super-linearly at low column densities before
approaching a more linear growth at higher column densities. This behavior reflects the low abundance of this species in gas of low column density and high CO abundance. It makes it a good probe of gas column density for regions of relatively high column density. 

The dominant formation mechanism of N$_2$H$^+$ molecule is
\begin{equation}
\mathrm{H_3^+} + \mathrm{N_2} \rightarrow \mathrm{N_2H^+} + \mathrm{H_2}.
\end{equation}
When the CO abundance is close to 10$^4$ which is a typical value in MCs \citep{Wil94,van95}, H$_3^+$ mainly combines with CO to form HCO$^+$. Also, N$_2$H$^+$ is destroyed by combining with CO,
\begin{equation}
\mathrm{N_2H^+} + \mathrm{CO} \rightarrow \mathrm{N_2} + \mathrm{HCO^+}.
\end{equation}
Therefore, N$_2$H$^+$ line can be abundant in dense gas where CO is depleted from the gas phase \citep{Ber97,Aik01,Lee03,Lee04,Tat08}. Therefore, $I_\mathrm{tot}$ of the N$_2$H$^+$ line represents the column density of the dense gas if we assume that the N$_2$H$^+$ line is optically thin. 

The $I_\mathrm{tot}$ of the N$_2$H$^+$ line is proportional to the amount of cold and dense gas along the line of sight if the N$_2$H$^+$ line is optically thin. The mean $I_\mathrm{tot}$ value at a given $N_\mathrm{H_2}$ in the ISF region is slightly higher than that of the L1641 and Ophiuchus regions. This result implies that the amount of the dense gas along the line of sight in the ISF region is slightly larger than that in the L1641 and Ophiuchus cloud. \citet{Tat08} also mentioned that the abundance of N$_2$H$^+$ decreases toward the south in ISF. Therefore, the difference in $I_\mathrm{tot}$ of N$_2$H$^+$ also can be explained with an abundance difference between the regions. 

\subsection{Variation of Velocity Dispersion with Column Density} \label{Sec_dv}
The right-hand panels of Figures \ref{fig_line_cold_CO} and \ref{fig_line_cold_HC} display the RMS velocity dispersion (moment 2; $\sigma_V$) versus $N_\mathrm{H_2}$. The dispersion depends strongly on $N_\mathrm{H_2}$, increasing by factors of 3-5, depending on the lines and regions. Weak lines can produce unrealistically small values of $\sigma_V$, but we have tried to avoid that by using the moment masking method to make the $\sigma_V$ map and by adopting the average values weighted by the inverse of the square of the uncertainty. Consequently, we believe that the strong trends in $\sigma_V$ versus $N_\mathrm{H_2}$ are real.

Two simple cloud models can be adopted to describe the increasing $N_\mathrm{H_2}$ toward the center from the outer region: (1) a structure with a varying depth, like a cylinder with a uniform density located along the sky plane, and (2) a structure with a uniform depth, such as a slab with varying density, as illustrated in Figure \ref{fig_cloud_models}. The observed linewidth is determined by sampling a turbulent velocity field along the line of sight (y-direction). If the linewidth correlates with path length, as expected from the turbulent velocity field \citep{Lar81,Sol87,Hey97,Kle00}, the $\sigma_V$ increases with $N_\mathrm{H_2}$ in the cylindrical model while that remains constant in the slab-like model. This result suggest that the cylinder cloud model can explain the low-$N_\mathrm{H_2}$ edges of the cloud with low-$\sigma_V$, unlike the slab-like cloud model.


\subsection{Line Luminosities and Luminosity Ratios}
The line luminosity of each line in the Orion A and Ophiuchus clouds is calculated by
\begin{equation}
L_\mathrm{line} = D^2\ \theta_\mathrm{pix}^2\ \sum^{N_\mathrm{pix}} I_\mathrm{tot.pix}
\end{equation}
where $D$, $\theta_\mathrm{pix}$, $N_\mathrm{pix}$, $I_\mathrm{tot.pix}$ are the distance, angular size of a pixel (20\arcsec), number of pixels, and line integrated intensity per pixel, respectively. The $I_\mathrm{tot.pix}$ is defined as 
\begin{equation}
I_\mathrm{tot.pix} = \frac{\theta_\mathrm{pix}^2}{2\pi\ \left(\frac{\theta_\mathrm{beam}}{2\sqrt{2ln(2)}}\right)^2}\ I_\mathrm{tot}
\end{equation}
where $\theta_\mathrm{beam}$ is the beam size. The distance to the Orion A cloud is assumed to be 416.3~pc, which is the average of the distances to the Orion nebular cluster, L1641, and L1647 \citep[389, 417, and 443~pc, respectively][]{Kou18}. For the Ophiuchus cloud, we adopted 137~pc \citep{Ort17}.

The value of $N_\mathrm{pix}$ requires discussion. We have done the summation in two ways. In the first, only the pixels that in the moment-masked map of each line were included. Very weak emission extended over large areas can however contribute substantially to the line luminosity \citep{Eva20}. For a comparison to other galaxies where all the emission is included in a beam, we need to include that emission. We did that by including all pixels in the original spectral maps. The resulting values are denoted by ``unbiased" in Table \ref{tbl_lum_ratio}. The comparison of the two values confirms that regions where individual lines are not clearly detected nonetheless add substantial luminosity to some lines, especially HCN and HCO$^+$ in Ophiuchus. The drawback of including all pixels is that baseline subtraction must be very good to avoid systematic offsets; an example is the negative luminosity of N$_2$H$^+$ from the unbiased method. That line is clearly very concentrated with no contribution from very extended regions.

Table \ref{tbl_lum_ratio}  presents the ratios of the total line luminosities to that of the $^{13}$CO line, which has the highest line luminosity in both clouds; the total line luminosities of the $^{13}$CO line in the Orion A and Ophiuchus clouds are 290.3 and 10.9~K~km~s$^{-1}$~pc$^2$, respectively. The total line luminosities of the C$^{18}$O, HCN, HCO$^+$, and CS lines are all lower than 10\% of the $^{13}$CO luminosity. The line ratios follow similar patterns in the two clouds. However, as mentioned above, the luminosities derived from the moment-masked maps of HCO$^+$ and HCN are much smaller than those from unbiased maps, indicative of the extended weak HCO$^+$ and HCN emission. 

The $^{13}$CO-to-C$^{18}$O and HCO$^+$-to-HCN luminosity ratios are also given in Table \ref{tbl_lum_ratio}. The $^{13}$CO-to-C$^{18}$O and HCO$^+$-to-HCN luminosity ratios have been used to study the properties of galaxies \citep{Kri08,Jim17,Men20}. The $^{13}$CO-to-C$^{18}$O ratios  are much larger than those for starburst (3.4$\pm$0.9) and normal spiral galaxies \citep[6.0$\pm$0.9][]{Jim17}. \citet{Men20} found a $^{13}$CO-to-C$^{18}$O ratio of 2.5$\pm$0.6 with the stacked spectra of 24 galaxies. The weak detection of the C$^{18}$O line could cause a large uncertainty in the ratio. The observation of the C$^{18}$O line in galaxies with better sensitivities is needed to confirm this difference. 

The HCO$^+$-to-HCN ratio for galaxies has been used to distinguish the phenomena in galaxies, such as an active galactic nucleus (AGNs) and starburst. The HCO$^+$-to-HCN ratios for the AGN-dominated galaxies that were studied by \citet{Kri08} are about 0.66, increasing  to about 1.5 as the contribution of the starburst increases. The ratio in Orion A (1.2-1.5) is similar to that for starbursts, while that in Ophiuchus (1.0) is intermediate between those of AGN and those of starbursts.

\section{Summary} \label{Sec_sum}
We obtained large and homogeneous line maps of the Orion A and Ophiuchus clouds in six different molecular transitions as one of TRAO-KSPs, TIMES. Both clouds were mapped in $^{13}$CO $J$=1$-$0/C$^{18}$O $J$=1$-$0, HCN $J$=1$-$0/HCO$^+$ $J$=1$-$0, and N$_2$H$^+$ $J$=1$-$0/CS $J$=2$-$1 using the TRAO 13.7~m telescope. The areas of mapped regions were 8.7 deg$^2$ toward the Orion A and 3.9 deg$^2$ toward the Ophiuchus clouds with $\sim$ 50~$\arcsec$ beam size. We discussed the physical and chemical environments traced by the observed lines in both clouds. The main results are summarized as follows:  

\begin{enumerate}
\item The observed $^{13}$CO line traces relatively diffuse gas in the MCs. For the Orion A cloud, there are a large scale north-south velocity gradient and complex velocity structures. For the Ophiuchus cloud, the L1688 and L1709 regions have different velocities, and generally show random motions. The Ophiuchus cloud is kinematically quiescent compared to the Orion A cloud. 
\item The C$^{18}$O line traces high column density regions, which are potentially the birthplace of the stars. The emission line maps show clumpy structures, and their spatial distribution is well correlated with that of the young embedded protostars. 
\item The N$_2$H$^+$ line traces cold and dense clumps/cores in the observed clouds. These clumps/cores seem to be embedded in the clumps that are revealed by the C$^{18}$O line. 
\item The CS $J=$2$-$1\ line traces the broadest range of the column density (over a one order of magnitude), making it a good probe of column density in MCs.
\item The HCN and HCO$^+$ lines trace the gas affected by the active star-forming activities. In the Orion A cloud, both emission lines coexist in ISF, the active star-forming clusters (L1641-N, -C, -S cluster) and nearby the HH objects. In the Ophiuchus cloud, however, the HCN line is mainly detected toward Oph-A and -B while the HCO$^+$ line is emitted from Oph-C, -E, and -F. 
\item The high velocity wing structures are marginally detected in the C$^{18}$O line spectrum obtained toward the OMC-1. The N$_2$H$^+$ lines do not have high velocity wings.
\item The velocity dispersions all increase strongly with column density, suggesting that the edges of the cloud are largely defined by small path length, not just low volume density.
\item The $^{13}$CO-to-C$^{18}$O line luminosity ratios for the Orion A and Ophiuchus clouds are much larger than that of starburst galaxies while the HCO$^+$-to-HCN ratios are comparable to that of the starburst galaxies. 
\end{enumerate}

In a companion paper (Yun et al., submitted), we use the TIMES data to explore the relationship between turbulence and star formation activity in MCs. We apply principal component analysis \citep[the PCA;][]{Hey97,Bru13}, which is one of the statistical methods used to derive the low-order velocity structure function, to the spectral maps presented in this paper. The uniform coverage, sensitivity and range of gas tracers included in the TIMES program  are ideal for studying MC kinematics and turbulence.

\section*{Acknowledgment}
This work was supported by the National Research Foundation of Korea (NRF) grant funded by the Korea government (MSIT) (grant number 2021R1A2C1011718).

\clearpage

\begin{figure}
\epsscale{1.0}
\plotone{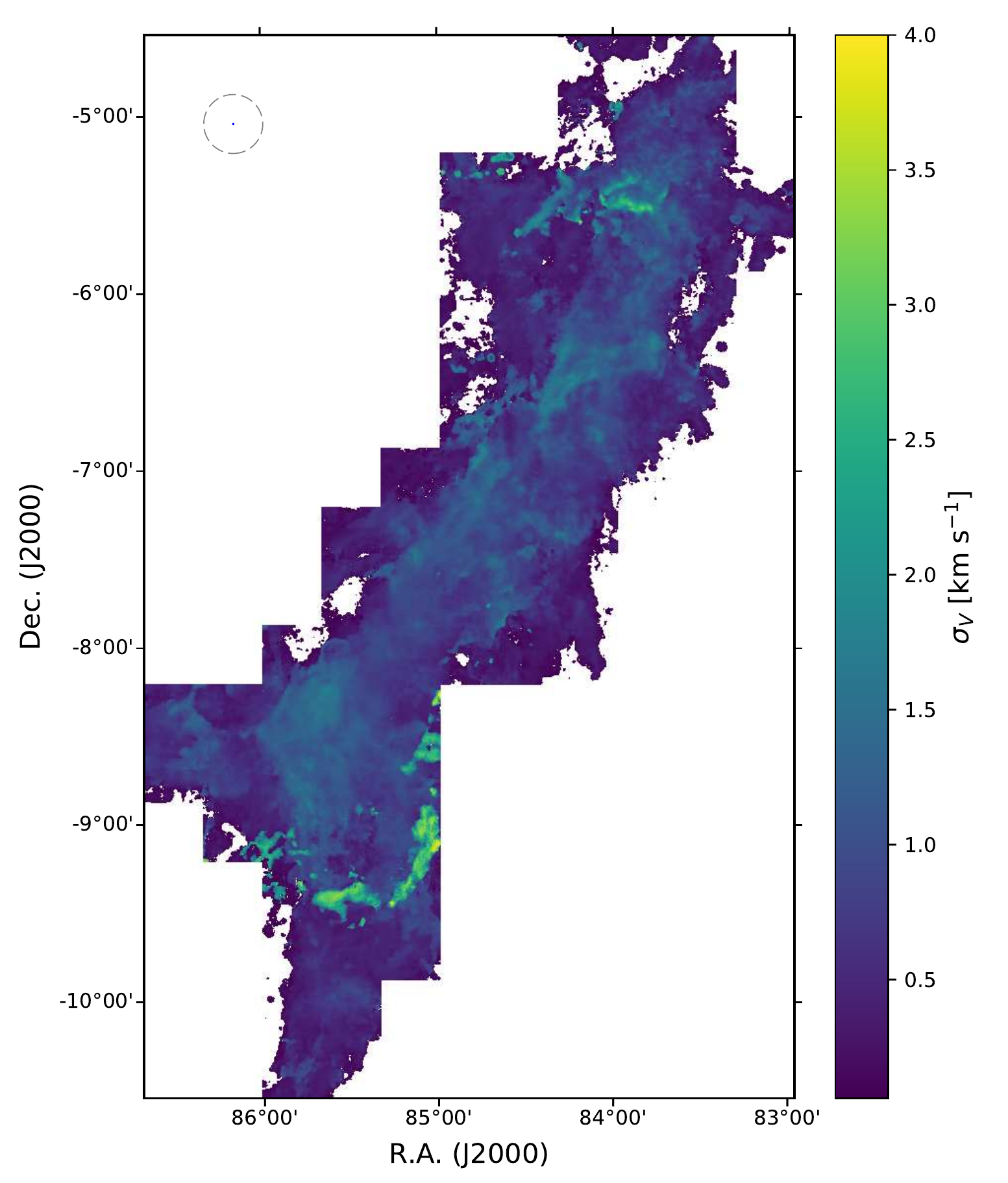}
\caption{The moment 2 (the equivalent of velocity dispersion; $\sigma_V$) map for the $^{13}$CO line in the Orion A cloud. \label{fig_Ori_13CO_mmt2}}
\end{figure}

\begin{figure}
\epsscale{1.0}
\plotone{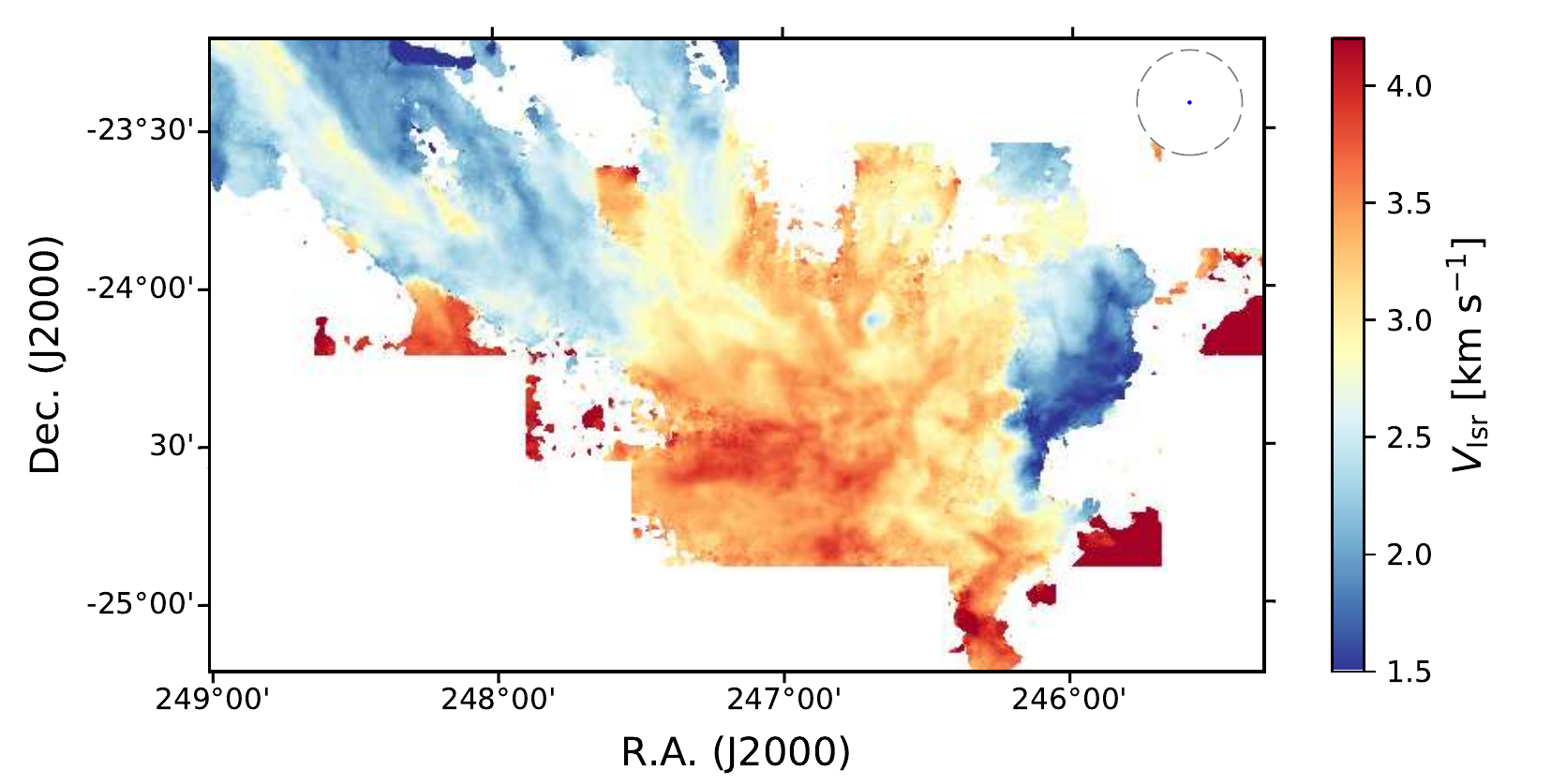}
\caption{Same as Figure \ref{fig_Ori_13CO_mmt1} except for the Ophiuchus cloud. \label{fig_Oph_13CO_mmt1}}
\end{figure}

\begin{figure}
\epsscale{1.0}
\plotone{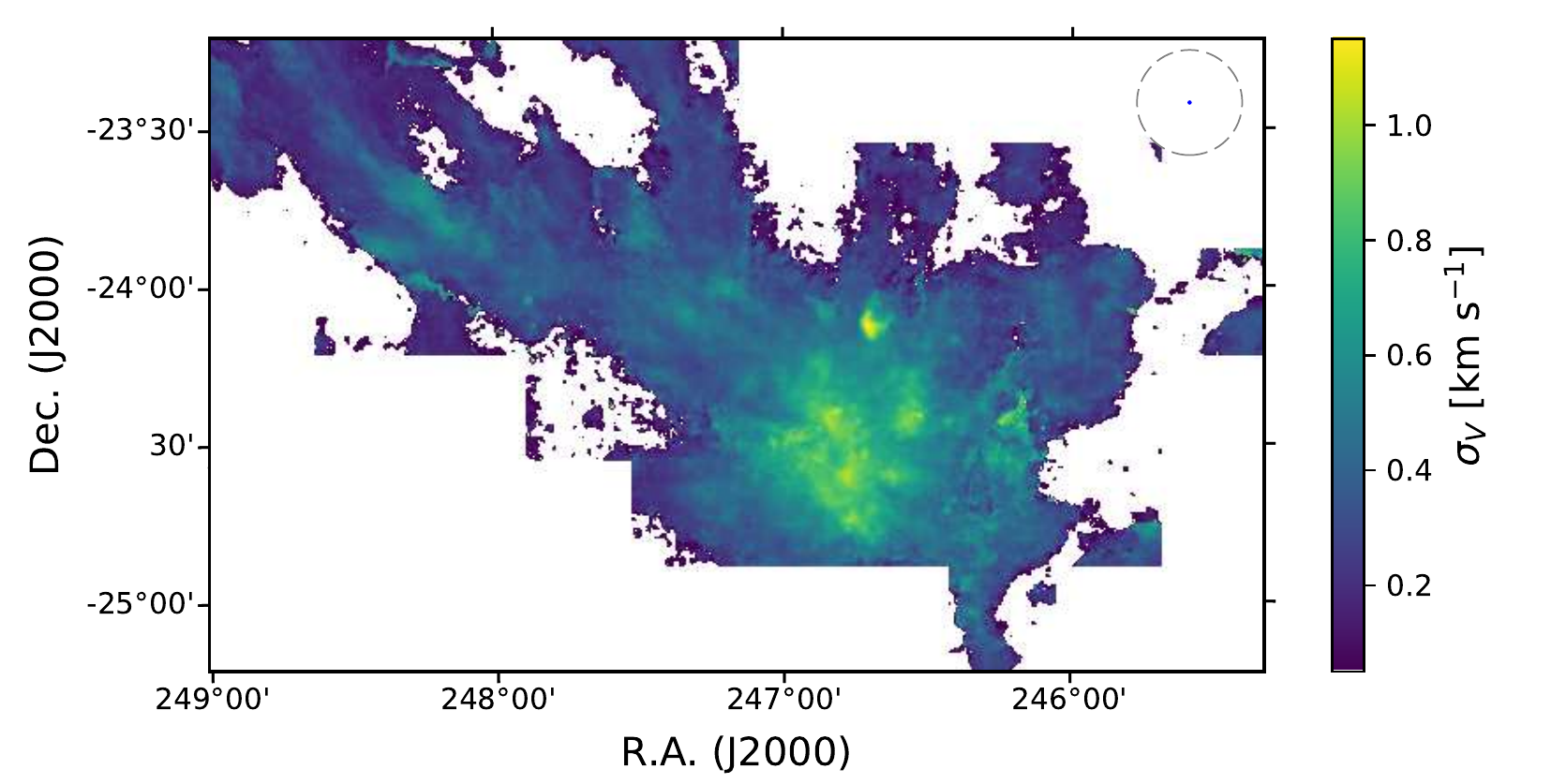}
\caption{Same as Figure \ref{fig_Ori_13CO_mmt2} except for the Ophiuchus cloud. \label{fig_Oph_13CO_mmt2}}
\end{figure}

\begin{figure}
\epsscale{0.75}
\plotone{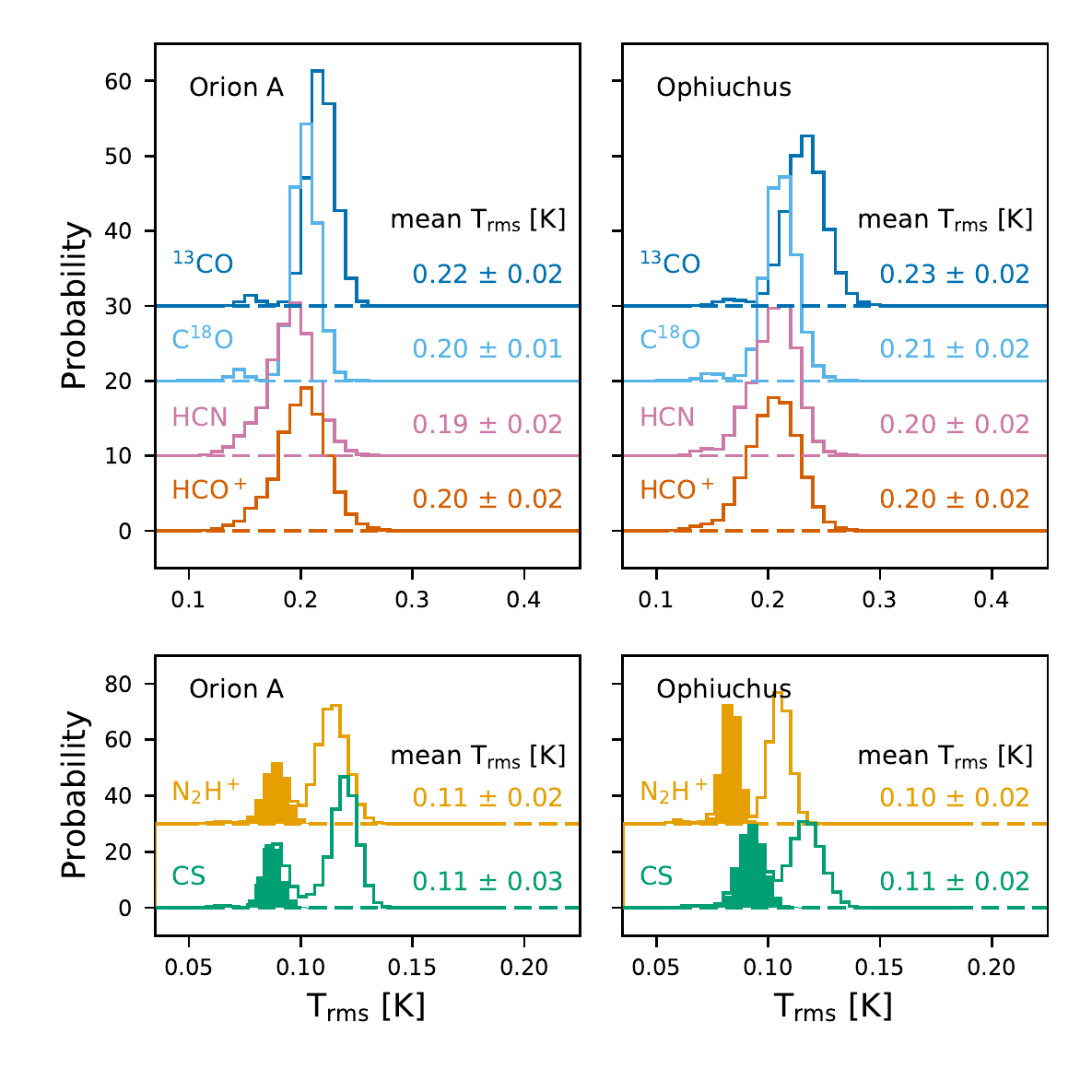}
\caption{PDFs of the $T_\mathrm{rms}$ for the observed maps in the Orion A (left panels) and Ophiuchus (right panels) clouds. The top panels shows the PDFs for the $^{13}$CO, C$^{18}$O, HCN, and HCO$^+$ maps, and the lower panels shows those for the N$_2$H$^+$ and CS maps. The $T_\mathrm{rms}$ PDFs for the N$_2$H$^+$ and CS lines toward the representative star-forming regions (orange dotted lines on Figure \ref{fig_cloud_2MASS}) are presented with the filled histograms. The mean and standard deviation for each PDF are summarized on the right side of each PDF. \label{fig_rms_PDF}}
\end{figure}

\begin{figure}
\epsscale{1.0}
\plotone{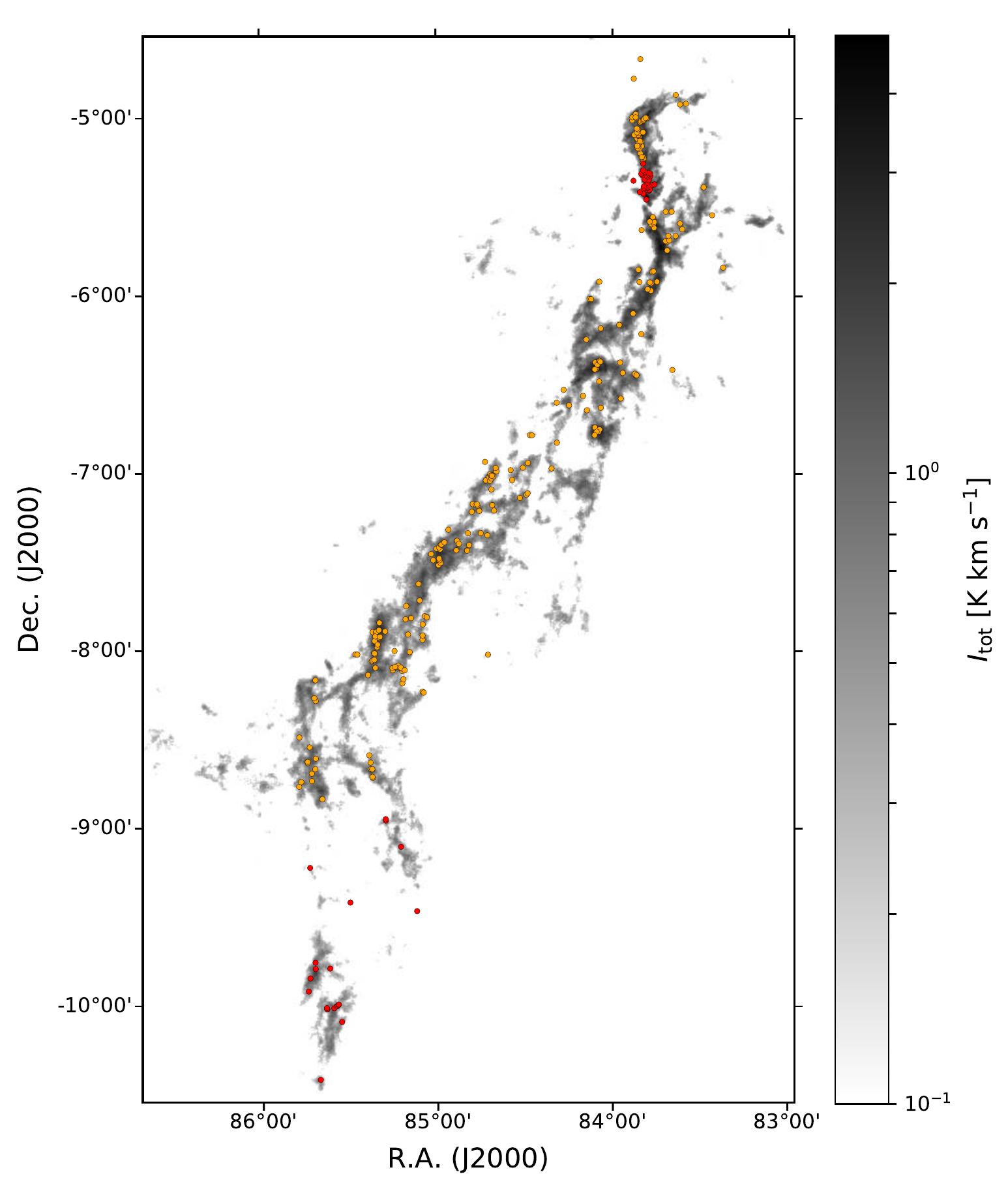}
\caption{Distribution of young embedded protostars in the Orion A cloud. The background image is the $I_\mathrm{tot}$ map of C$^{18}$O. The Class 0/I young stellar objects (YSOs) and flat spectrum sources, identified with \textit{Herschel} space observatory \citep{Fur16}, is presented in orange circles. Since \citet{Fur16} did not provide YSOs near the Orion Nebula and the southernmost part of the cloud, we adopt the $Spitzer$ YSO catalog provided by \citet{Meg12} (red circles). \label{fig_Ori_ctlg}}
\end{figure}

\begin{figure}
\epsscale{0.5}
\plotone{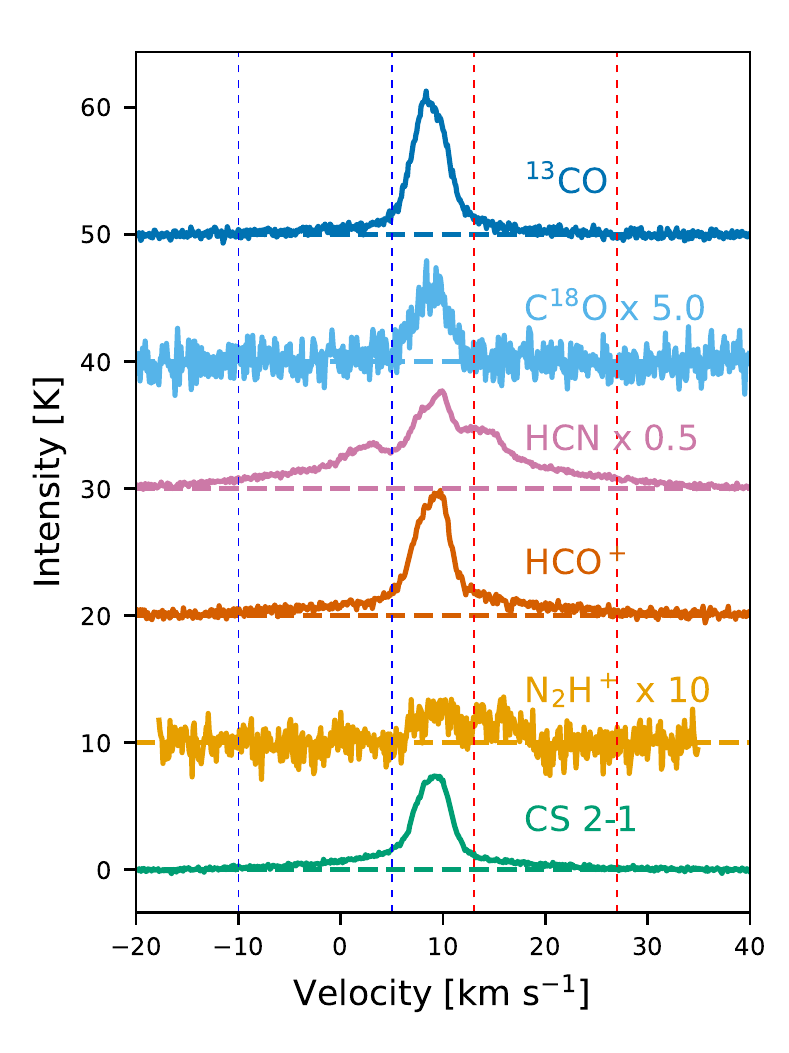}
\caption{Observed lines toward Orion KL. The red and blue dotted vertical lines indicate the velocity ranges where the red (from $-$10 to $+$5~km~s$^{-1}$) and blue (from $+$13 to $+$27~km~s$^{-1}$) wing structures are presented. \label{fig_KL_lines}}
\end{figure}

\begin{figure}
\epsscale{1.0}
\plotone{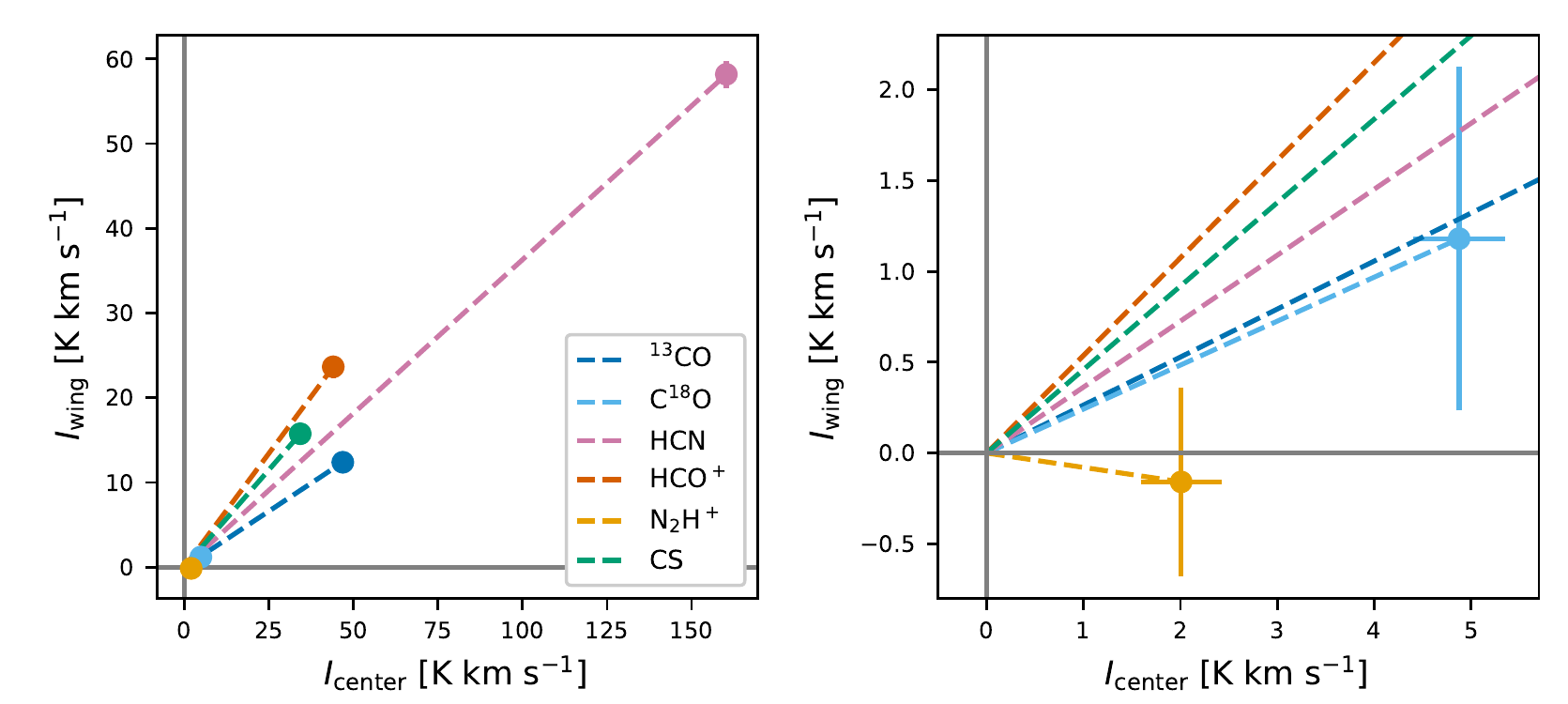}
\caption{Left: Comparison between the integrated intensities for the central peak ($I_\mathrm{center}$) and broad wing structures ($I_\mathrm{wing}$) of the Orion KL line spectra presented in Figure \ref{fig_KL_lines}. The 3-$\sigma$ error ranges are presented in the errorbars, however their sizes are similar or smaller than the symbol size. The gray solid lines indicates the position of the origin, and each dashed line presents the straight line from the origin to each data point. Right: Zoom-in on the origin of the diagram. \label{fig_KL_ratio}}
\end{figure}

\begin{figure}
\epsscale{1.0}
\plotone{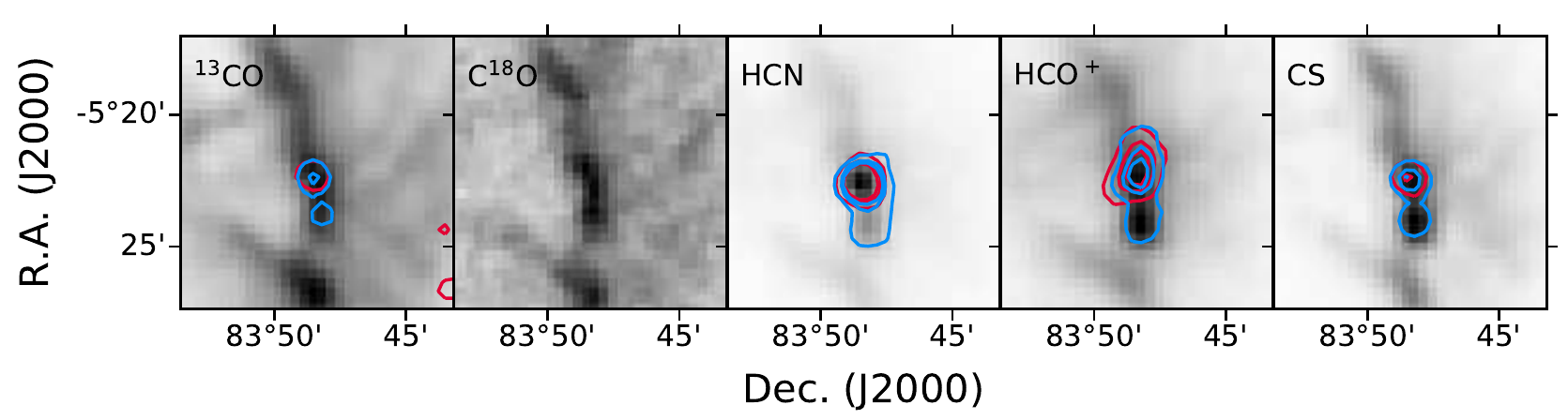}
\caption{Spatial distributions of the broad wing emission. In each panel, the red and blue contours indicate the integrated intensities for the red- and blue-shifted wings. The contour levels are 2, 6, and 9~K~km~s$^{-1}$. The background image exhibit the map of $I_\mathrm{center}$. \label{fig_broad_map}}
\end{figure}

\begin{figure}
\epsscale{1.0}
\plotone{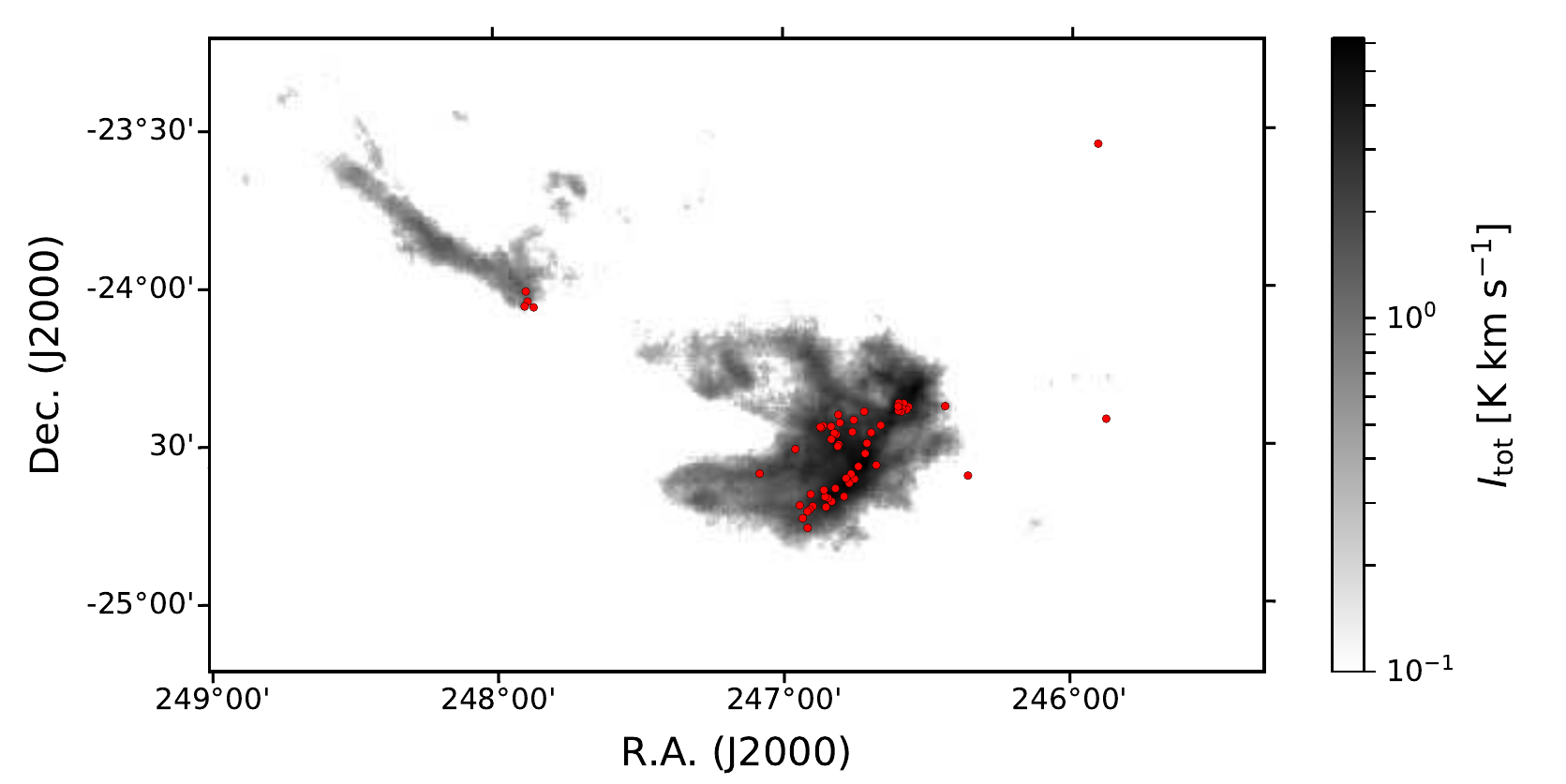}
\caption{Distribution of young embedded protostars in the Ophiuchus cloud. The background image is the $I_\mathrm{tot}$ map of C$^{18}$O, and the YSO catalog provided by \citet{Dun15} is presented in red circles. \label{fig_Oph_ctlg}}
\end{figure}

\begin{figure}
\epsscale{1.0}
\plotone{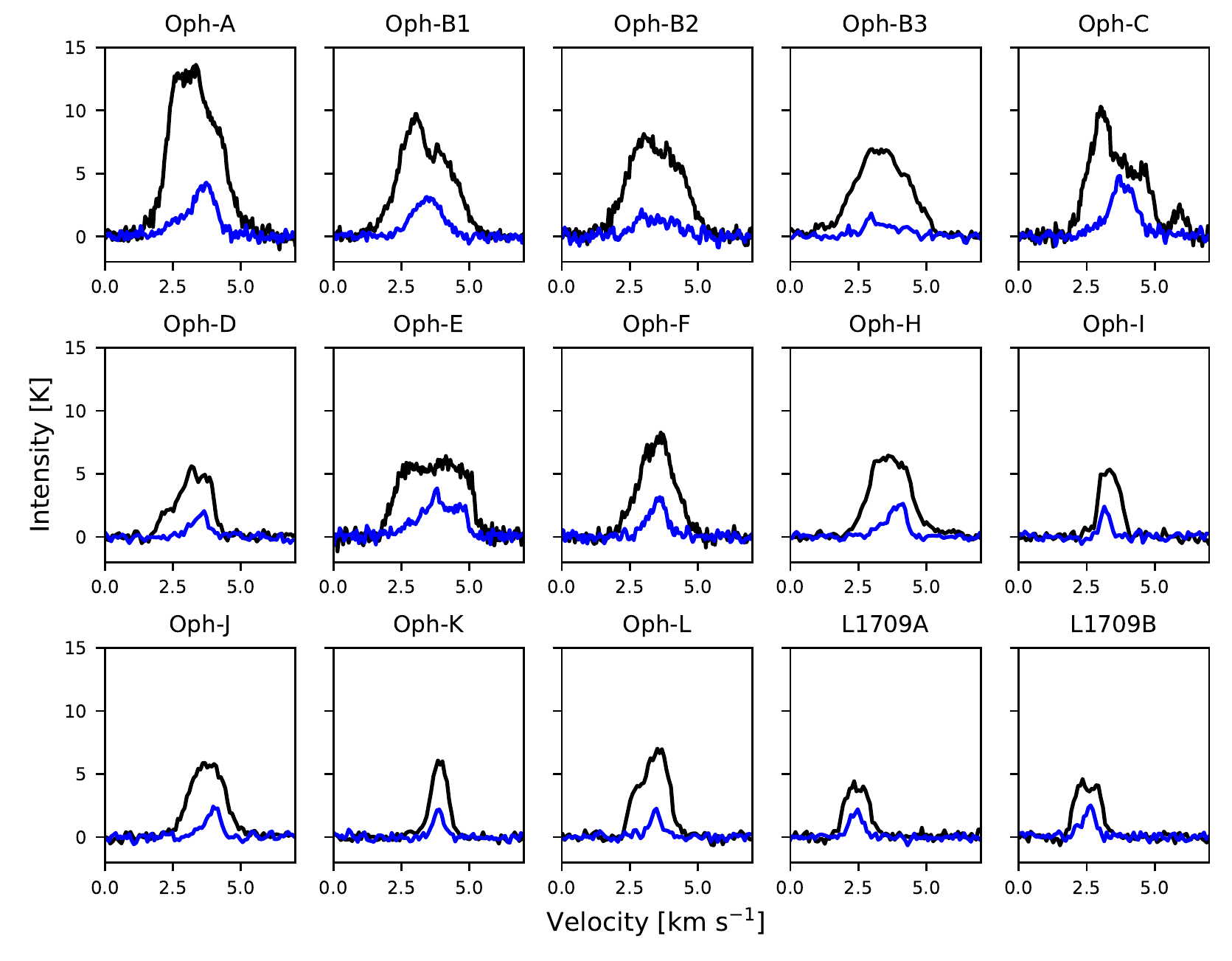}
\caption{Observed $^{13}$CO (black) and C$^{18}$O (blue) lines toward the dense cores \citep{Lor90,Pan17}. \label{fig_Oph_lines}}
\end{figure}

\begin{figure}
\epsscale{1.0}
\plotone{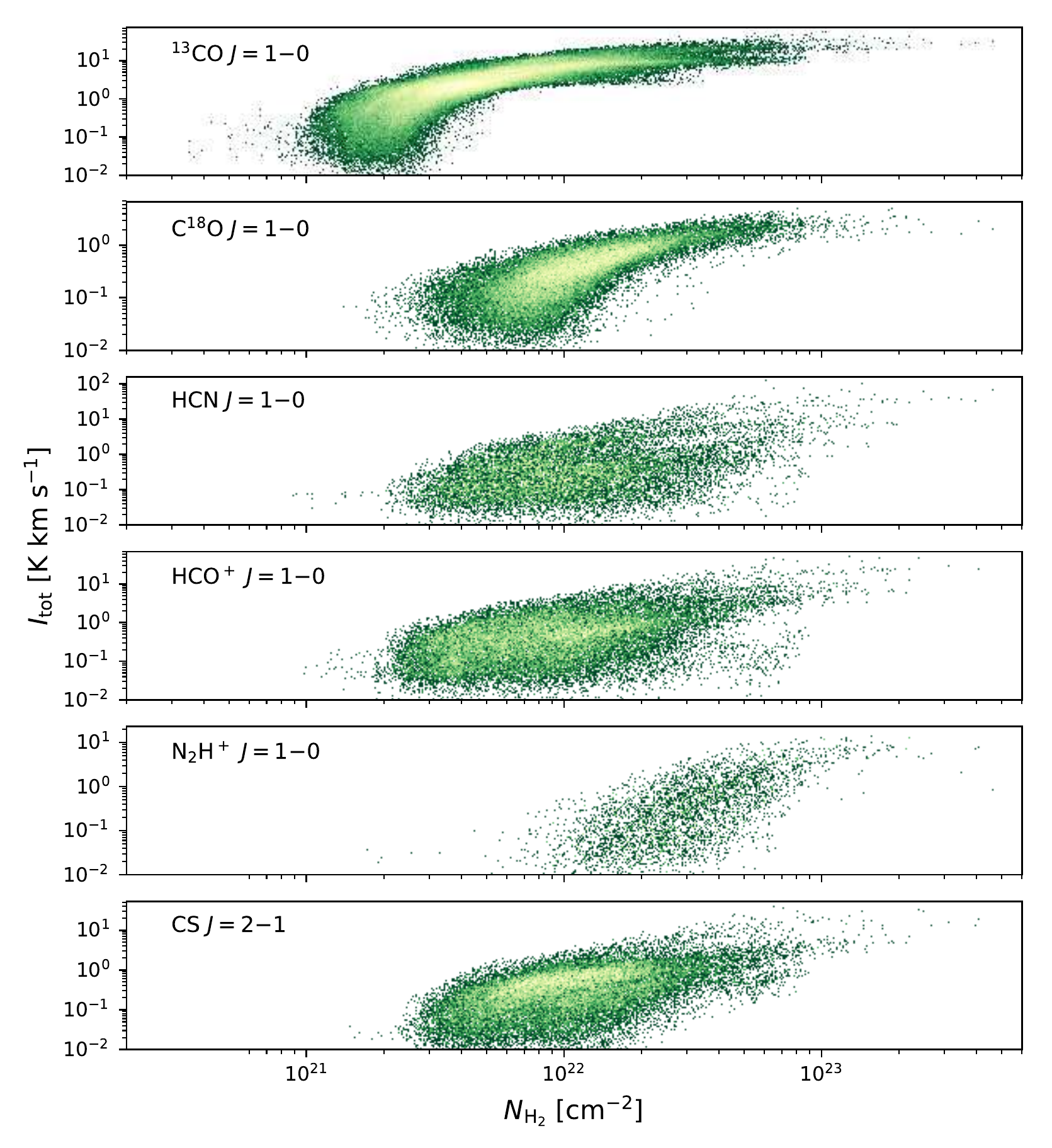}
\caption{Comparison of $I_\mathrm{tot}$ and column density ($N_\mathrm{H_2}$) in the Orion A cloud. From the top- to bottom-panels, the diagrams show for the $^{13}$CO, C$^{18}$O, HCN, HCO$^+$, N$_2$H$^+$, and CS lines. \label{fig_line_cold_Ori}}
\end{figure}

\begin{figure}
\epsscale{1.0}
\plotone{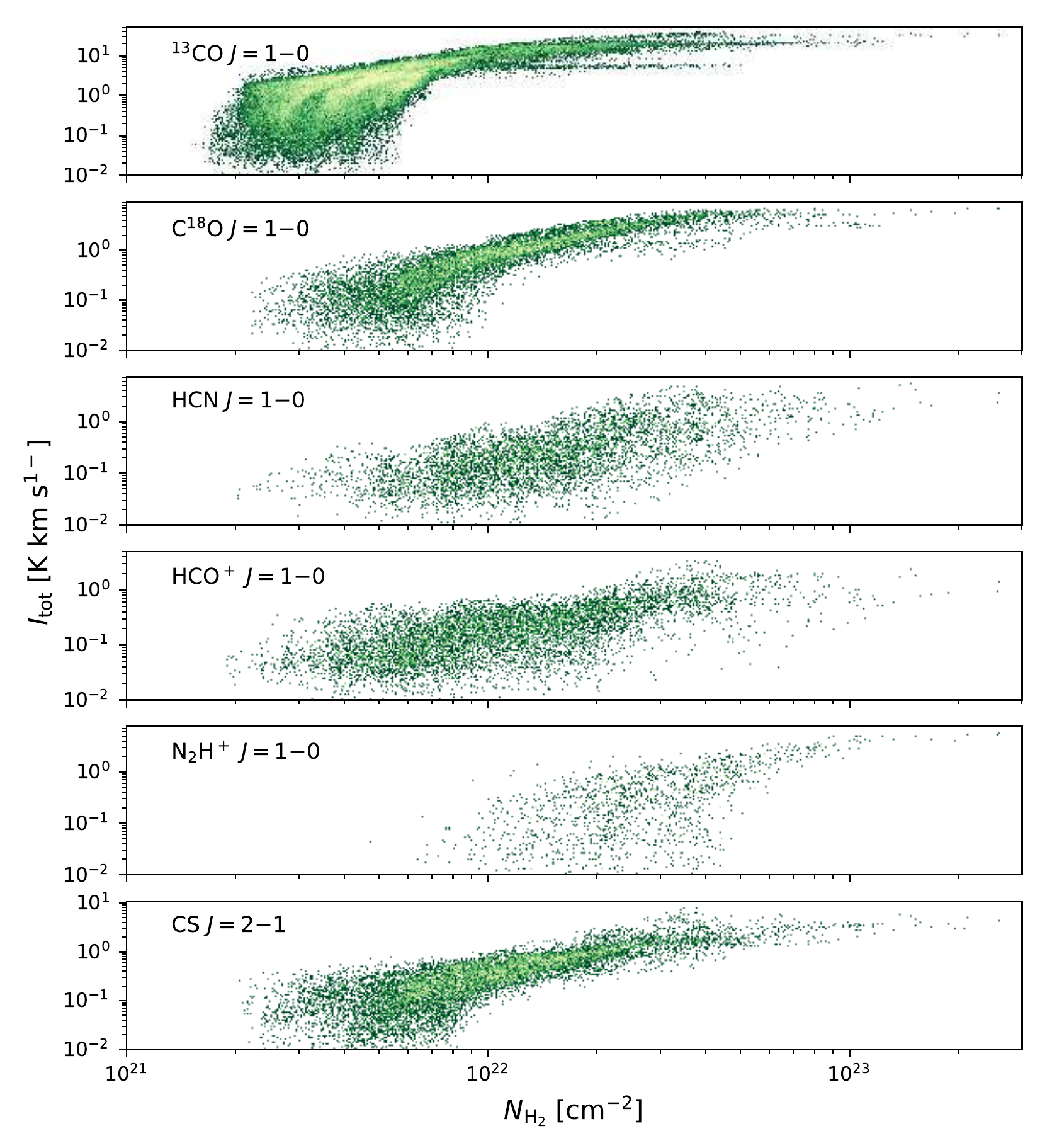}
\caption{Same as Figure \ref{fig_line_cold_Ori} but for the Ophiuchus cloud. \label{fig_line_cold_Oph}}
\end{figure}

\begin{figure}
\epsscale{1.0}
\plotone{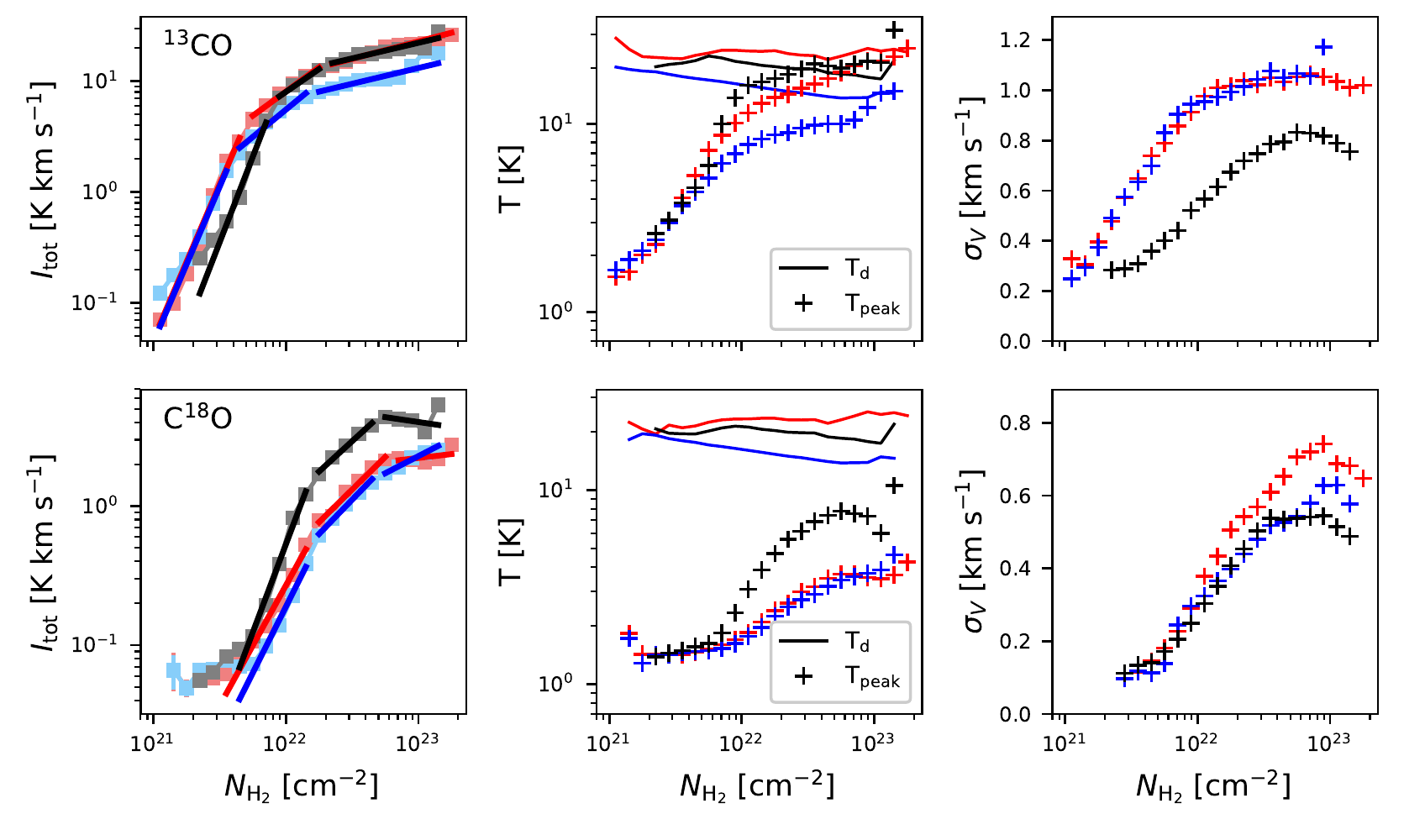}
\caption{Variation of means of $I_\mathrm{tot}$ (left), $T_\mathrm{peak}$ (middle), and $\sigma_V$ (right) for the $^{13}$CO (top) and C$^{18}$O (bottom) lines. The red, blue, and black colors indicates the results for the ISF, L1641, and Ophiuchus regions. The mean $I_\mathrm{tot}$ values and their 1-$\sigma$ error ranges are presented in square symbols and errorbars with weak colors. The red, blue, black solid lines indicates the power-law fitting results. In the middle panel, the mean $T_\mathrm{peak}$ values are presented in the plus symbols, and the variations of mean $T_\mathrm{d}$ values are also exhibited in solid lines. \label{fig_line_cold_CO}}
\end{figure}

\begin{figure}
\epsscale{1.0}
\plotone{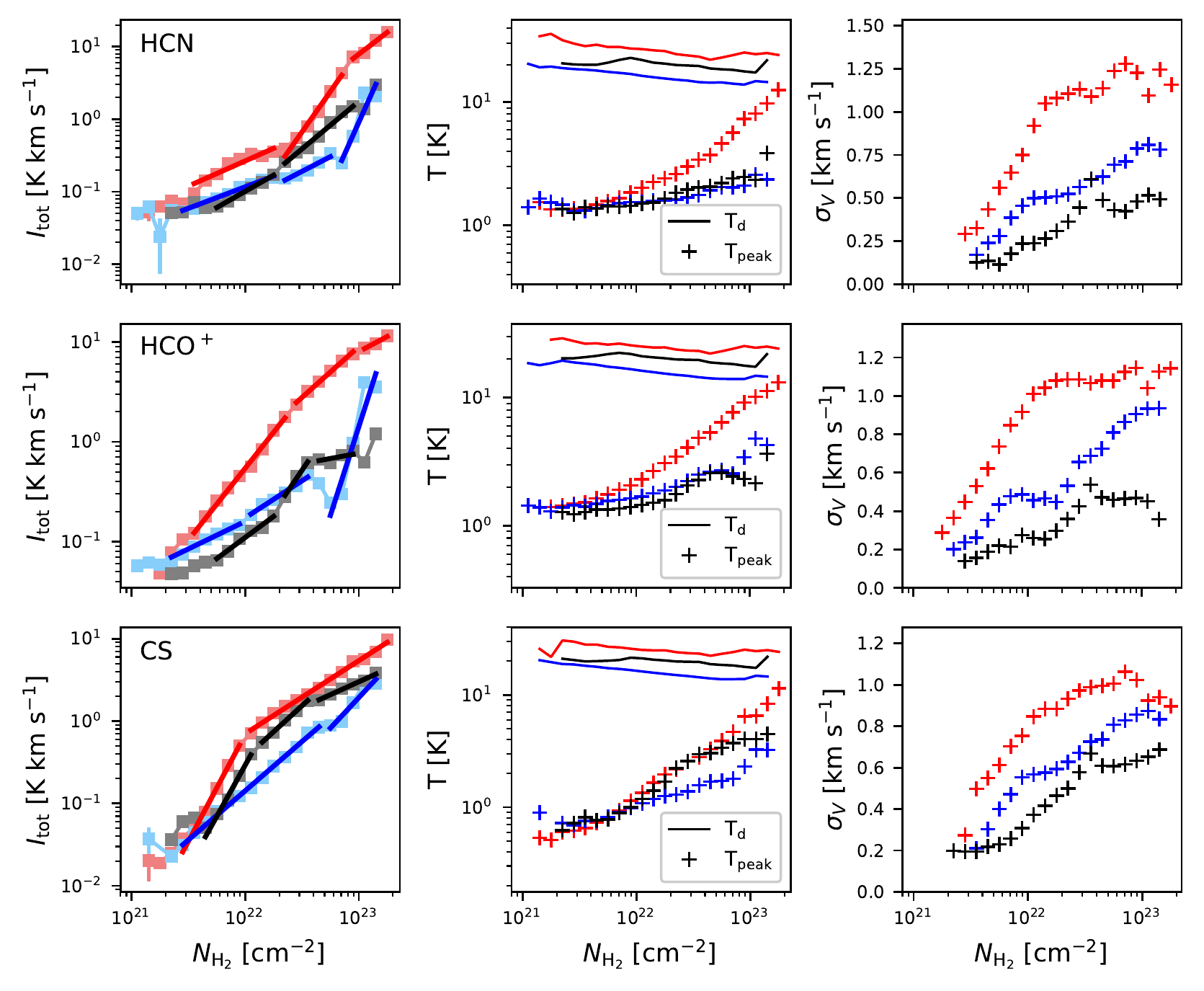}
\caption{Same as Figure \ref{fig_line_cold_CO} but for the HCN (top), HCO$^+$ (middle), and CS (bottom) lines  \label{fig_line_cold_HC}}
\end{figure}

\begin{figure}
\epsscale{0.5}
\plotone{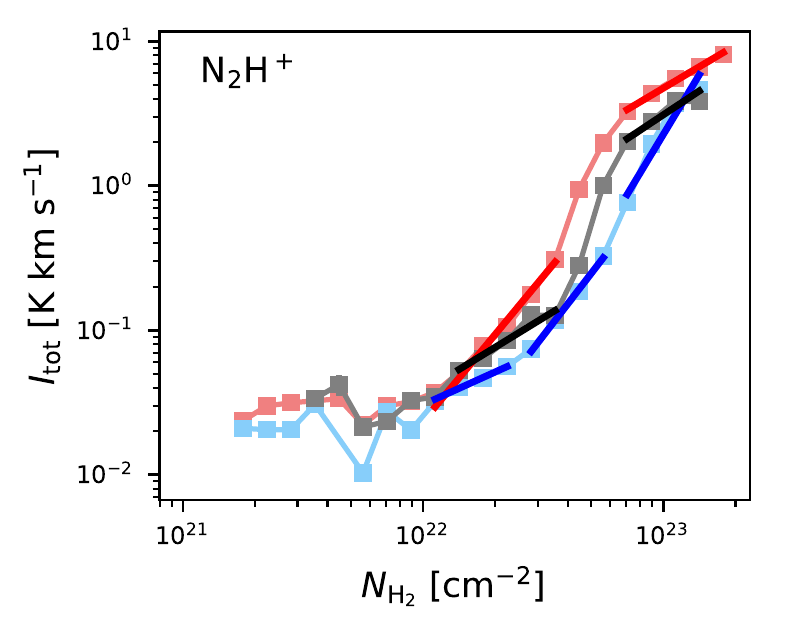}
\caption{Same as the left panel of Figure \ref{fig_line_cold_CO} but for the N$_2$H$^+$ line.  \label{fig_line_cold_NH}}
\end{figure}

\begin{figure}
\epsscale{1.0}
\plotone{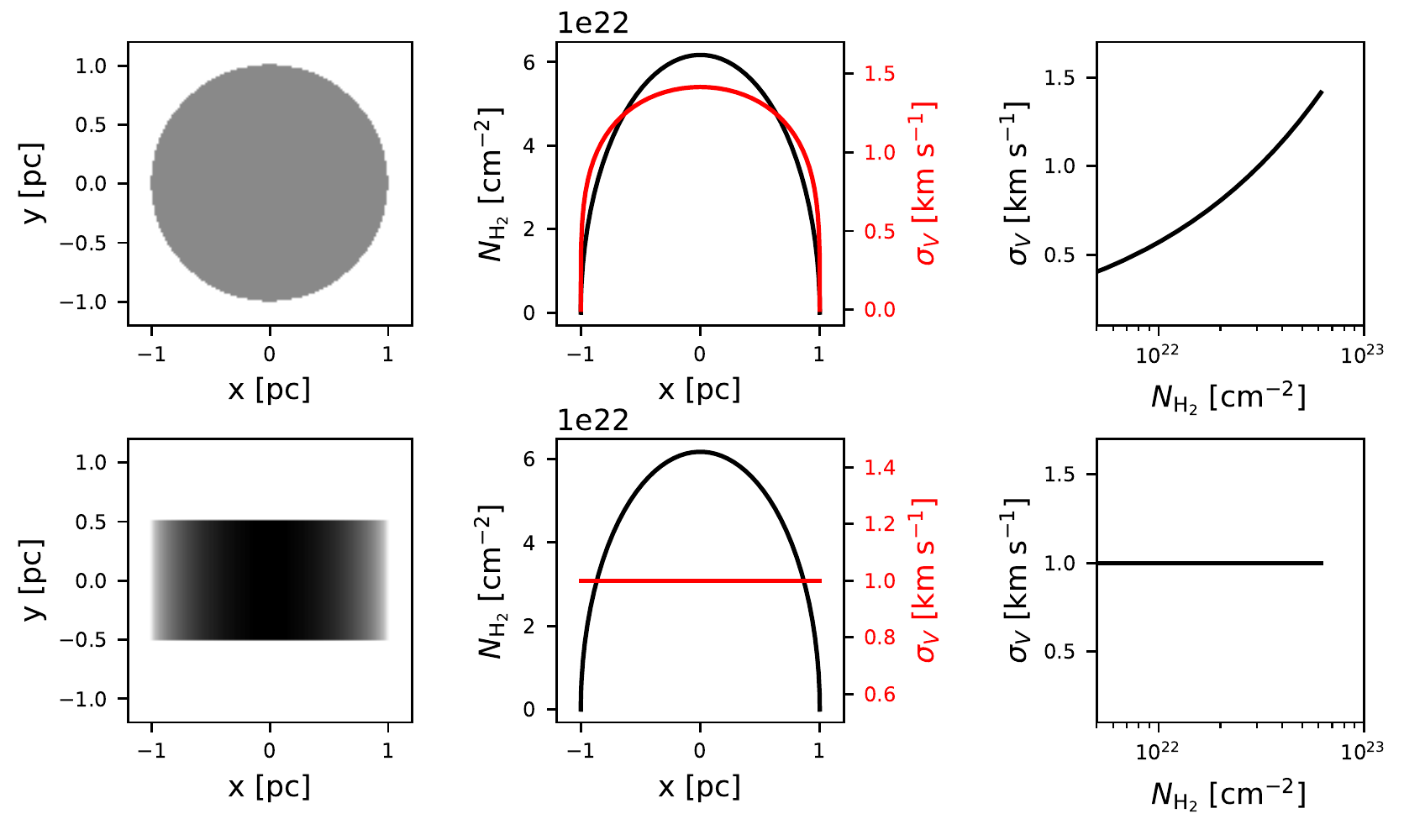}
\caption{Cloud models with varying $N_\mathrm{H_2}$. The top and bottom panels describe the cylindrical and the slab-like cloud models, respectively. The z-axis in the model clouds are located along the plane of sky, and thus the y-axis is the line of sight. The left panels show the cross-sections of the model clouds assuming that we observe the clouds along the y-axis. The middle panels present the variation of $N_\mathrm{H_2}$ (black) and $\sigma_V$ (red) along the x-axis. The right panels show the relation between $\sigma_V$ and $N_\mathrm{H_2}$ for the model clouds. A uniform density of 10$^4$~cm$^{-3}$ is assumed for the cylindrical model. The density structure within the slap-like model is adjusted to have the $N_\mathrm{H_2}$ distribution the same as that of the cylindrical model (see the middle panels). The $\sigma_V$ for a given x is derived assuming the burger spectrum ($\sigma_V \propto l^{0.5}$), which describes the velocity field of shock-dominated turbulence. \label{fig_cloud_models}}
\end{figure}

\begin{deluxetable}{lcccccc}
\tablecolumns{6}
\tabletypesize{\scriptsize}
\tablecaption{Properties of the observation in each line\label{tbl_lines}}
\tablewidth{0pt}
\tablehead{
  \colhead{Line}&\colhead{Rest frequency}&\colhead{Velocity resolution}&\colhead{Beam size\tablenotemark{b}}&\colhead{$n_\mathrm{crit}$\tablenotemark{a}}&\colhead{Beam efficiency\tablenotemark{b}}\\
  \colhead{}&\colhead{[GHz]}&\colhead{[km~s$^{-1}$]}&\colhead{[arcsec]}&\colhead{[cm$^{-3}$]}&\colhead{[\%]}
}
\startdata
$^{13}$CO $J$=1$-$0&	110.201&	0.0838&	49.0&	1 $\times$ 10$^{3}$& 46 $\pm$ 2\\
C$^{18}$O $J$=1$-$0&	109.782&	0.0833&	49.1&	1 $\times$ 10$^{3}$& 46 $\pm$ 2\\
HCN $J$=1$-$0&			88.631&		0.1032&	56.0&	2 $\times$ 10$^{6}$& 45 $\pm$ 3\\
HCO$^+$ $J$=1$-$0&		89.188&		0.1026&	55.7&	3 $\times$ 10$^{5}$& 46 $\pm$ 3\\
N$_2$H$^+$ $J$=1$-$0&	93.173&		0.0982&	54.1&	2 $\times$ 10$^{5}$& 47 $\pm$ 2\\
CS $J$=2$-$1&			97.980&		0.0934&	52.0&	3 $\times$ 10$^{5}$& 48 $\pm$ 2\\
\enddata
\tablenotetext{a}{Critical densities for observed lines; From \citet{Ung97}}
\tablenotetext{b}{Beam sizes and efficiencies for the observed lines are derived using a linear interpolation method based on those provided by \citet{Jeo19}} 
\end{deluxetable}

\begin{deluxetable}{lcc}
\tablecolumns{3}
\tabletypesize{\scriptsize}
\tablecaption{Parameters for data reduction \label{tbl_vrng}}
\tablewidth{0pt}
\tablehead{
  \colhead{line}&\colhead{$V_\mathrm{win}$}&\colhead{$V_\mathrm{space}$}\\
  &\colhead{[km~s$^{-1}$]}&\colhead{[km~s$^{-1}$]}
}
\startdata
\sidehead{Orion A}
$^{13}$CO&	(0, 20)&	(-20, 40)\\
C$^{18}$O&  (0, 20)&	(-20, 40)\\
HCN&    	(-10, 30)&	(-50, 70)\\
HCO$^+$&	(-5, 25)&	(-35, 55)\\
N$_2$H$^+$&	(-5, 22)&	(-18, 35)\\
CS&	        (-5, 17)&	(-20, 40)\\
\sidehead{Ophiuchus}
$^{13}$CO&	(-1, 8)&	(-7, 14)\\
C$^{18}$O&	(-1, 8)&	(-10, 17)\\
HCN&	    (-8, 13)&	(-29, 34)\\
HCO$^+$&	(-1, 8)&	(-10, 17)\\
N$_2$H$^+$&	(-10, 15)&	(-35, 40)\\
CS&	        (-1, 8)&	(-10, 17)\\
\sidehead{Near Orion KL\tablenotemark{a}}
$^{13}$CO&	(-13, 30)&	\\
C$^{18}$O&  (-5, 25)&	\\
HCN&	    (-30, 50)&	\\
HCO$^+$&	(-20, 40)&	\\
CS&	        (-15, 38)&	\\
\enddata
\tablenotetext{a}{Velocity windows ($V_\mathrm{win}$) for the line spectra with broad wing structures. Velocity spaces ($V_\mathrm{space}$) for these lines are the same as that of the Orion A cloud.}
\end{deluxetable}

\begin{deluxetable}{lcc}
\tablecolumns{3}
\tabletypesize{\scriptsize}
\tablecaption{The velocity ranges for the central peak and broad wings \label{tbl_vrng_wing}}
\tablewidth{0pt}
\tablehead{
  \colhead{line}&\colhead{Central peak}&\colhead{Blue/red wings}\\
  &\colhead{[km~s$^{-1}$]}&\colhead{[km~s$^{-1}$]}
}
\startdata
$^{13}$CO&	(5.0, 13.0)&	(-11.0, 5.0)/(13.0, 29.0)\\
C$^{18}$O&  (5.0, 13.0)&	(-11.0, 5.0)/(13.0, 29.0)\\
HCN&    	(0.2, 20.1)&	(-15.8, 0.2)/(20.1, 36.1)\\
HCO$^+$&	(5.0, 13.0)&	(-11.0, 5.0)/(13.0, 29.0)\\
N$_2$H$^+$&	(-1.7, 21.2)&	(-17.7, -1.7)/(21.2, 37.2)\\
CS&	        (5.0, 13.0)&	(-11.0, 5.0)/(13.0, 29.0)\\
\enddata
\end{deluxetable}

\begin{deluxetable}{lcccc}
\tablecolumns{5}
\tabletypesize{\scriptsize}
\tablecaption{Power-law fitting results \label{tbl_fit}}
\tablewidth{0pt}
\tablehead{
  \colhead{Region}&\colhead{Line}&\colhead{$\log(N_\mathrm{H_2})$}&\colhead{$\alpha$}&\colhead{$\log(b)$}
}
\startdata
ISF&        $^{13}$CO&  21.0-21.7&   2.79$\pm$0.24&  -59.99$\pm$5.22\\
ISF&        $^{13}$CO&  21.7-22.3&   0.87$\pm$0.06&  -18.30$\pm$1.36\\
ISF&        $^{13}$CO&  22.3-23.3&   0.31$\pm$0.03&  -5.86$\pm$0.58\\
L1641&      $^{13}$CO&  21.0-21.6&   2.80$\pm$0.14&  -60.18$\pm$3.00\\
L1641&      $^{13}$CO&  21.6-22.2&   0.98$\pm$0.09&  -20.92$\pm$1.89\\
L1641&      $^{13}$CO&  22.2-23.2&   0.28$\pm$0.03&  -5.43$\pm$0.64\\
Ophiuchus&  $^{13}$CO&  21.3-21.9&   3.09$\pm$0.21&  -66.78$\pm$4.59\\
Ophiuchus&  $^{13}$CO&  21.9-22.3&   0.82$\pm$0.07&  -17.17$\pm$1.55\\
Ophiuchus&  $^{13}$CO&  22.3-23.2&   0.28$\pm$0.04&  -5.11$\pm$0.79\\
ISF&        C$^{18}$O&  21.5-22.2&   1.73$\pm$0.14&  -38.57$\pm$3.12\\
ISF&        C$^{18}$O&  22.2-22.8&   0.94$\pm$0.08&  -21.01$\pm$1.78\\
ISF&        C$^{18}$O&  21.8-23.3&   0.11$\pm$0.12&  -2.21$\pm$2.72\\
L1641&      C$^{18}$O&  21.6-22.2&   1.90$\pm$0.16&  -42.60$\pm$3.45\\
L1641&      C$^{18}$O&  22.2-22.7&   0.99$\pm$0.06&  -22.14$\pm$1.38\\
L1641&      C$^{18}$O&  22.7-23.2&   0.49$\pm$0.04&  -10.95$\pm$0.89\\
Ophiuchus&  C$^{18}$O&  21.6-22.2&   2.53$\pm$0.18&  -55.99$\pm$3.92\\
Ophiuchus&  C$^{18}$O&  22.2-22.7&   0.87$\pm$0.06&  -19.20$\pm$1.40\\
Ophiuchus&  C$^{18}$O&  22.7-23.2&  -0.14$\pm$0.16&  3.83$\pm$3.74\\
ISF&        HCN&        21.5-22.3&   0.69$\pm$0.11&  -15.86$\pm$2.32\\
ISF&        HCN&        22.3-22.9&   2.24$\pm$0.14&  -50.47$\pm$3.10\\
ISF&        HCN&        22.9-23.3&   1.21$\pm$0.15&  -26.84$\pm$3.45\\
L1641&      HCN&        21.4-22.3&   0.59$\pm$0.03&  -13.95$\pm$0.72\\
L1641&      HCN&        22.3-22.8&   0.81$\pm$0.11&  -18.95$\pm$2.38\\
L1641&      HCN&        22.8-23.2&   3.50$\pm$0.83&  -80.52$\pm$19.07\\
Ophiuchus&  HCN&        21.7-22.3&   0.89$\pm$0.03&  -20.63$\pm$0.56\\
Ophiuchus&  HCN&        22.3-23.0&   1.31$\pm$0.09&  -29.88$\pm$2.00\\
ISF&        HCO$^+$&    21.5-22.4&   1.44$\pm$0.04&  -31.87$\pm$0.94\\
ISF&        HCO$^+$&    22.4-23.0&   1.02$\pm$0.04&  -22.46$\pm$0.96\\
ISF&        HCO$^+$&    23.0-23.3&   0.60$\pm$0.11&  -13.01$\pm$2.44\\
L1641&      HCO$^+$&    21.3-22.0&   0.55$\pm$0.03&  -12.91$\pm$0.69\\
L1641&      HCO$^+$&    22.0-22.6&   0.74$\pm$0.05&  -17.06$\pm$1.03\\
L1641&      HCO$^+$&    22.7-23.2&   3.55$\pm$0.70&  -81.47$\pm$16.08\\
Ophiuchus&  HCO$^+$&    21.7-22.3&   0.84$\pm$0.07&  -19.52$\pm$1.44\\
Ophiuchus&  HCO$^+$&    22.3-22.6&   1.68$\pm$0.18&  -38.17$\pm$4.13\\
Ophiuchus&  HCO$^+$&    22.6-23.0&   0.21$\pm$0.20&  -4.91$\pm$4.64\\
ISF&        CS&         21.4-22.0&   2.59$\pm$0.08&  -57.12$\pm$1.75\\
ISF&        CS&         22.0-23.3&   0.89$\pm$0.02&  -19.67$\pm$0.47\\
L1641&      CS&         21.4-22.7&   1.19$\pm$0.03&  -27.10$\pm$0.68\\
L1641&      CS&         22.7-23.2&   1.48$\pm$0.28&  -33.80$\pm$6.40\\
Ophiuchus&  CS&         21.6-22.1&   2.49$\pm$0.13&  -55.23$\pm$2.83\\
Ophiuchus&  CS&         22.1-22.6&   1.26$\pm$0.04&  -28.25$\pm$0.93\\
Ophiuchus&  CS&         22.6-23.2&   0.63$\pm$0.02&  -13.90$\pm$0.53\\
ISF&        N$_2$H$^+$& 22.0-22.6&   2.01$\pm$0.11&  -45.76$\pm$2.55\\
ISF&        N$_2$H$^+$& 22.8-23.3&   0.99$\pm$0.05&  -22.01$\pm$1.14\\
L1641&      N$_2$H$^+$& 22.0-22.4&   0.75$\pm$0.05&  -18.08$\pm$1.10\\
L1641&      N$_2$H$^+$& 22.4-22.8&   2.16$\pm$0.10&  -49.65$\pm$2.26\\
L1641&      N$_2$H$^+$& 22.8-23.2&   2.76$\pm$0.46&  -63.10$\pm$10.58\\
Ophiuchus&  N$_2$H$^+$& 22.1-22.6&   1.02$\pm$0.19&  -23.87$\pm$4.17\\
Ophiuchus&  N$_2$H$^+$& 22.8-23.2&   1.10$\pm$0.24&  -24.85$\pm$5.44\\
\enddata
\end{deluxetable}

\begin{deluxetable}{lccc}
\tablecolumns{3}
\tabletypesize{\scriptsize}
\tablecaption{The $N_\mathrm{H_2}$ regimes with $\alpha\sim$1.0 in ISF\label{tbl_kauff}}
\tablewidth{0pt}
\tablehead{
  &\multicolumn{2}{c}{From \citet{kau17}}&\colhead{This work}\\
  \colhead{line}&\colhead{$\log(A_\mathrm{V})$}&\colhead{$\log(N_\mathrm{H_2})$\tablenotemark{a}}&\colhead{$\log(N_\mathrm{H_2})$\tablenotemark{b}}\\
  &\colhead{[mag]}&\colhead{[cm$^{-2}$]}&\colhead{[cm$^{-2}$]}
}
\startdata
$^{13}$CO&	0.7-1.0&	21.7-22.0&  21.7-22.3\\
C$^{18}$O&  1.0-1.3&	22.0-22.3&  22.2-22.8\\
HCN&    	0.7-1.4&	21.7-22.4&  22.9-23.3\\
N$_2$H$^+$&	1.8-2.0&	22.8-23.0&  22.8-23.3\\
\enddata
\tablenotetext{a}{The column density is derived via an equation, $A_\mathrm{V}/\mathrm{mag} = N_\mathrm{H_2}/9.4\times10^{20}~cm^{-2}$ \citep{kau17}.}
\tablenotetext{b}{The $N_\mathrm{H_2}$ regimes in Table \ref{tbl_fit} are repeated for a comparison.}
\end{deluxetable}

\begin{deluxetable}{lcccccccc}
\tablecolumns{9}
\tabletypesize{\scriptsize}
\tablecaption{Total line luminosities relative to $L_\mathrm{^{13}CO}$  and luminosity ratios \label{tbl_lum_ratio}}
\tablewidth{0pt}
\tablehead{
  \colhead{Cloud}& 
  \colhead{$L_\mathrm{^{13}CO}$}& \colhead{$L_\mathrm{C^{18}O}$}& \colhead{$L_\mathrm{HCN}$}& \colhead{$L_\mathrm{HCO^+}$}& \colhead{$L_\mathrm{N_2H^+}$}& \colhead{$L_\mathrm{CS}$}& \colhead{$L_\mathrm{^{13}CO}/L_\mathrm{C^{18}O}$}& \colhead{$L_\mathrm{HCO^+}/L_\mathrm{HCN}$}
}
\startdata
Orion A&    1.000&  0.039&  0.025&  0.036&  0.005&  0.030&  25.673&  1.458\\
Orion A (unbiased)&    1.000&  0.049&  0.030&  0.036&  0.005&  0.040&  20.475&  1.179\\
Ophiuchus&  1.000&  0.076&  0.009&  0.009&  0.004&  0.029&  13.180&  0.987\\
Ophiuchus (unbiased)&  1.000&  0.089&  0.055&  0.056&  -0.001&  0.035&  11.236&  1.007\\
\enddata
\end{deluxetable}
\clearpage

\appendix
\renewcommand\thefigure{\thesection.\arabic{figure}}
\section{Moment 0, 1, and 2 maps for the Orion A cloud} \label{App_mmtmaps_Ori}
\setcounter{figure}{0}
\begin{figure}
\epsscale{1.0}
\plotone{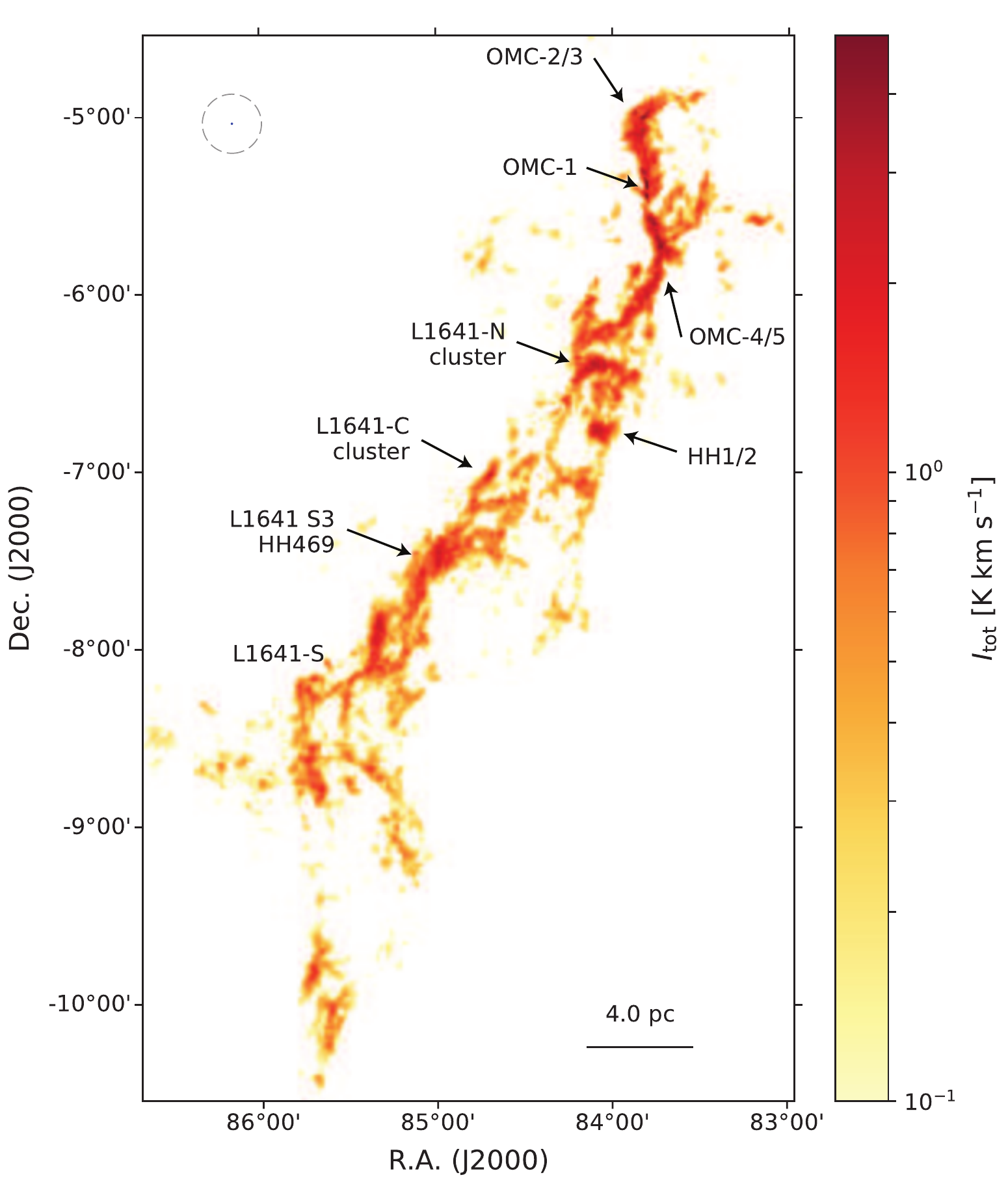}
\caption{The same as Figure \ref{fig_13CO_Ori} but for the C$^{18}$O line. We annotated the C$^{18}$O map with the names of the associated sources \citep{Dav09,Mei16,Mai16}. \label{fig_Ori_C18O}}
\end{figure}

\begin{figure}
\epsscale{1.0}
\plotone{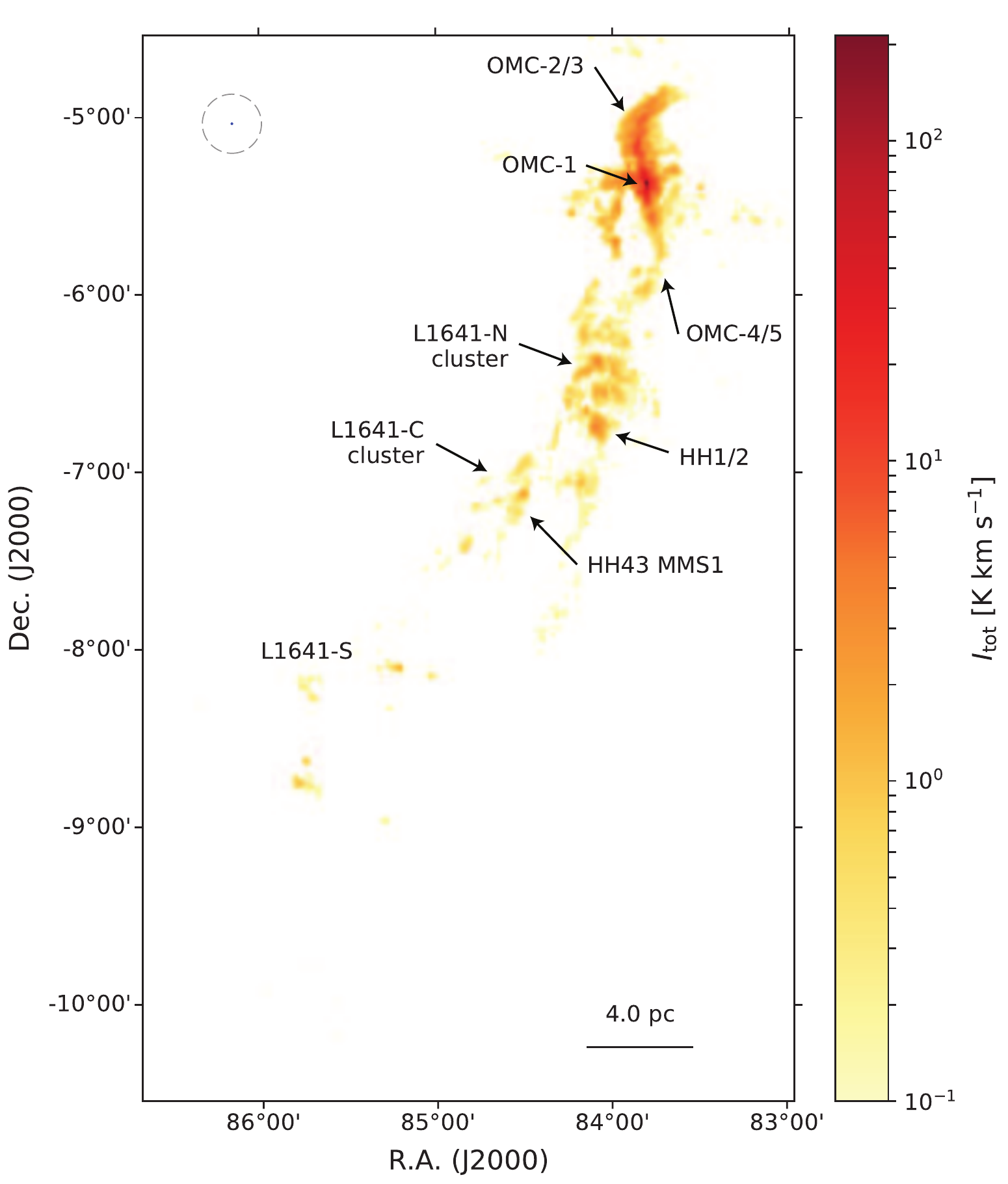}
\caption{The same as Figure \ref{fig_13CO_Ori} but for the the HCN line. We annotated the map with the names of the associated sources \citep{Dav09,Mei16,Mai16}. \label{fig_Ori_HCN}}
\end{figure}

\begin{figure}
\epsscale{1.0}
\plotone{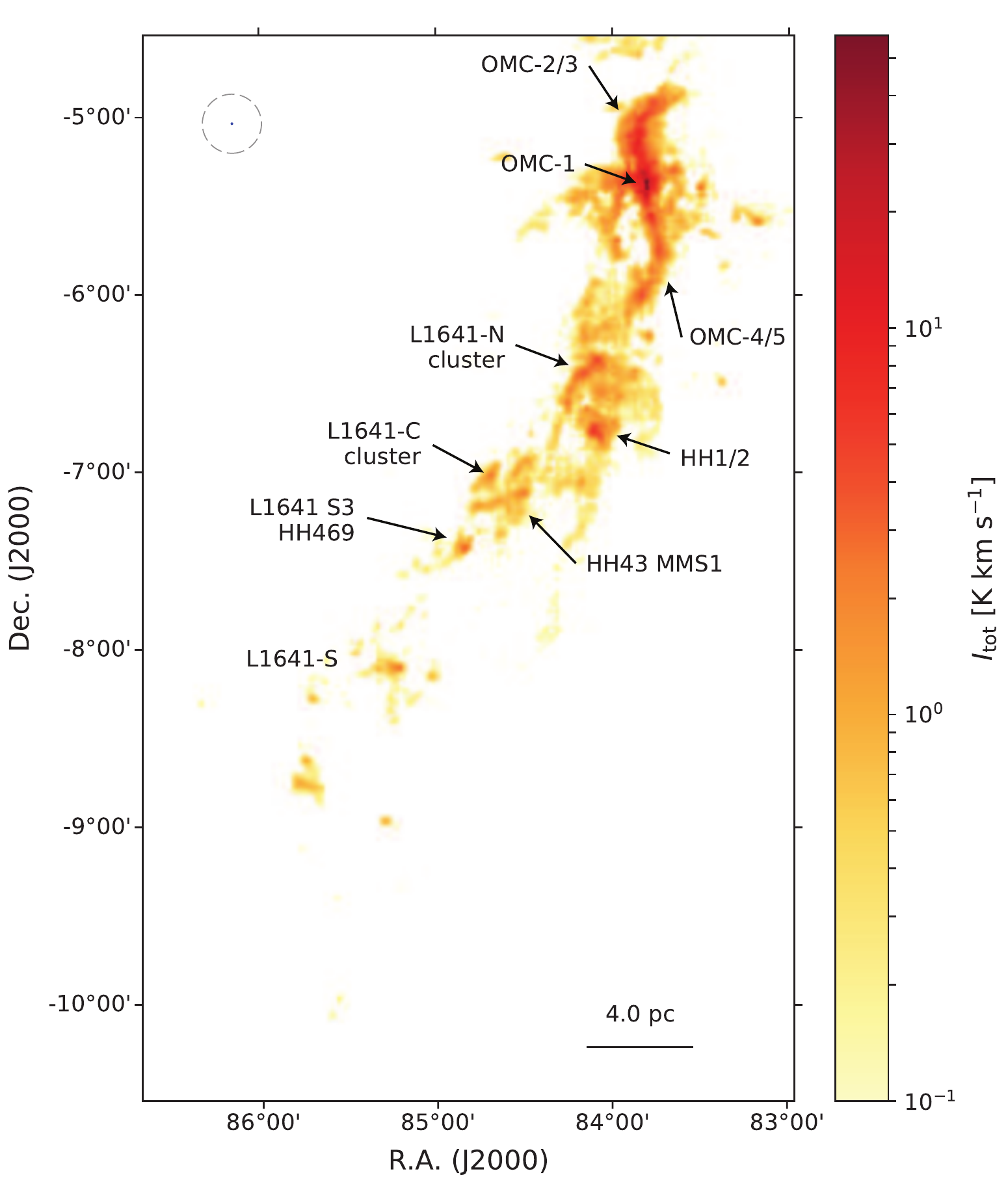}
\caption{The same as Figure \ref{fig_13CO_Ori} but for the the HCO$^+$ line. We annotated the map with the names of the associated sources \citep{Dav09,Mei16,Mai16}. \label{fig_Ori_HCOp}}
\end{figure}

\begin{figure}
\epsscale{1.0}
\plotone{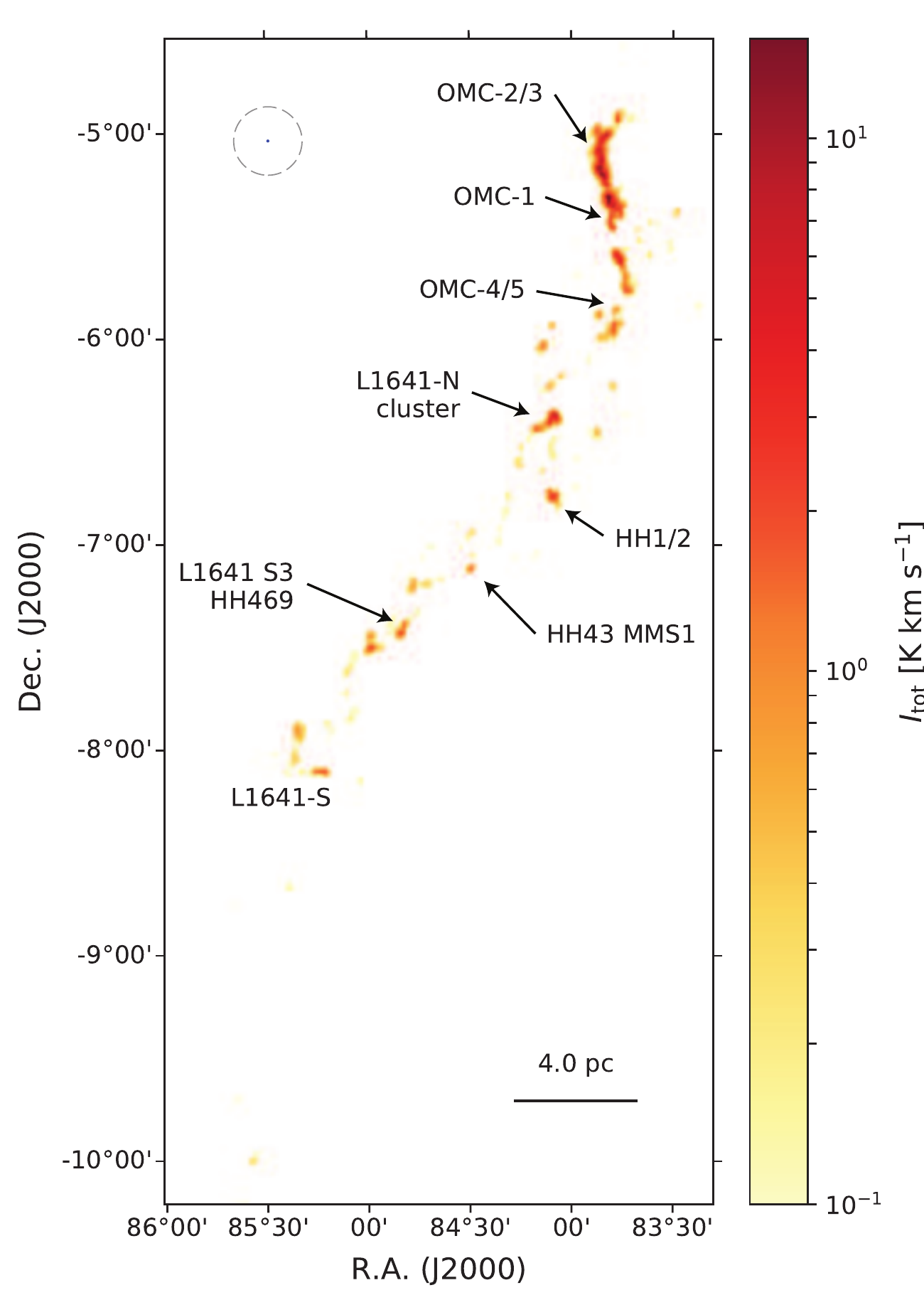}
\caption{The same as Figure \ref{fig_13CO_Ori} but for the the N$_2$H$^+$ line. We annotated the map with the names of the associated sources \citep{Dav09,Mei16,Mai16}. \label{fig_Ori_N2Hp}}
\end{figure}

\begin{figure}
\epsscale{1.0}
\plotone{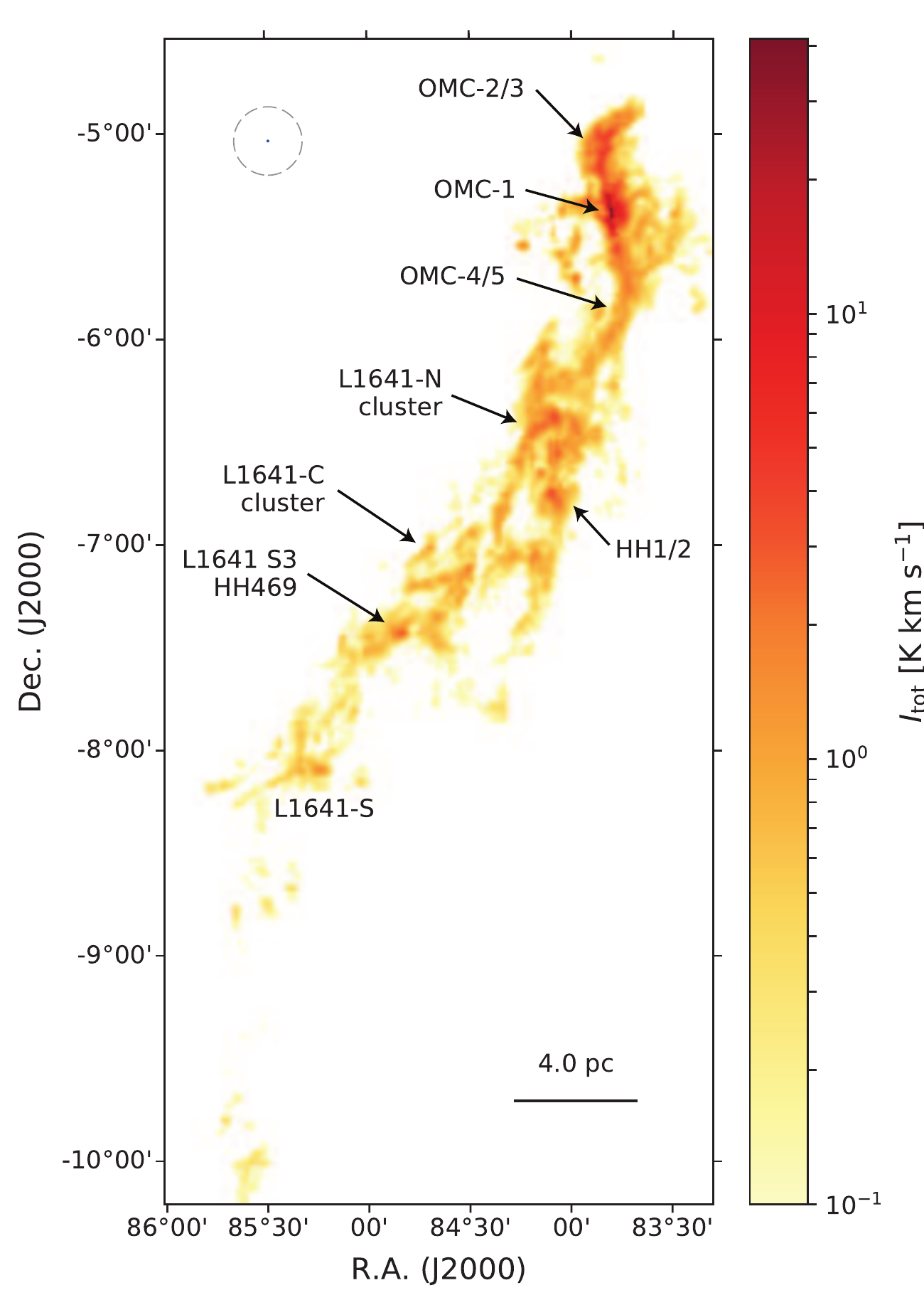}
\caption{The same as Figure \ref{fig_13CO_Ori} but for the the CS line. We annotated the map with the names of the associated sources \citep{Dav09,Mei16,Mai16}. \label{fig_Ori_CS}}
\end{figure}

\begin{figure}
\epsscale{1.0}
\plotone{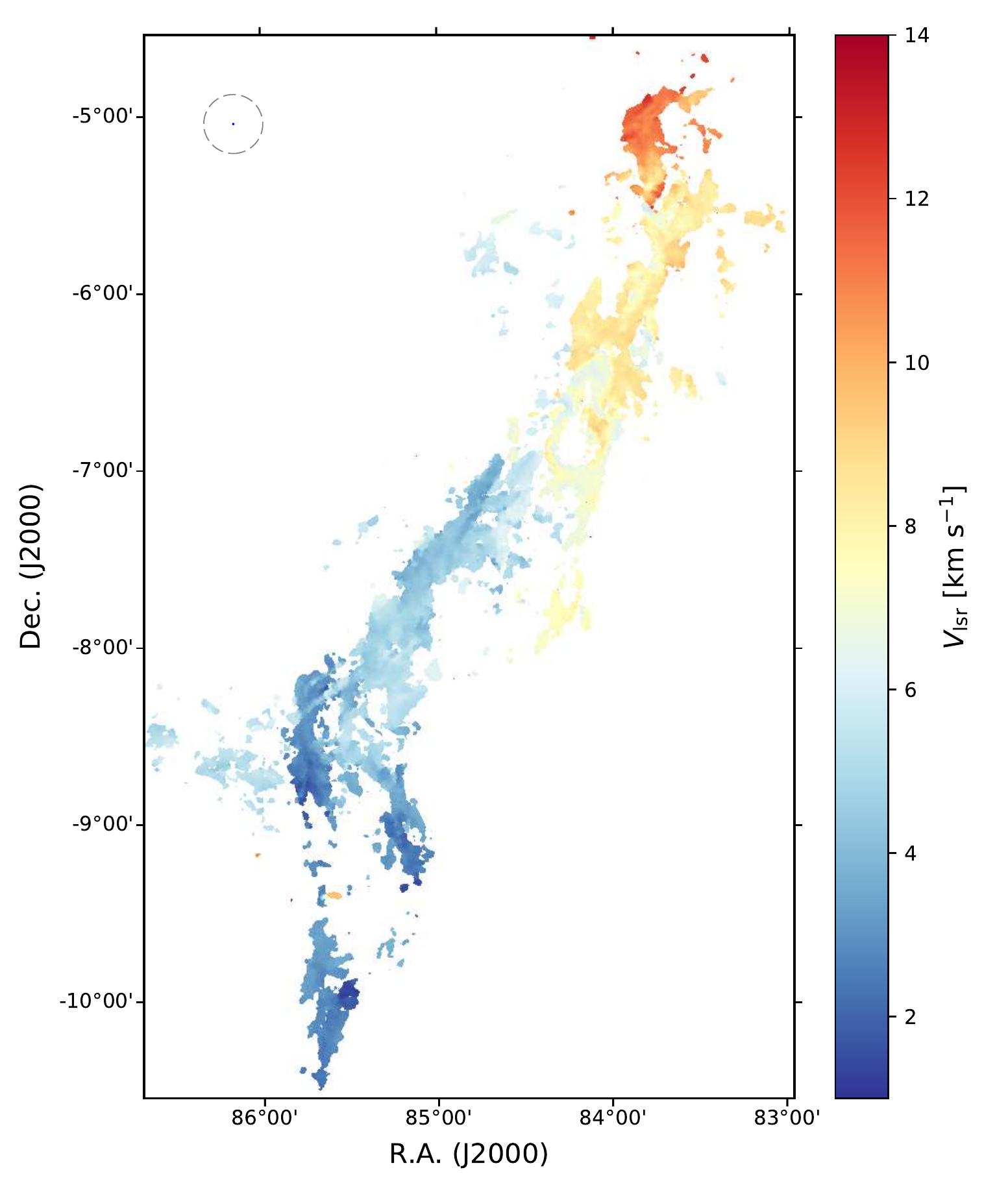}
\caption{Same as Figure \ref{fig_Ori_13CO_mmt1} but for the C$^{18}$O line. The color scale of the $V_\mathrm{lsr}$ map is the same as that of the $^{13}$CO line (Figure \ref{fig_Ori_C18O_mmt1}). \label{fig_Ori_C18O_mmt1}}
\end{figure}

\begin{figure}
\epsscale{1.0}
\plotone{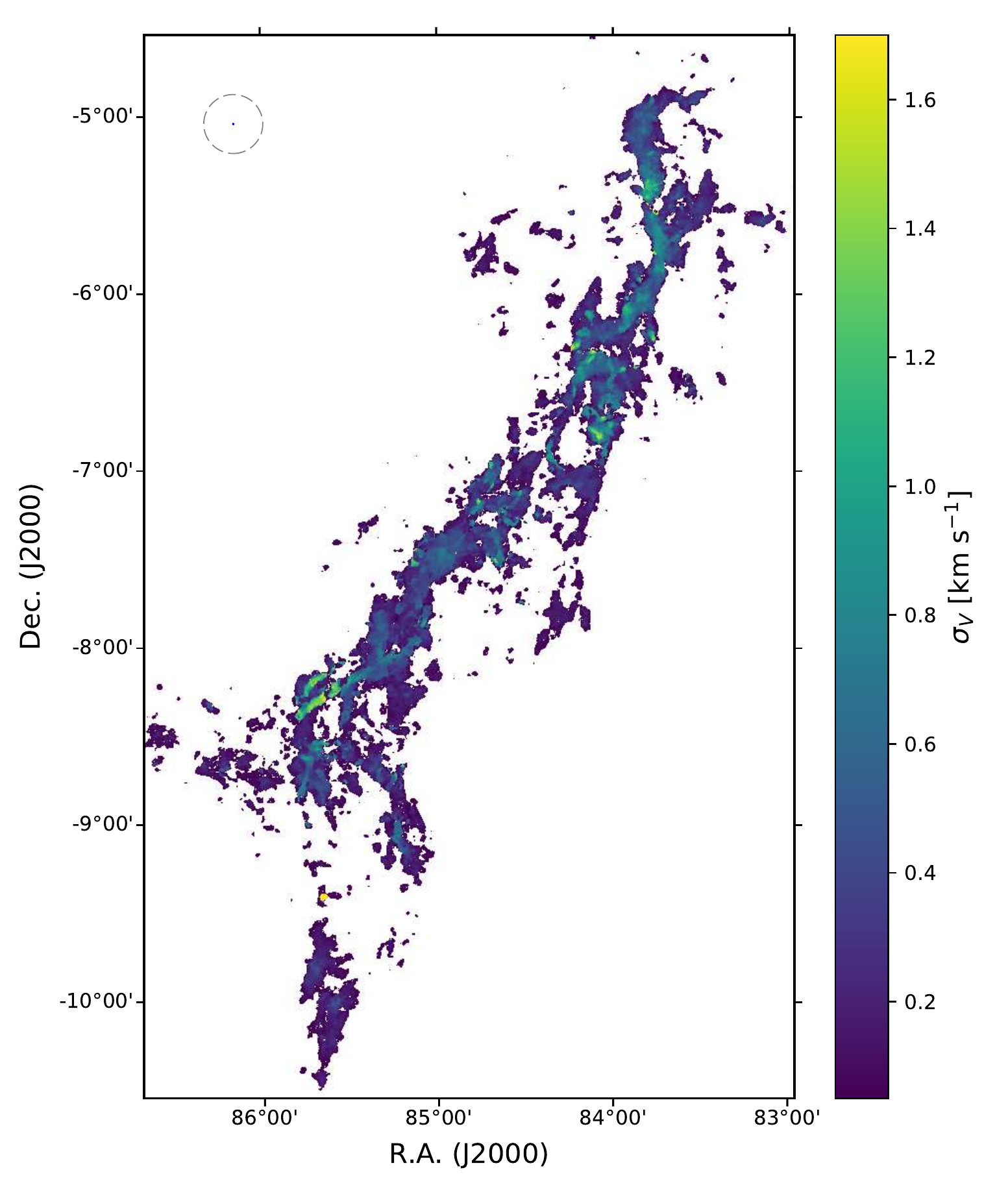}
\caption{Same as Figure \ref{fig_Ori_13CO_mmt2} but for the C$^{18}$O line. \label{fig_Ori_C18O_mmt2}}
\end{figure}

\begin{figure}
\epsscale{1.0}
\plotone{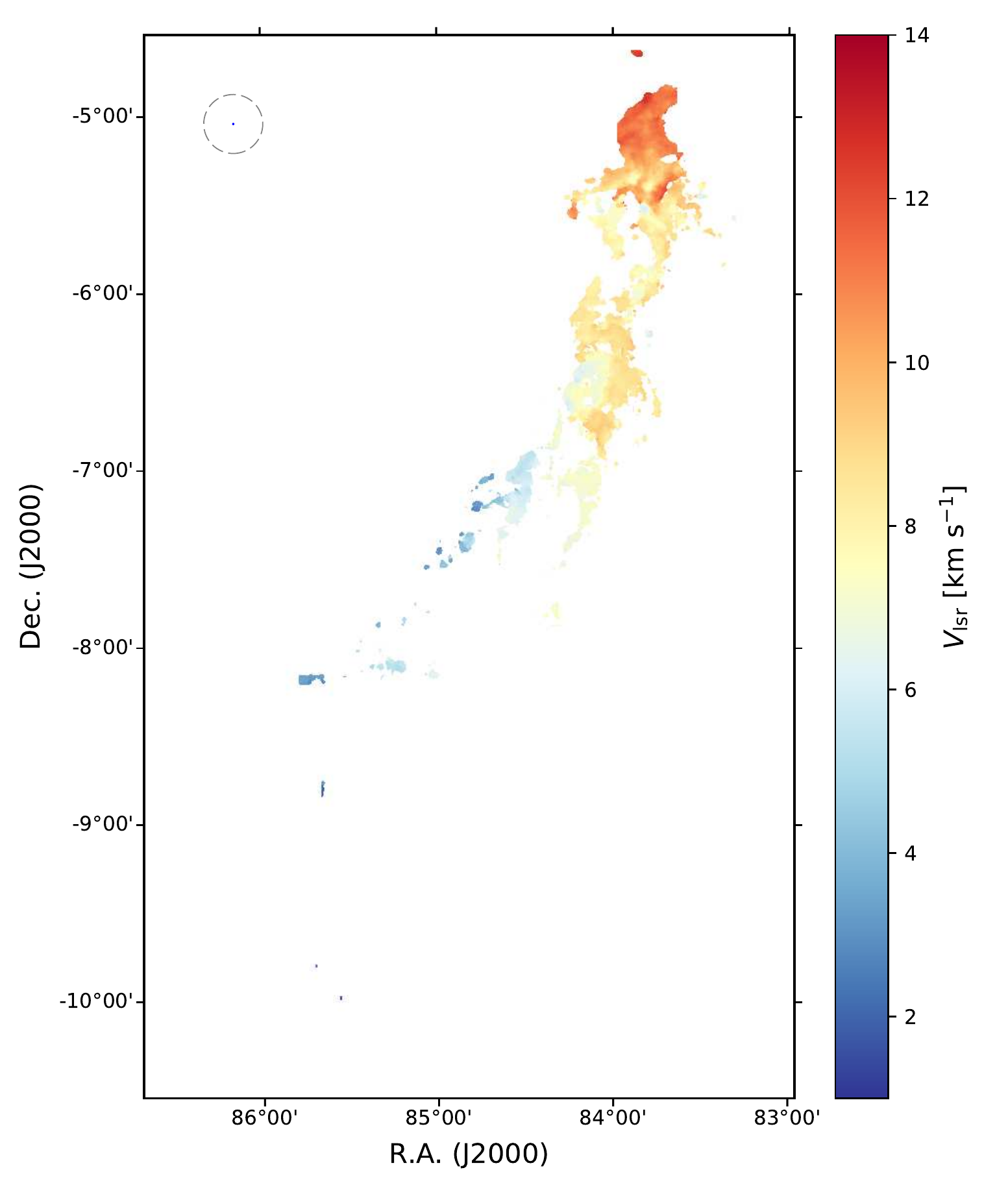}
\caption{Same as Figure \ref{fig_Ori_13CO_mmt1} but for the HCN line. \label{fig_Ori_HCN_mmt1}}
\end{figure}

\begin{figure}
\epsscale{1.0}
\plotone{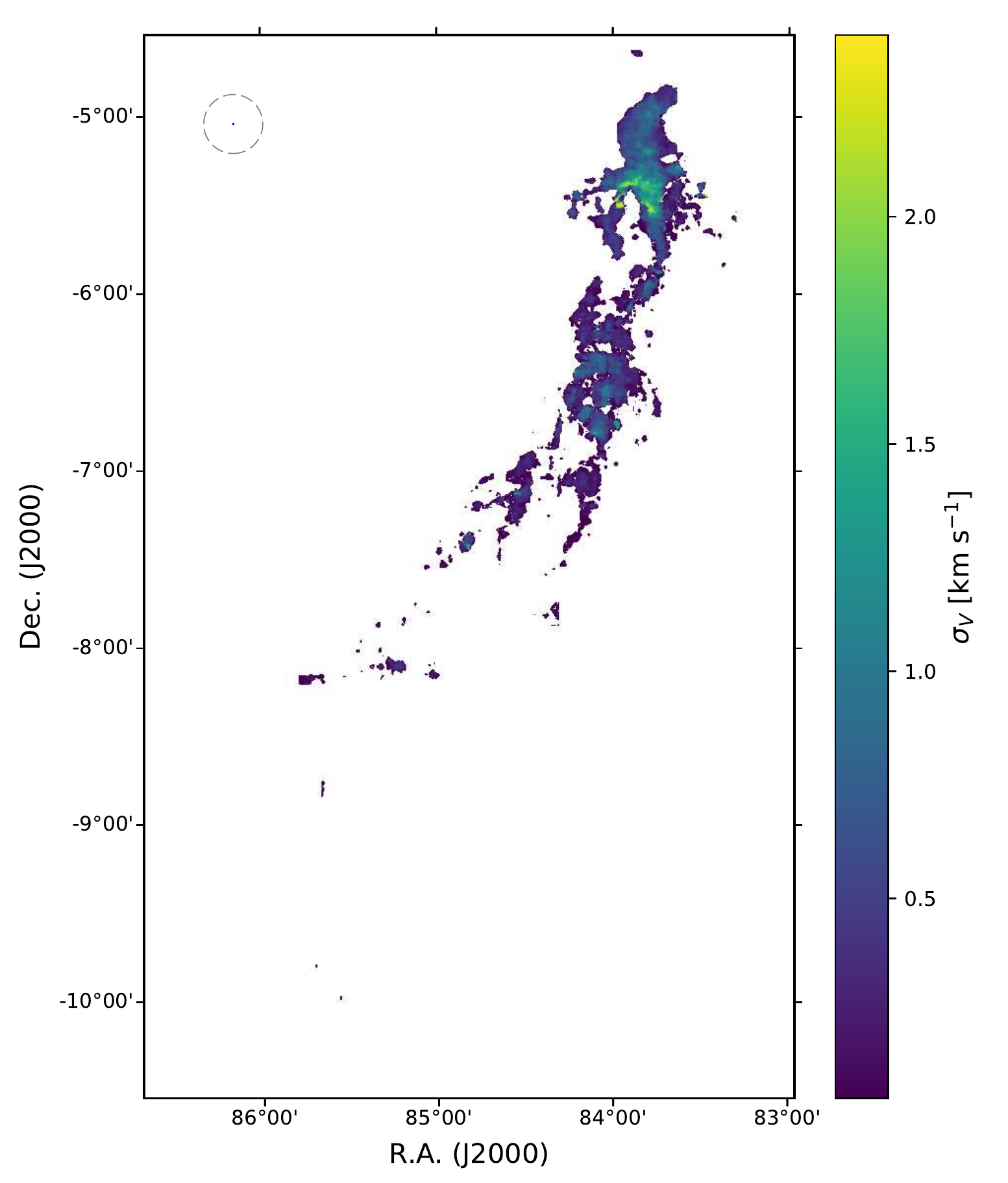}
\caption{Same as Figure \ref{fig_Ori_13CO_mmt2} but for the HCN line. Note that the $\sigma_V$ of the HCN line near the ISF would be underestimated (see Section \ref{Sec_anal} for more details). \label{fig_Ori_HCN_mmt2}}
\end{figure}

\begin{figure}
\epsscale{1.0}
\plotone{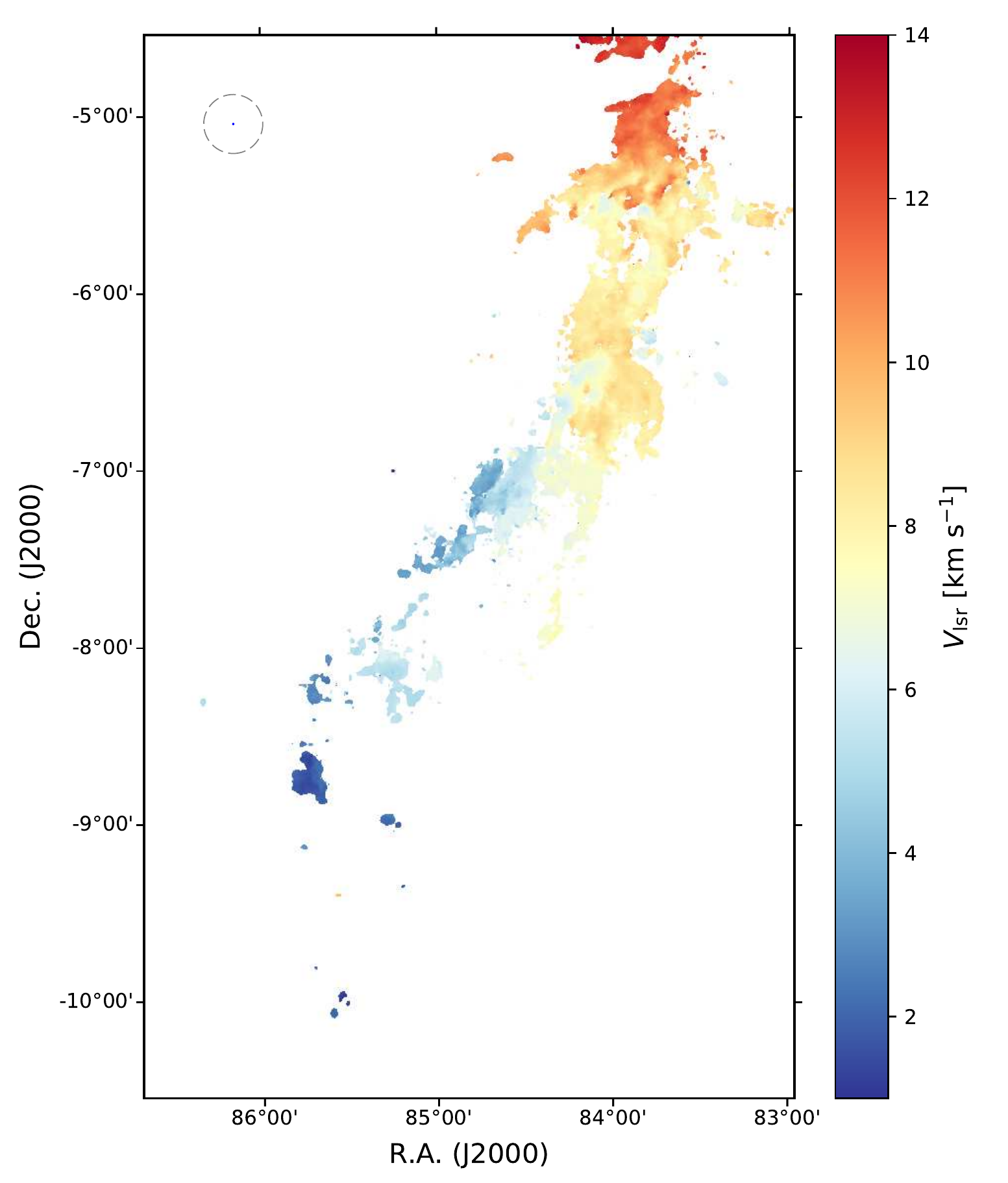}
\caption{Same as Figure \ref{fig_Ori_13CO_mmt1} but for the HCO$^+$ line. \label{fig_Ori_HCOp_mmt1}}
\end{figure}

\begin{figure}
\epsscale{1.0}
\plotone{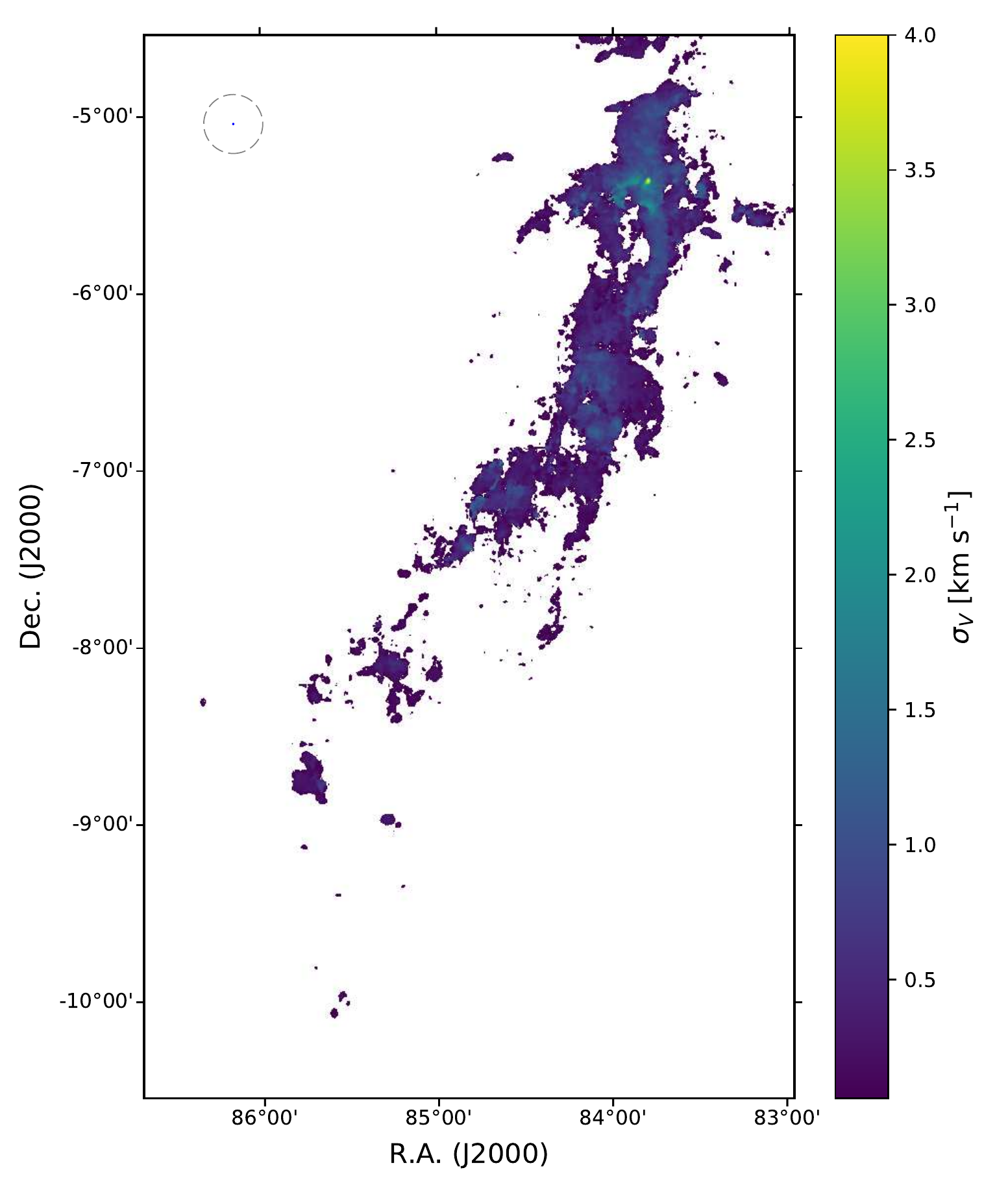}
\caption{Same as Figure \ref{fig_Ori_13CO_mmt2} but for the HCO$^+$ line. \label{fig_Ori_HCOp_mmt2}}
\end{figure}

\begin{figure}
\epsscale{1.0}
\plotone{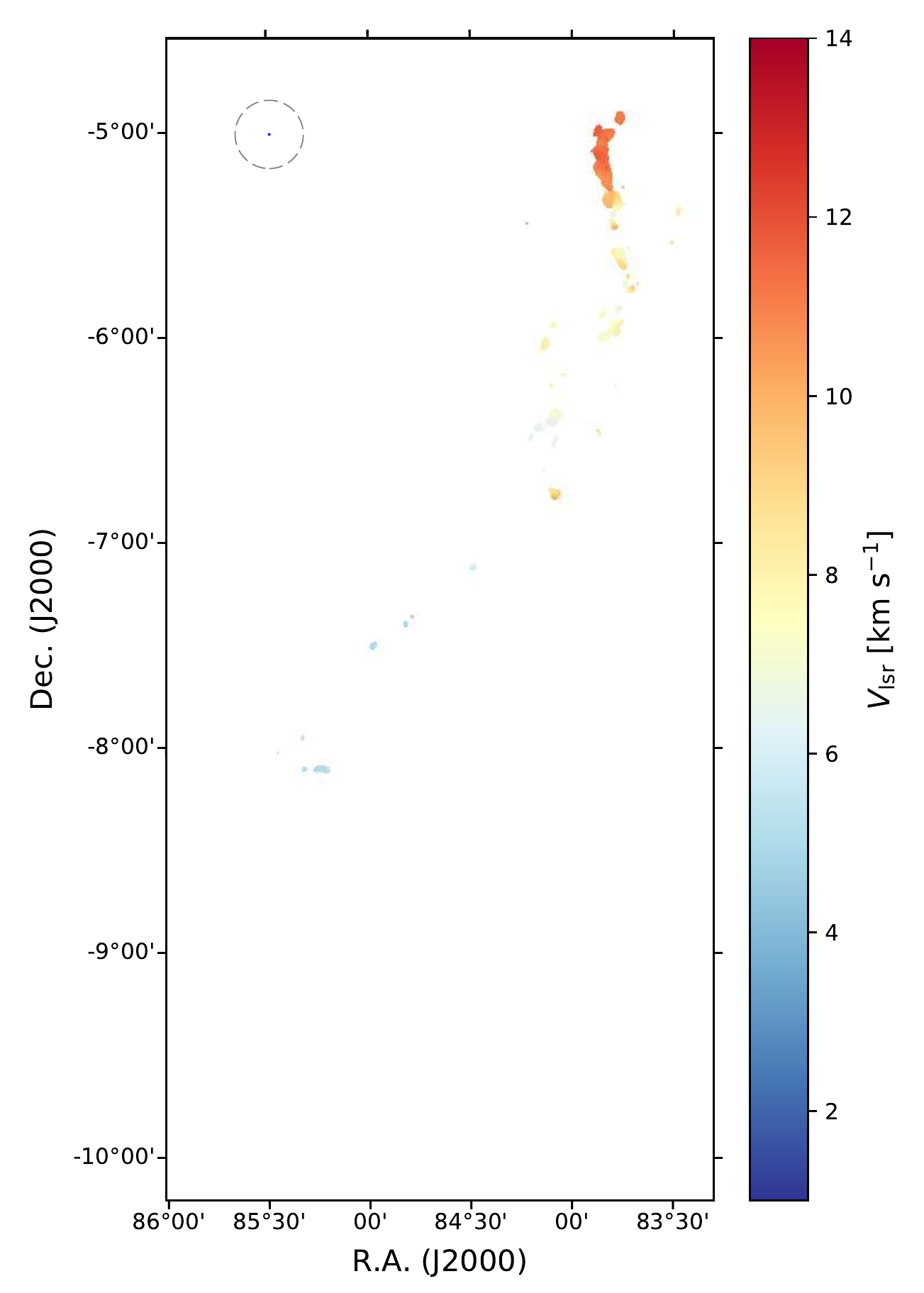}
\caption{Same as Figure \ref{fig_Ori_13CO_mmt1} but for the N$_2$H$^+$ line. \label{fig_Ori_N2Hp_mmt1}}
\end{figure}

\begin{figure}
\epsscale{1.0}
\plotone{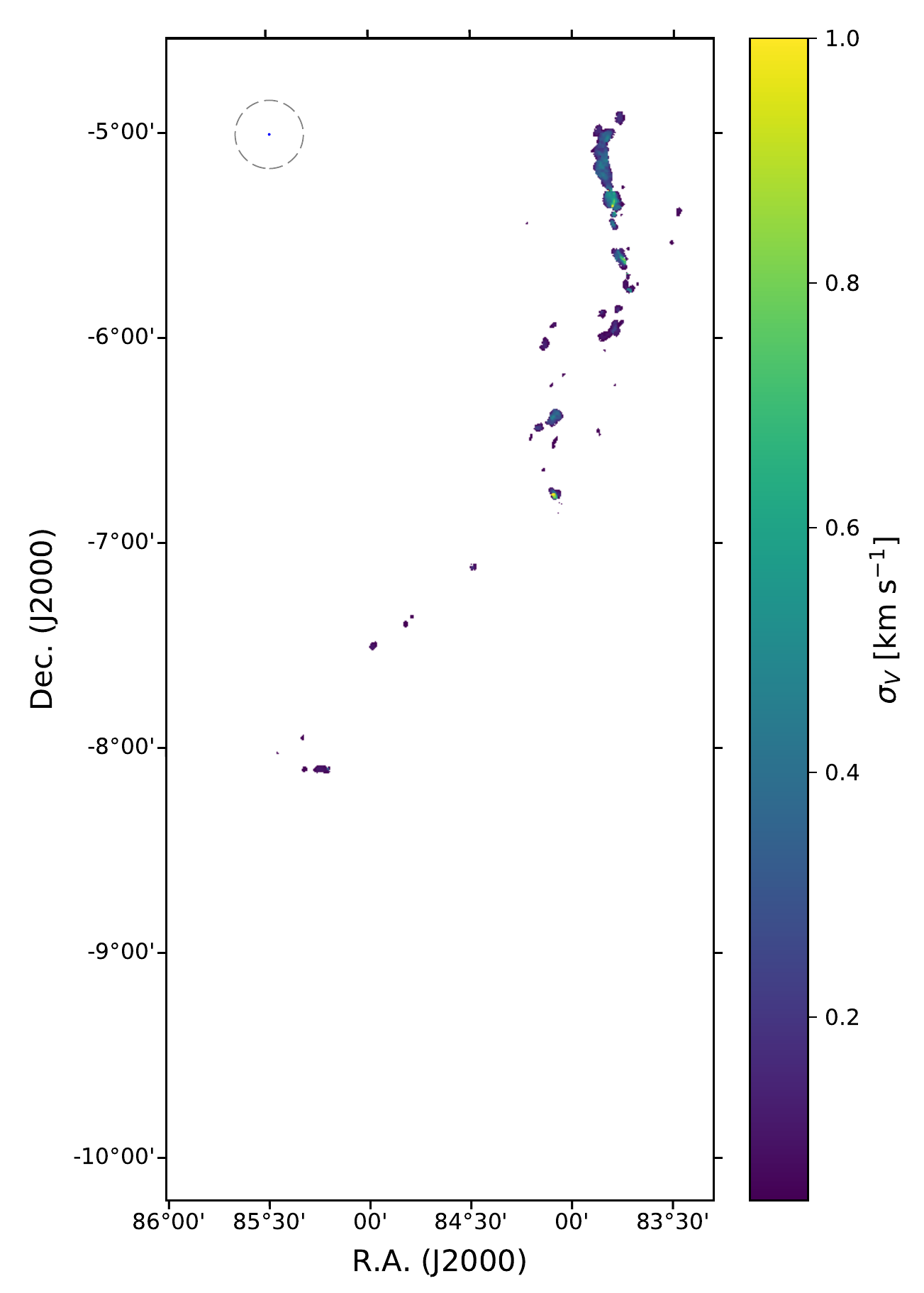}
\caption{Same as Figure \ref{fig_Ori_13CO_mmt2} but for the N$_2$H$^+$ line. \label{fig_Ori_N2Hp_mmt2}}
\end{figure}

\begin{figure}
\epsscale{1.0}
\plotone{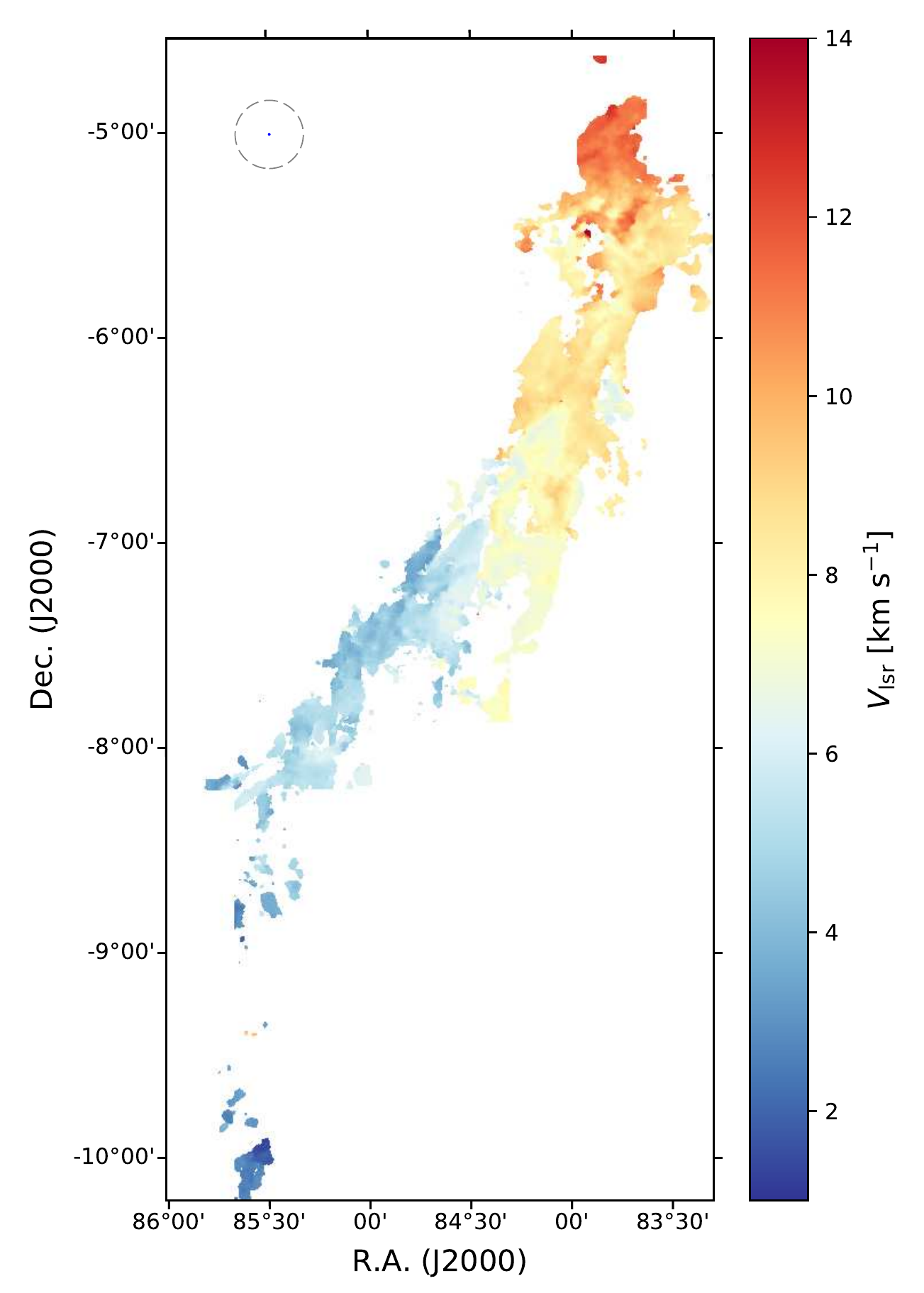}
\caption{Same as Figure \ref{fig_Ori_13CO_mmt1} but for the CS line. \label{fig_Ori_CS_mmt1}}
\end{figure}

\begin{figure}
\epsscale{1.0}
\plotone{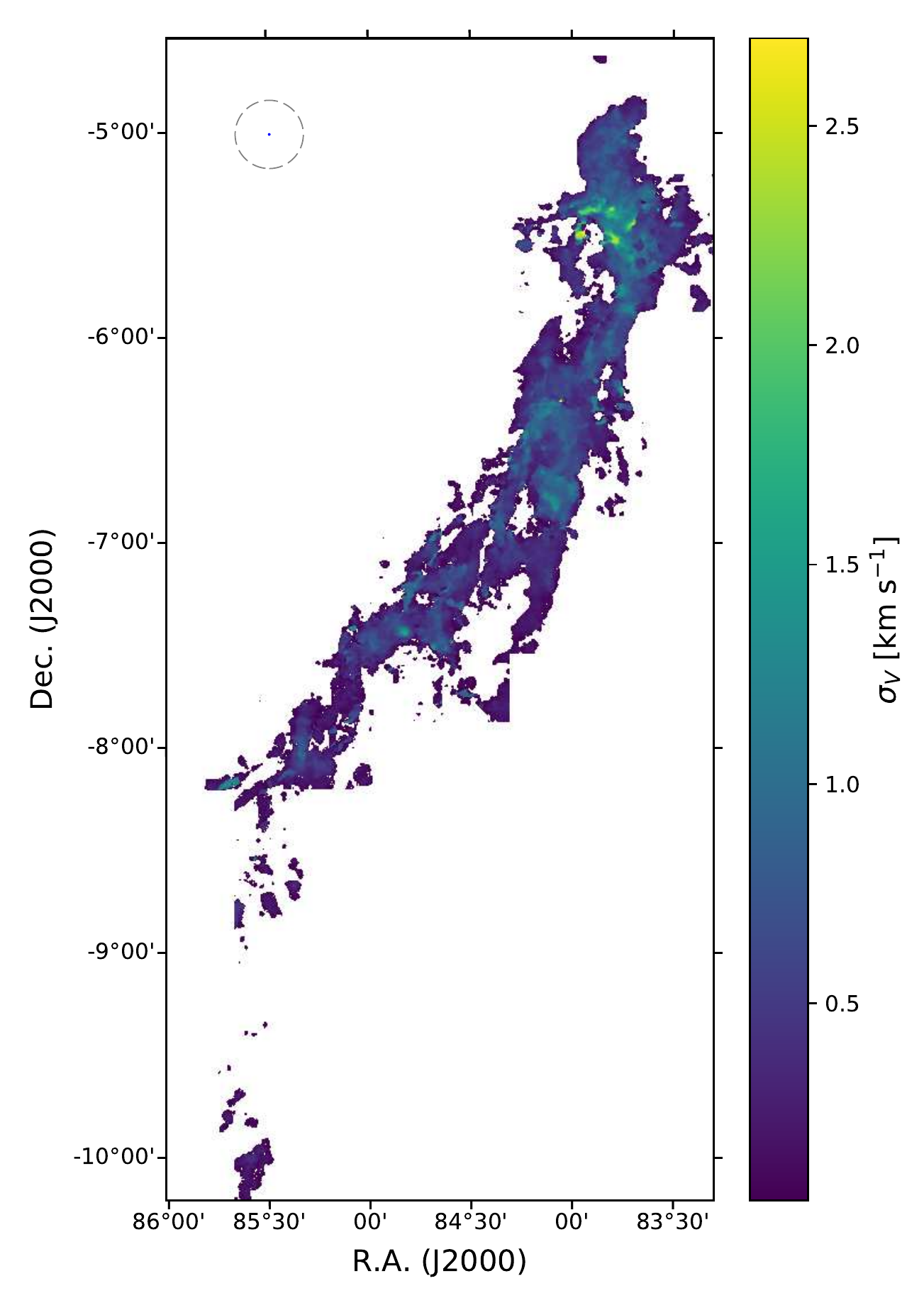}
\caption{Same as Figure \ref{fig_Ori_13CO_mmt2} but for the CS line. \label{fig_Ori_CS_mmt2}}
\end{figure}
\clearpage

\section{Moment 0, 1, and 2 maps for the Ophiuchus cloud} \label{App_mmtmaps_Oph}
\setcounter{figure}{0}
\begin{figure}
\epsscale{1.0}
\plotone{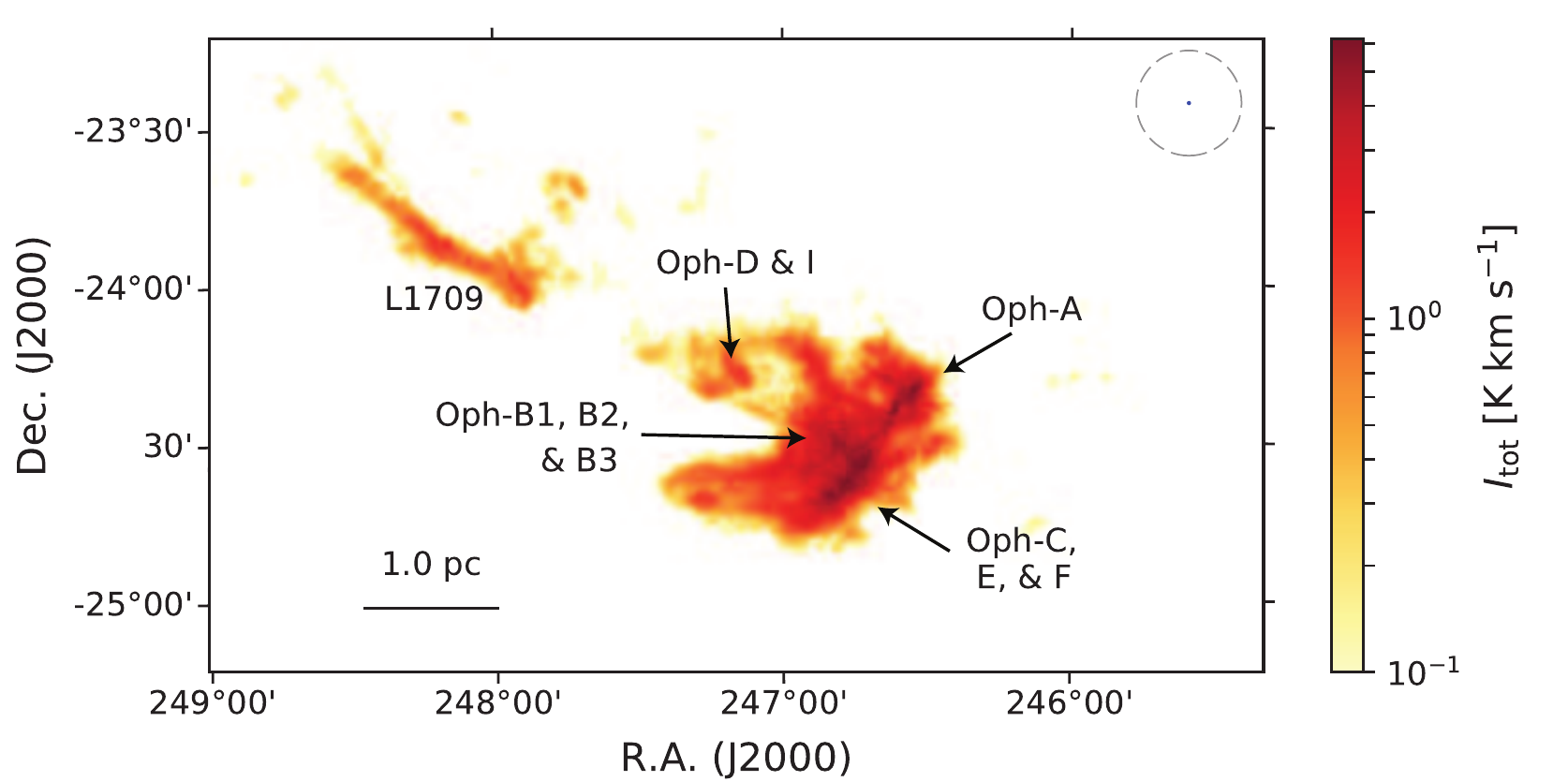}
\caption{Same as Figure \ref{fig_13CO_Oph} except for the C$^{18}$O line. The sources associated with the observed line is marked \citep{Lyn62,Lor89a,Lor90,Pan17}. \label{fig_Oph_C18O}}
\end{figure}

\begin{figure}
\epsscale{1.0}
\plotone{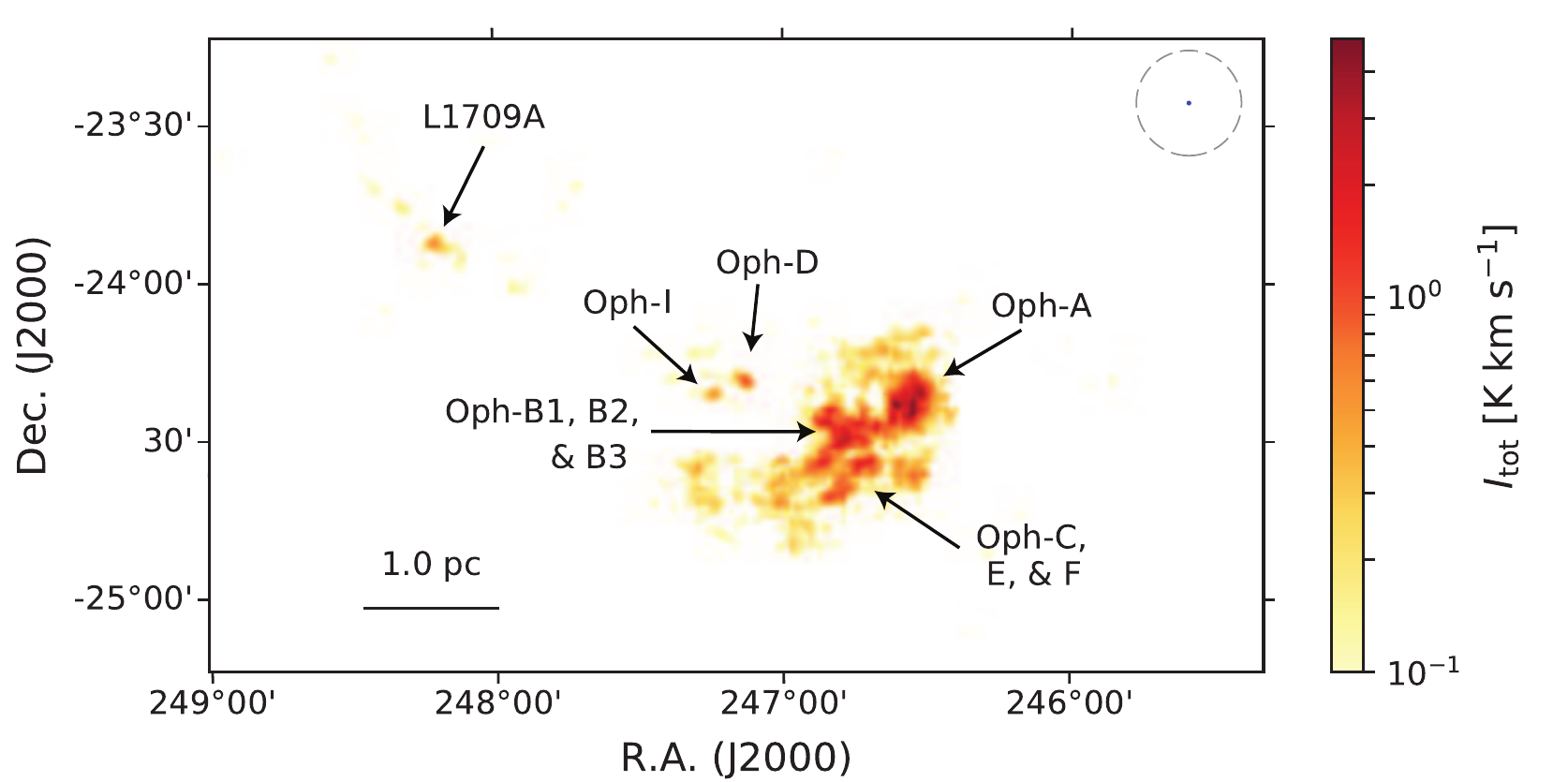}
\caption{Same as Figure \ref{fig_13CO_Oph} except for the HCN line. The sources associated with the observed line is marked \citep{Lyn62,Lor89a,Lor90,Pan17}. \label{fig_Oph_HCN}}
\end{figure}

\begin{figure}
\epsscale{1.0}
\plotone{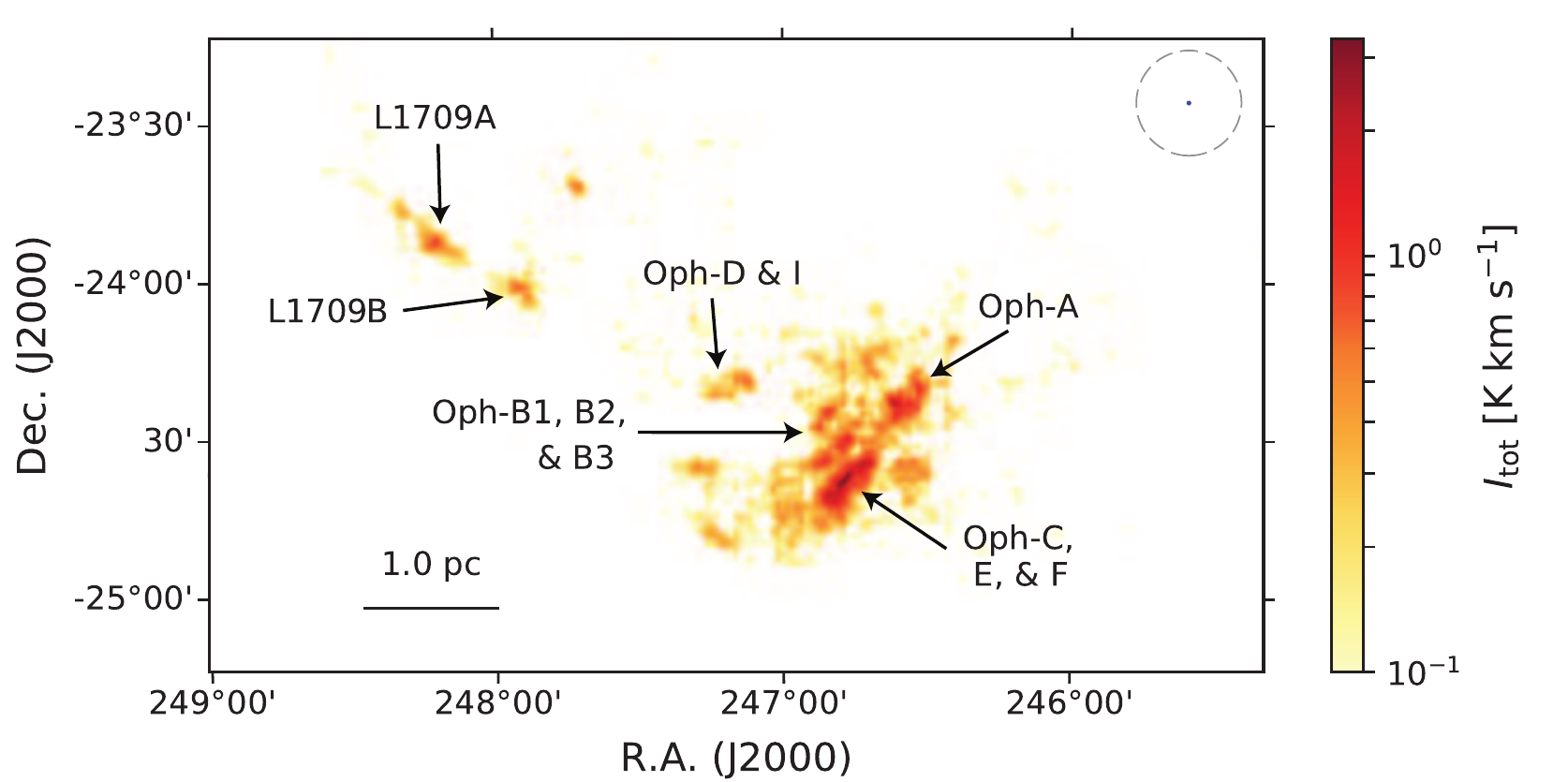}
\caption{Same as Figure \ref{fig_13CO_Oph} except for the HCO$^+$ line. The sources associated with the observed line is marked \citep{Lyn62,Lor89a,Lor90,Pan17}. \label{fig_Oph_HCOp}}
\end{figure}

\begin{figure}
\epsscale{1.0}
\plotone{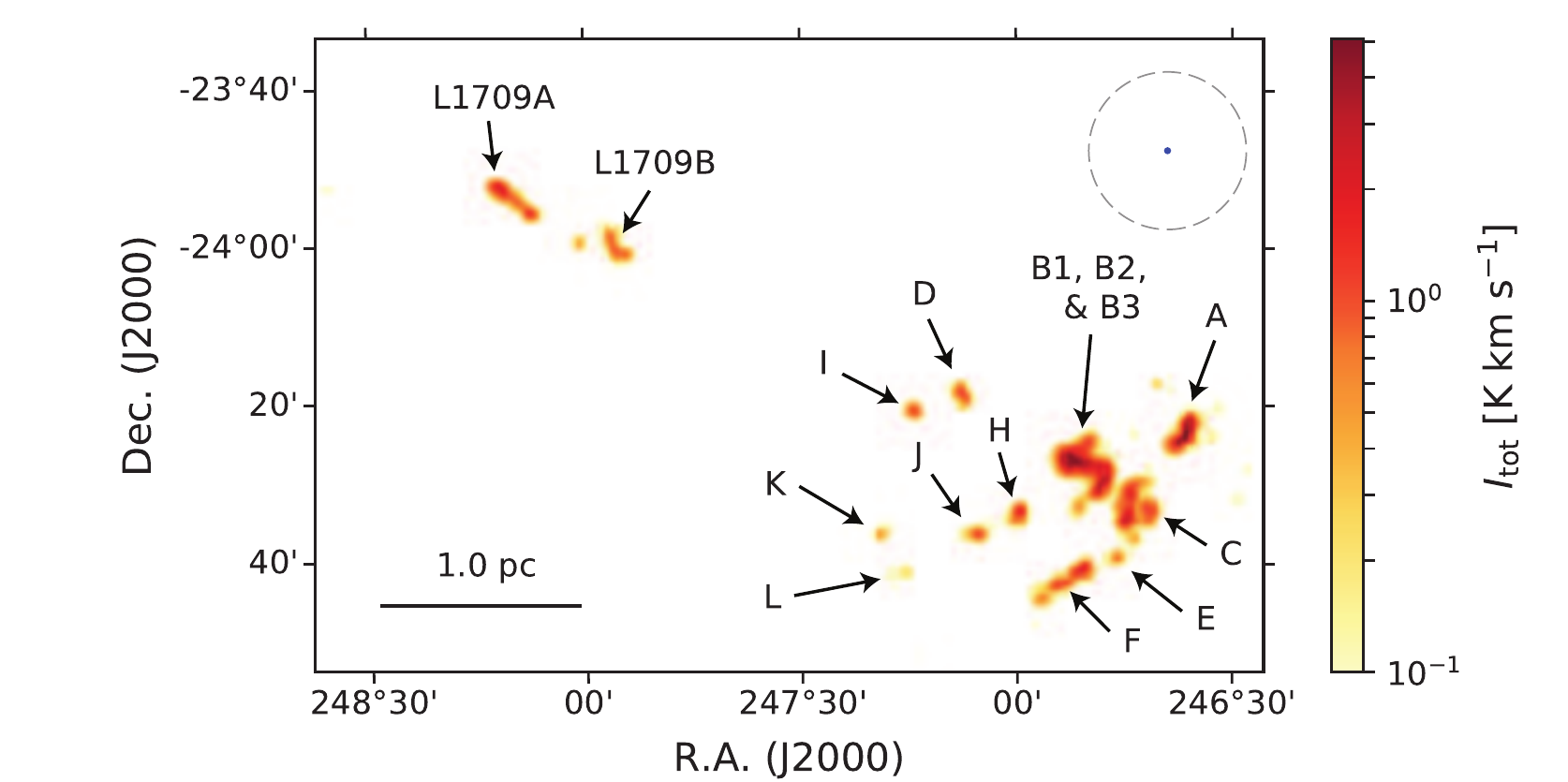}
\caption{Same as Figure \ref{fig_13CO_Oph} except for the N$_2$H$^+$ line. The names of the associated cores are marked \citep{Lyn62,Lor89a,Lor90,Pan17}. \label{fig_Oph_N2Hp}}
\end{figure}

\begin{figure}
\epsscale{1.0}
\plotone{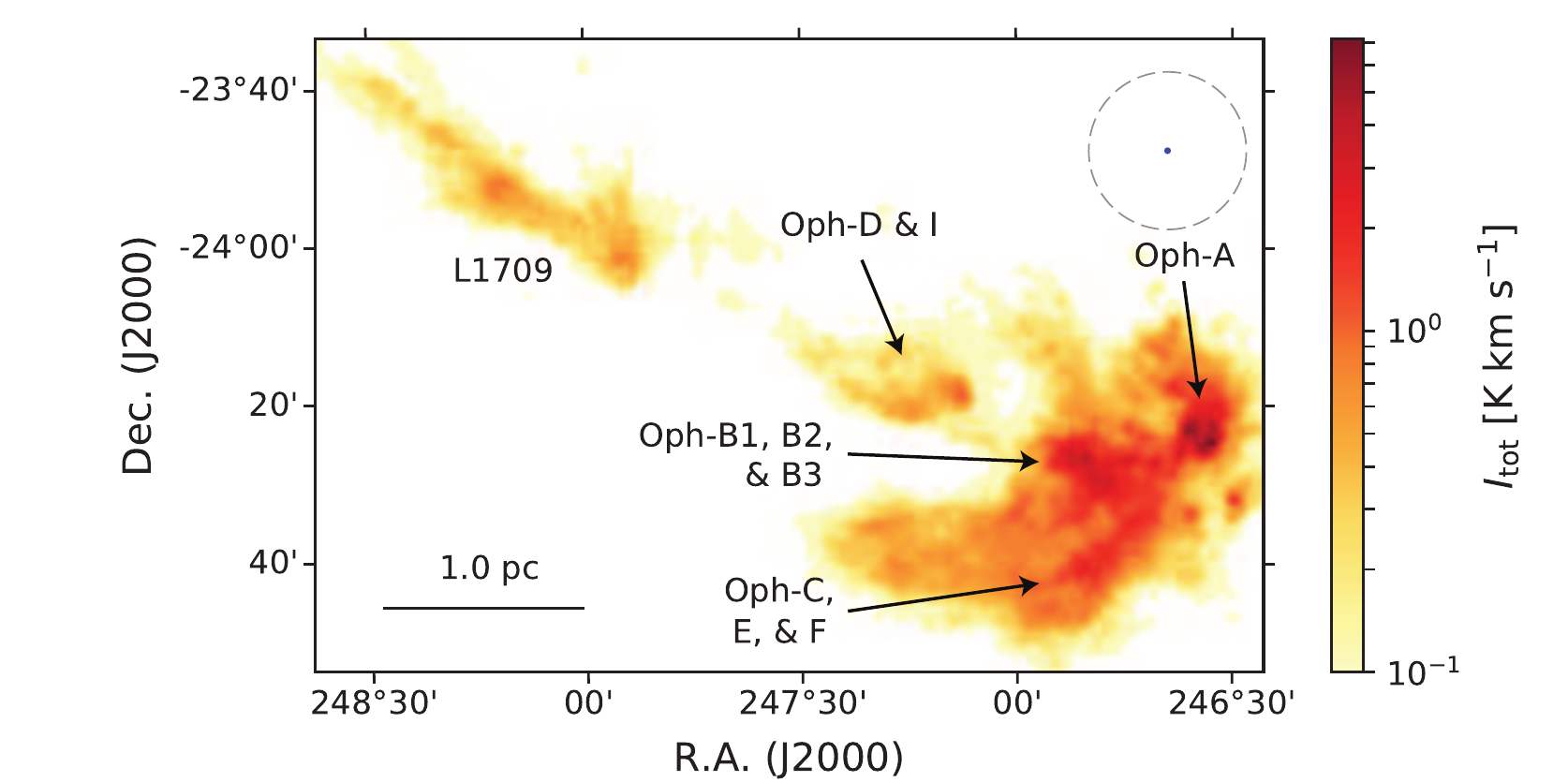}
\caption{Same as Figure \ref{fig_13CO_Oph} except for the CS line. The sources associated with the observed line is marked \citep{Lyn62,Lor89a,Lor90,Pan17}. \label{fig_Oph_CS}}
\end{figure}

\begin{figure}
\epsscale{1.0}
\plotone{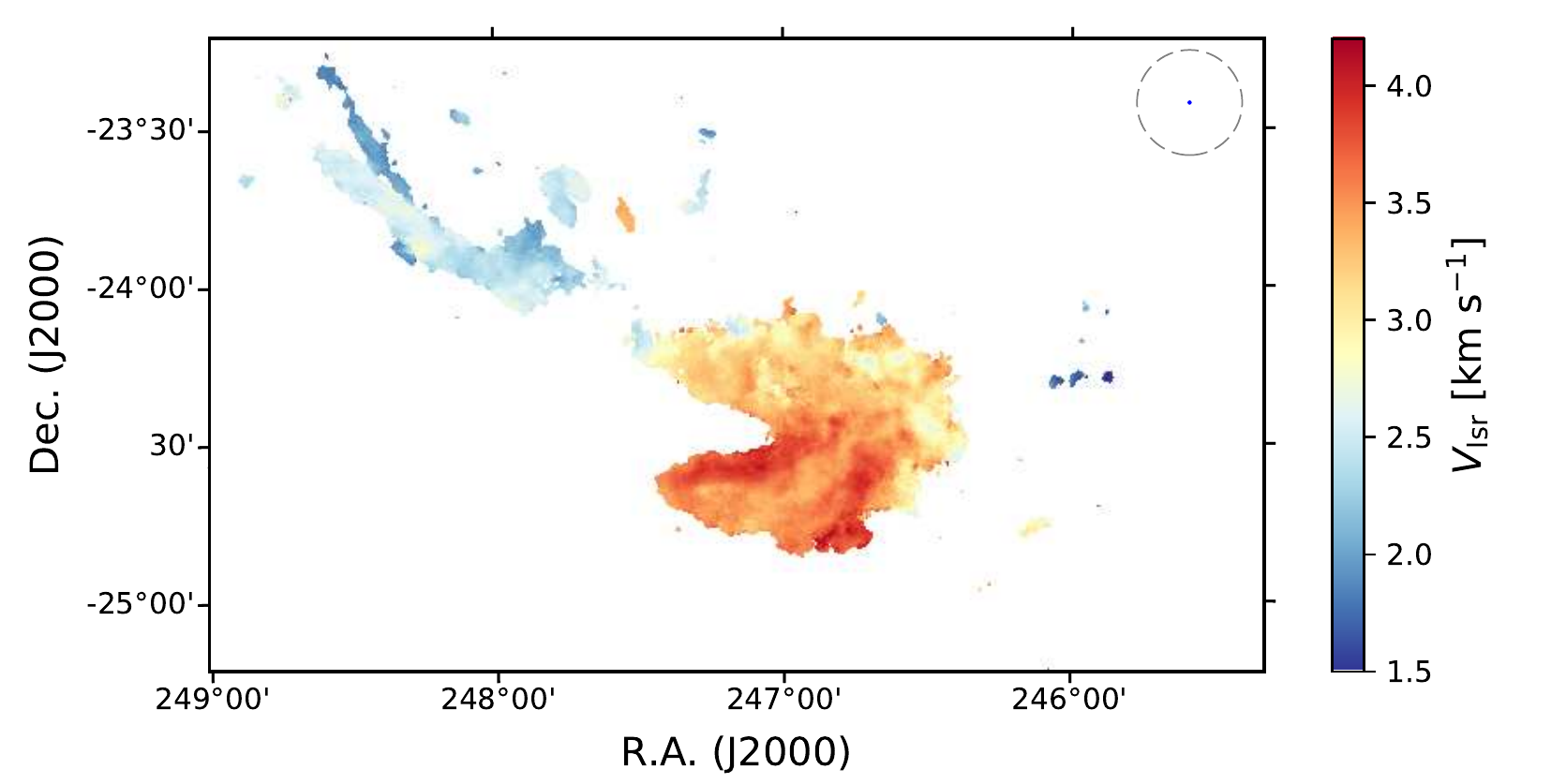}
\caption{Same as Figure \ref{fig_Oph_13CO_mmt1} except for the C$^{18}$O line. The color scale of the map is the same as that of the $^{13}$CO line (Figure \ref{fig_Oph_13CO_mmt1}). \label{fig_Oph_C18O_mmt1}}
\end{figure}

\begin{figure}
\epsscale{1.0}
\plotone{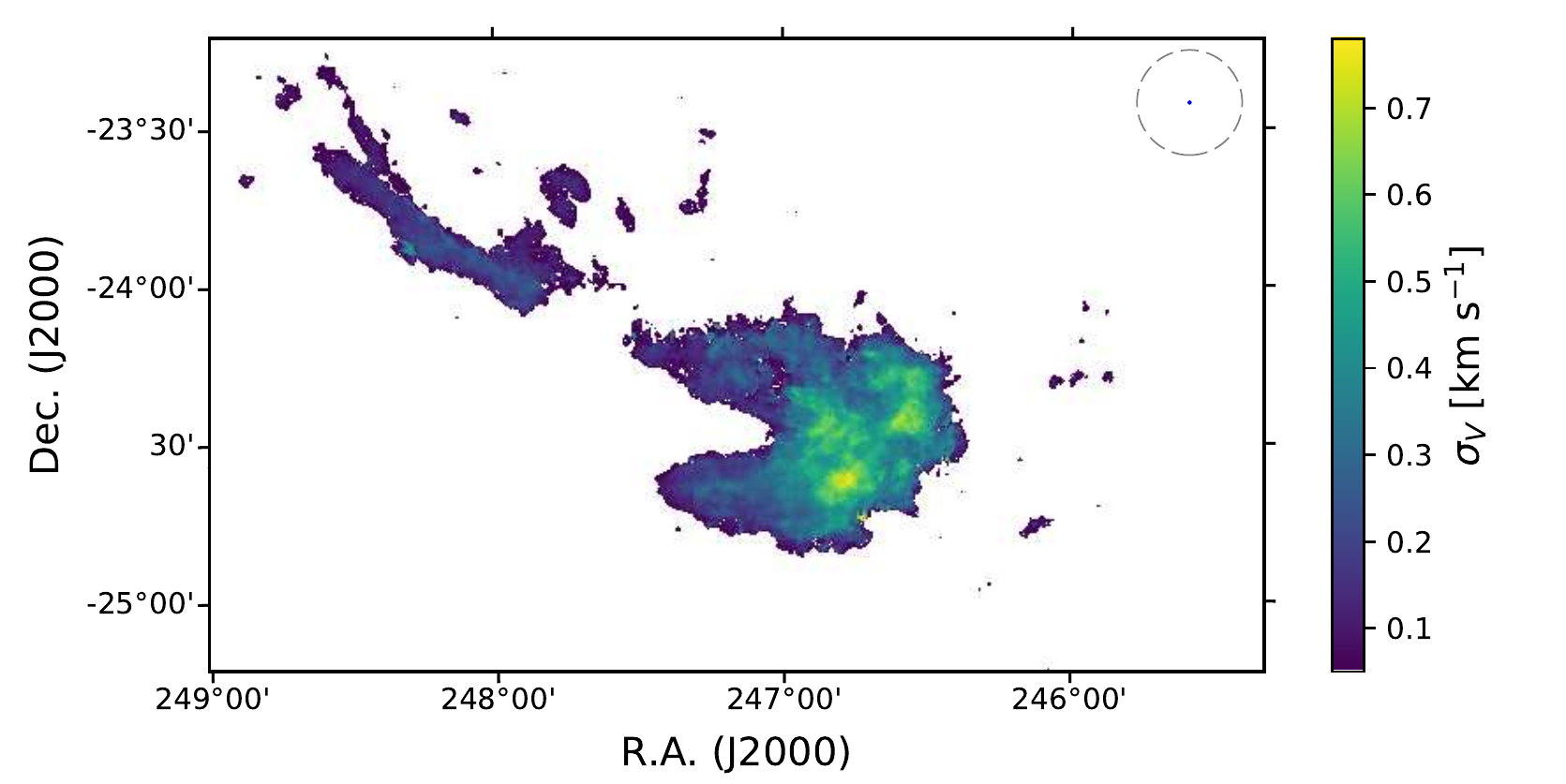}
\caption{Same as Figure \ref{fig_Oph_13CO_mmt2} except for the C$^{18}$O line. \label{fig_Oph_C18O_mmt2}}
\end{figure}

\begin{figure}
\epsscale{1.0}
\plotone{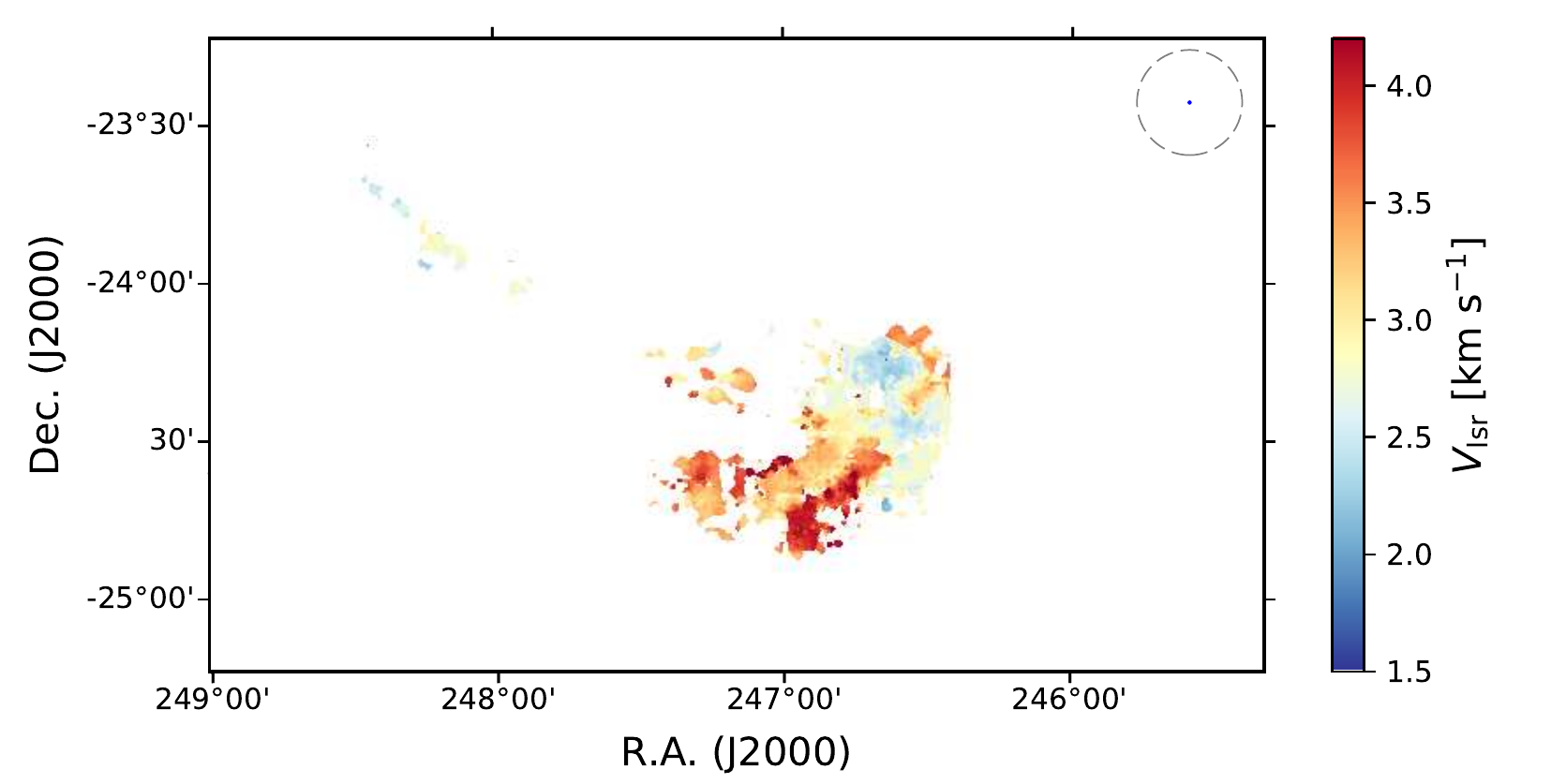}
\caption{Same as Figure \ref{fig_Oph_13CO_mmt1} but for the HCN line. \label{fig_Oph_HCN_mmt1}}
\end{figure}

\begin{figure}
\epsscale{1.0}
\plotone{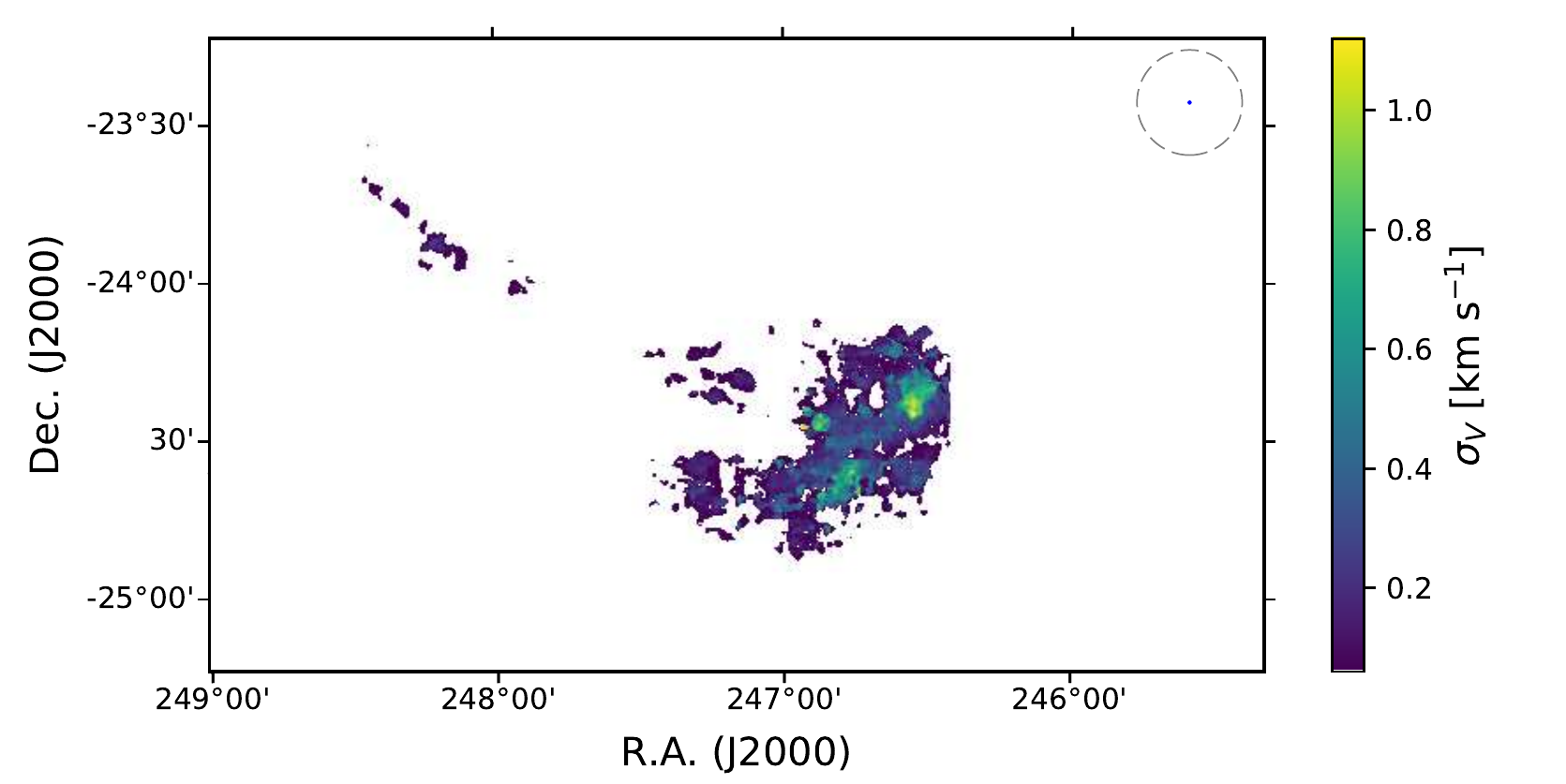}
\caption{Same as Figure \ref{fig_Oph_13CO_mmt2} but for the HCN line. \label{fig_Oph_HCN_mmt2}}
\end{figure}

\begin{figure}
\epsscale{1.0}
\plotone{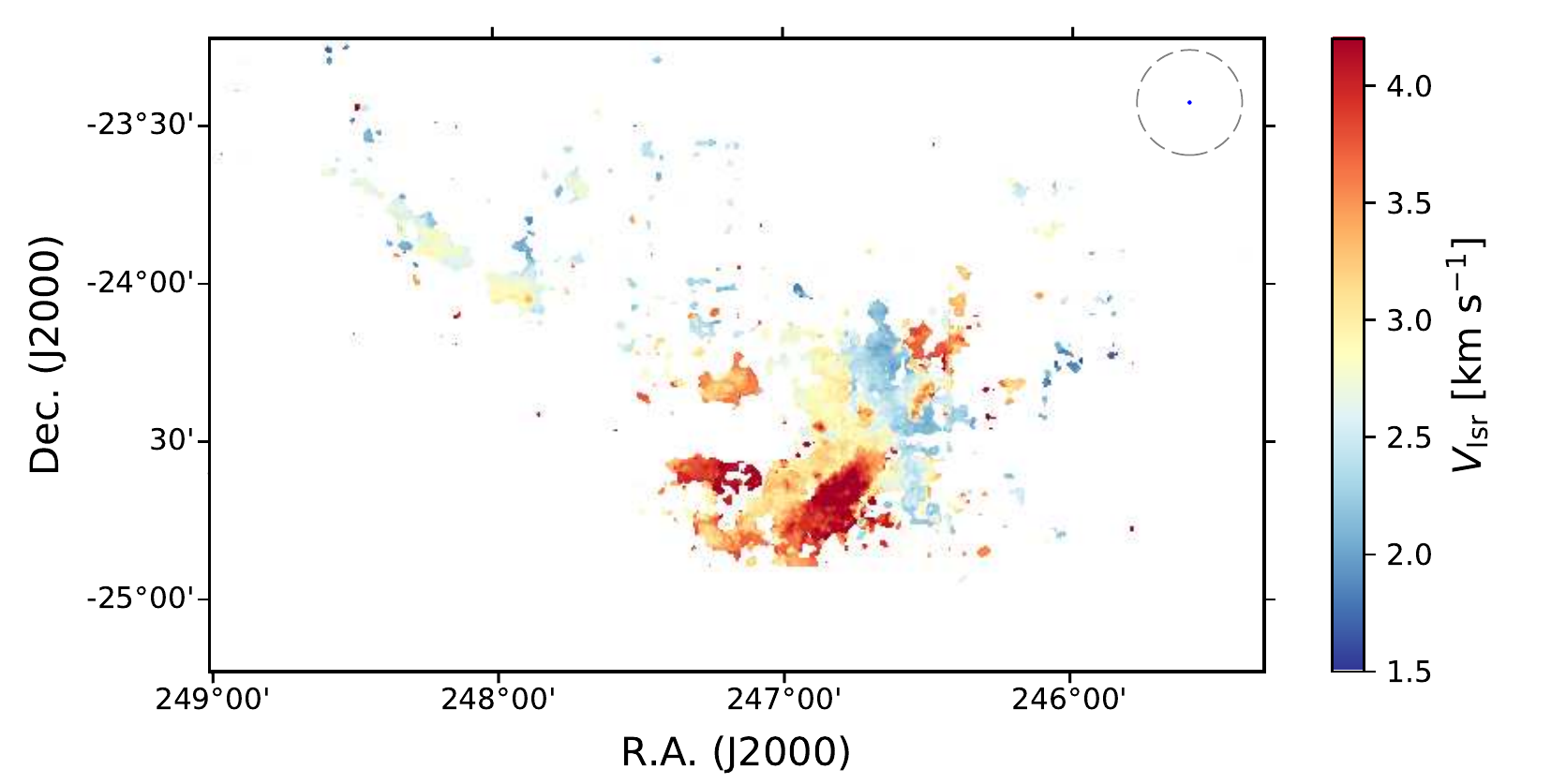}
\caption{Same as Figure \ref{fig_Oph_13CO_mmt1} but for the HCO$^+$ line. \label{fig_Oph_HCOp_mmt1}}
\end{figure}

\begin{figure}
\epsscale{1.0}
\plotone{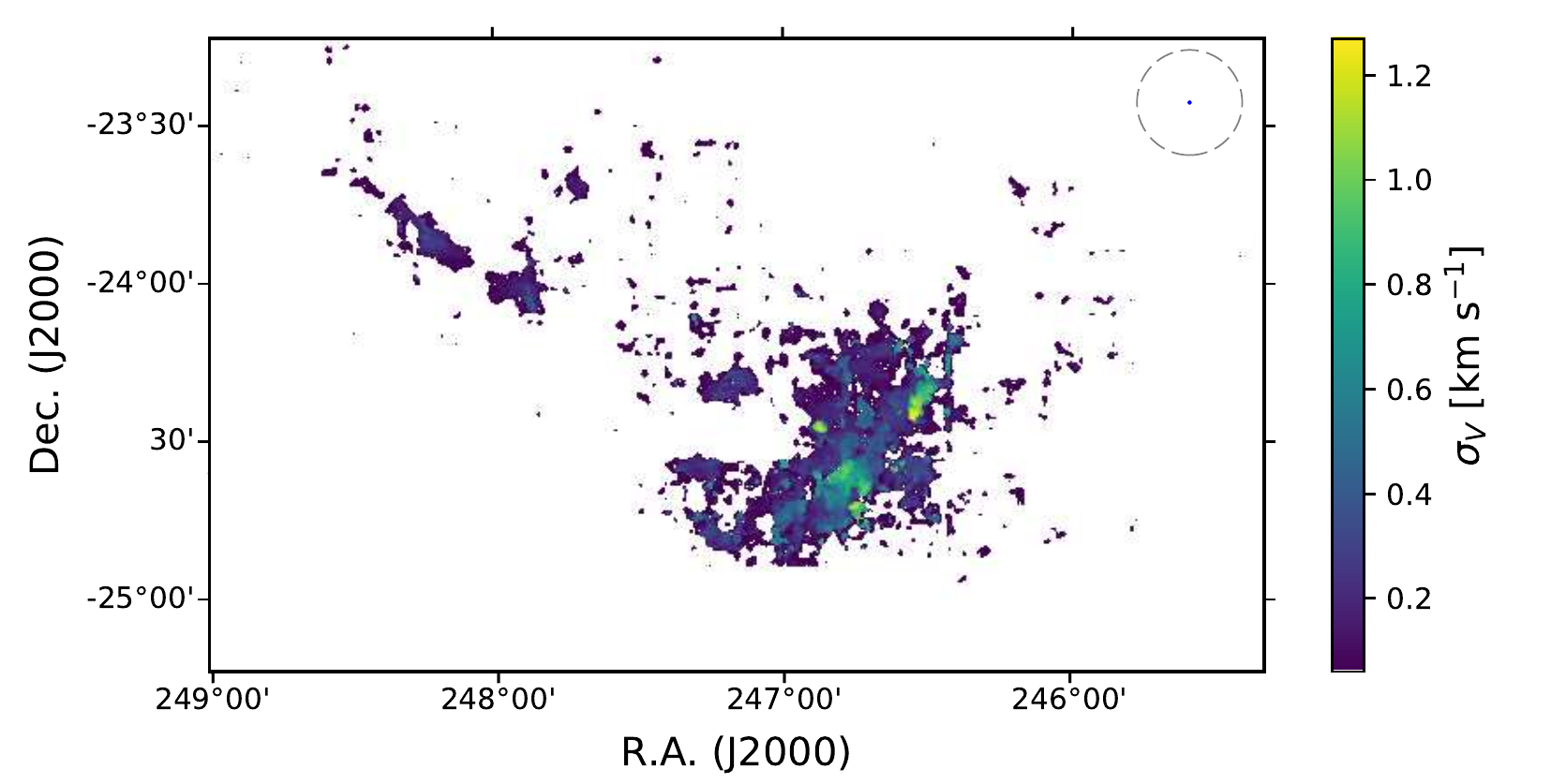}
\caption{Same as Figure \ref{fig_Oph_13CO_mmt2} but for the HCO$^+$ line. \label{fig_Oph_HCOp_mmt2}}
\end{figure}

\begin{figure}
\epsscale{1.0}
\plotone{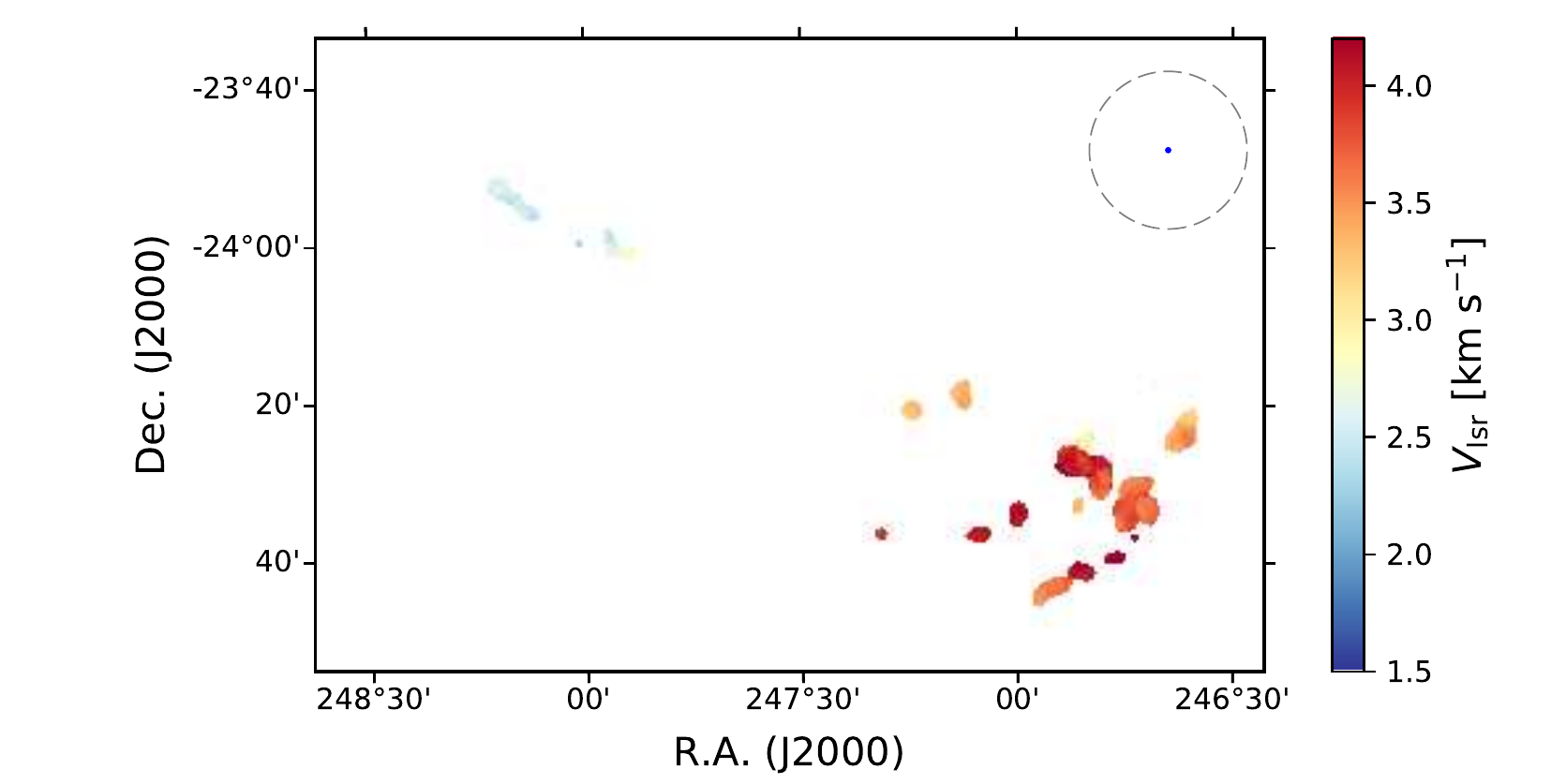}
\caption{Same as Figure \ref{fig_Oph_13CO_mmt1} but for the N$_2$H$^+$ line. \label{fig_Oph_N2Hp_mmt1}}
\end{figure}

\begin{figure}
\epsscale{1.0}
\plotone{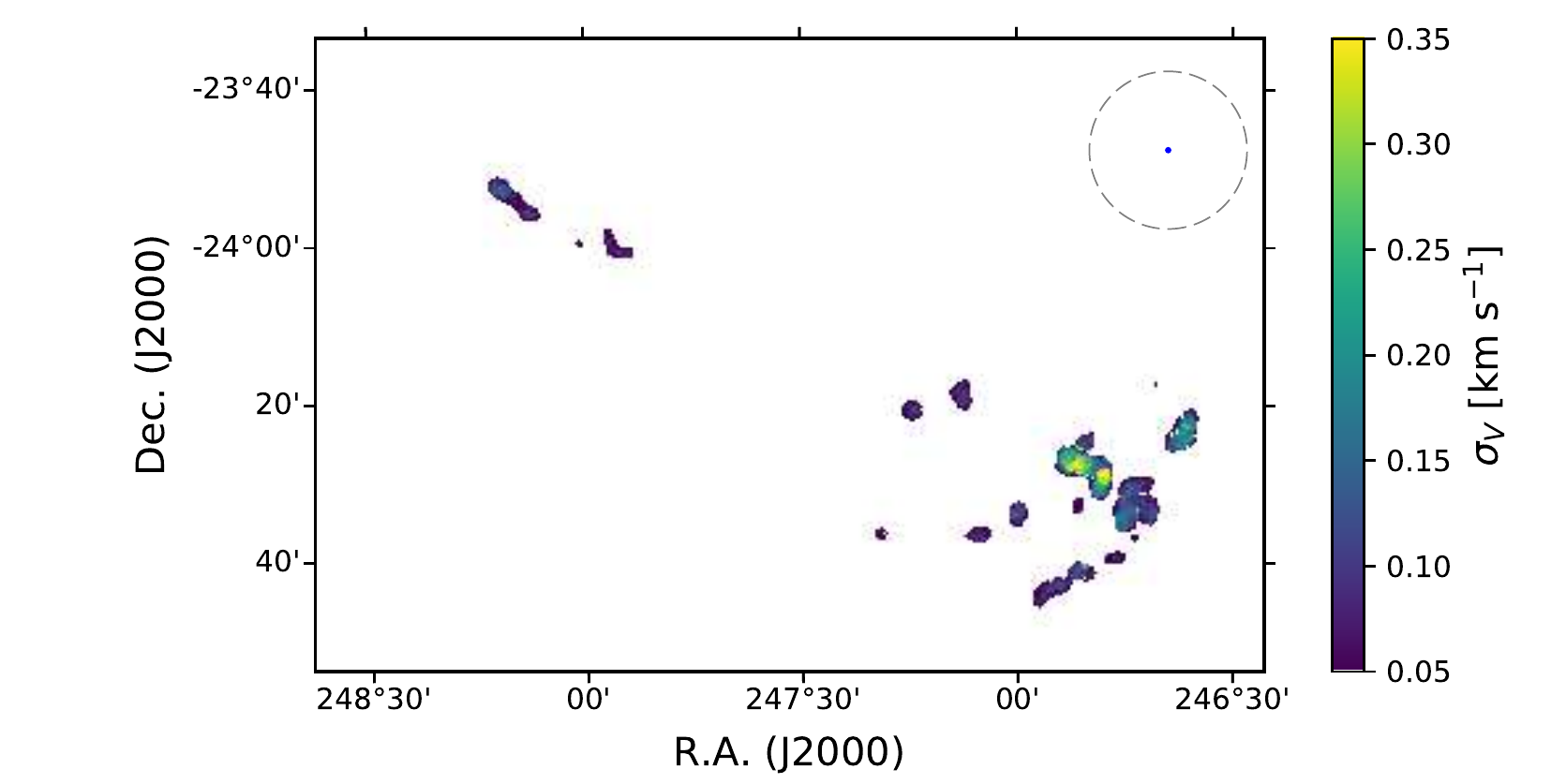}
\caption{Same as Figure \ref{fig_Oph_13CO_mmt2} but for the N$_2$H$^+$ line. \label{fig_Oph_N2Hp_mmt2}}
\end{figure}

\begin{figure}
\epsscale{1.0}
\plotone{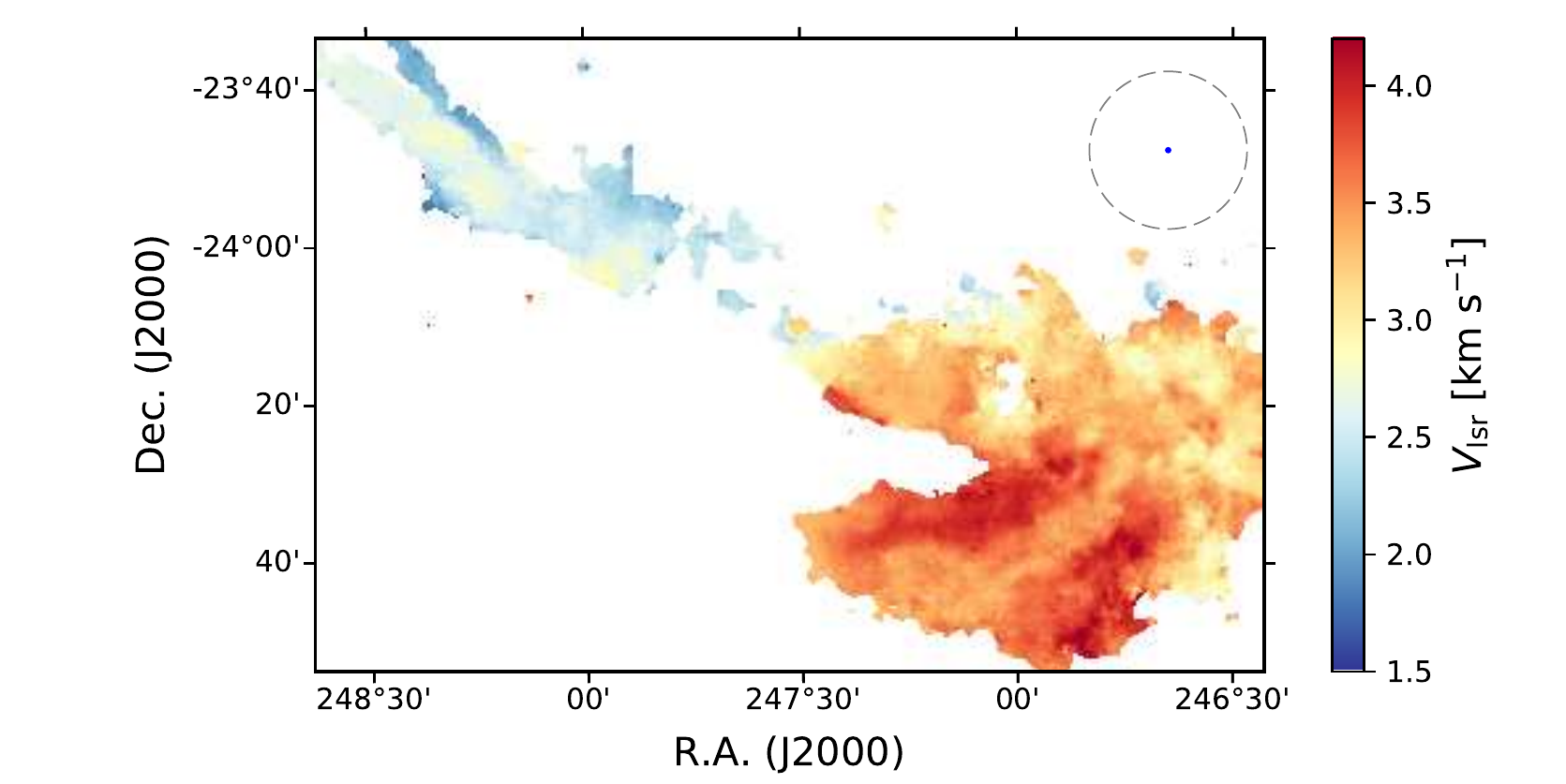}
\caption{Same as Figure \ref{fig_Oph_13CO_mmt1} but for the CS line. \label{fig_Oph_CS_mmt1}}
\end{figure}

\begin{figure}
\epsscale{1.0}
\plotone{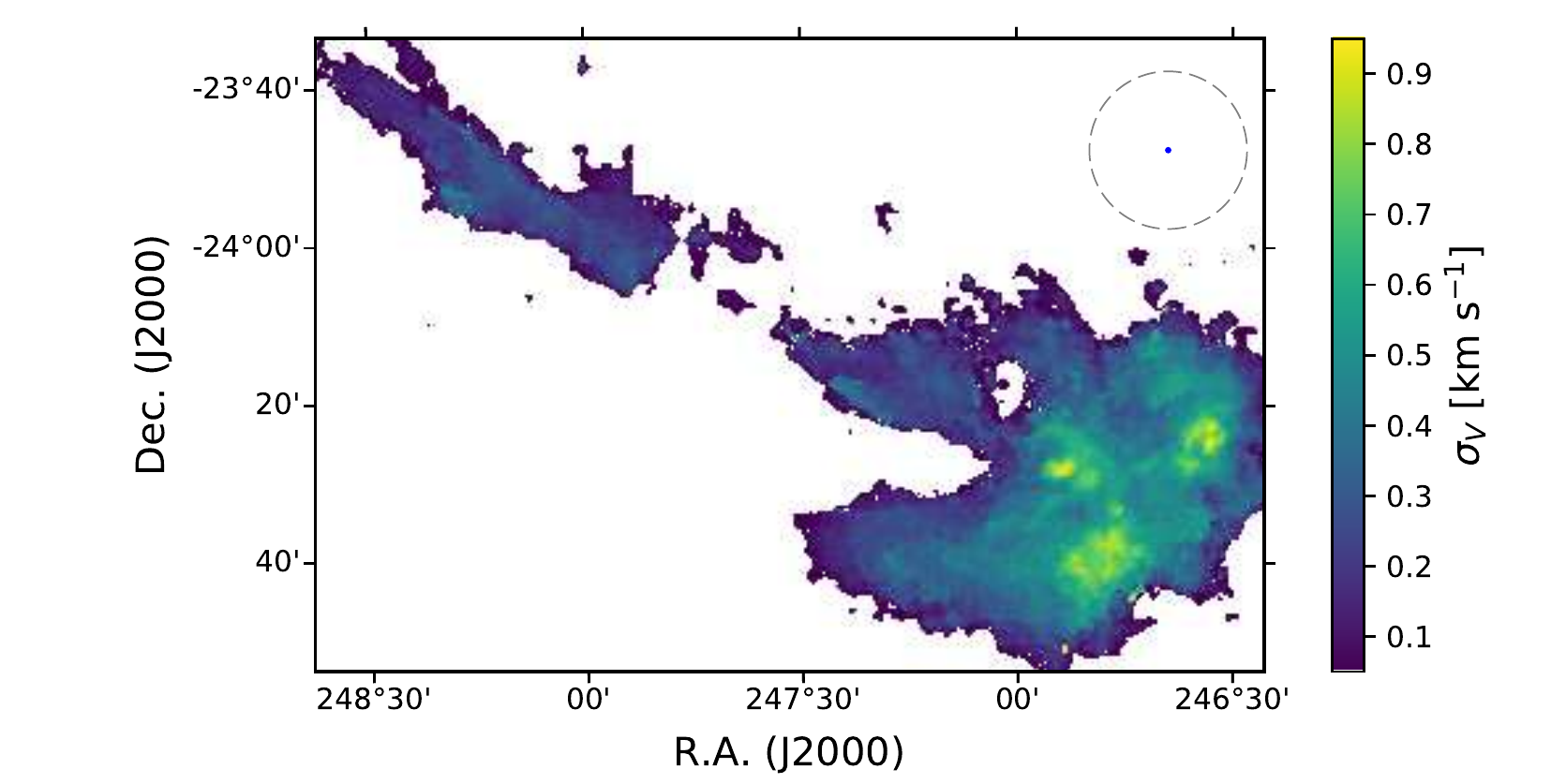}
\caption{Same as Figure \ref{fig_Oph_13CO_mmt2} but for the CS line. \label{fig_Oph_CS_mmt2}}
\end{figure}
\clearpage

\section{Multiple cloud components along the line of sight in the Orion A cloud} \label{App_highdv}
Figures \ref{fig_Ori_13CO_mmt2} and \ref{fig_Oph_13CO_mmt2} exhibit several small high-$\sigma_V$ regions. There are high-$\sigma_V$ regions in ISF and L1647 of the Orion A cloud. In the Ophiuchus cloud, there is a small high-$\sigma_V$ region in the northern part of L1688. In each cloud, these regions have $\sigma_V$ that are higher than the typical value of $\sigma_V$ in the other part of the cloud. 

Figures \ref{fig_ISF_PV} and \ref{fig_L1647_PV} show the position-velocity (PV) diagrams for these high-$\sigma_V$ regions in the Orion A cloud. And Figure \ref{fig_L1688_PV} presents the PV diagram for the high-$\sigma_V$ region in the Ophiuchus cloud. The PV diagrams demonstrate that there are two cloud components along the lines of sight at the high-$\sigma_V$ regions. Thus, the high-$\sigma_V$ values are caused by multiple gas components with different line of sight velocities. In the ISF region, the interactions between the MC and nearby sources, such as a foreground expending nebula and protostellar outflows, seems to produce these multiple cloud components \citep{Shi14,Kon18}. In the L1647 and L1688 regions, there might be foreground or background cloud components with different line of sight velocities.

\setcounter{figure}{0}
\begin{figure}
\epsscale{1.0}
\plotone{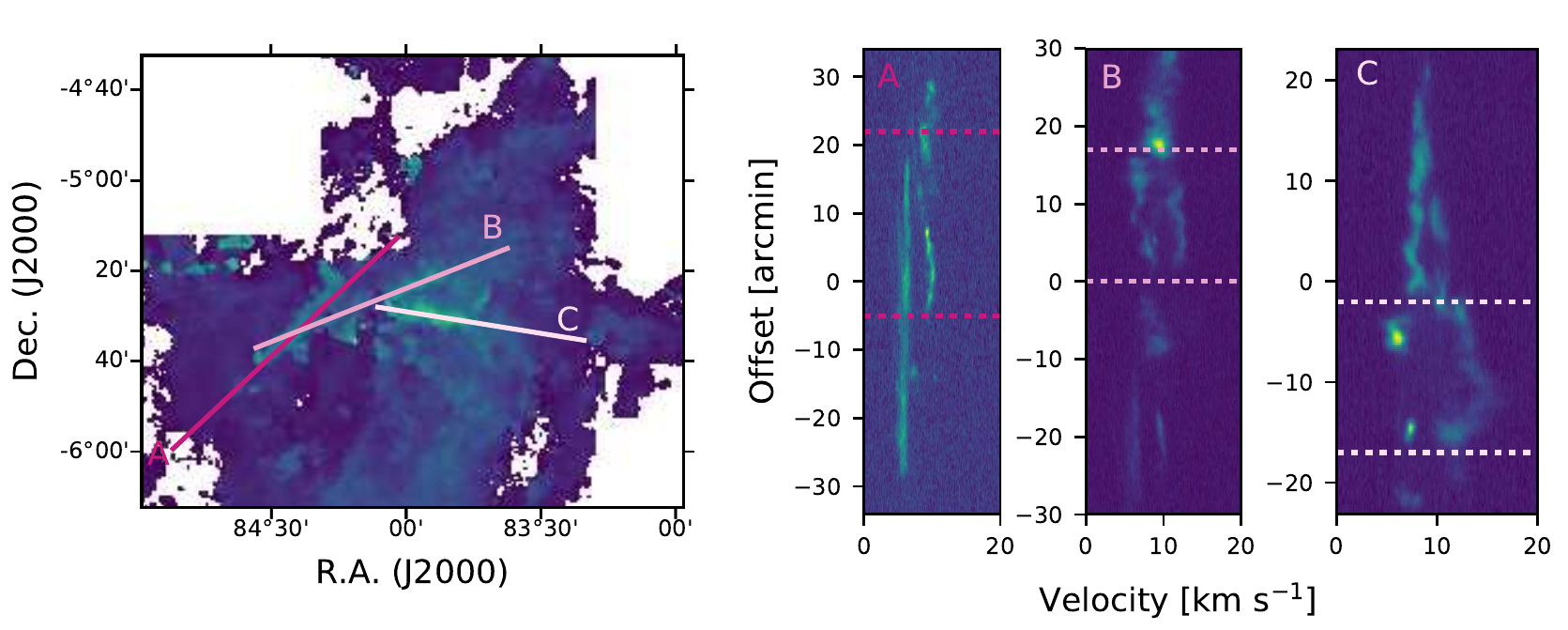}
\caption{Position-velocity (PV) diagrams for the high-$\sigma_V$ regions in ISF. The positions where the PV diagrams are extracted is presented with the solid lines on the $\sigma_V$ map} of $^{13}$CO (the first panel). The PV diagrams along the solid lines (A, B, and C) are presented on the second, third, and forth panels. The dotted horizontal lines in each PV diagram represents the position of the high-$\sigma_V$ regions on each line. The offset on the y-axis of the PV diagrams indicates a displacement from east to west on the line. \label{fig_ISF_PV}
\end{figure}

\begin{figure}
\epsscale{1.0}
\plotone{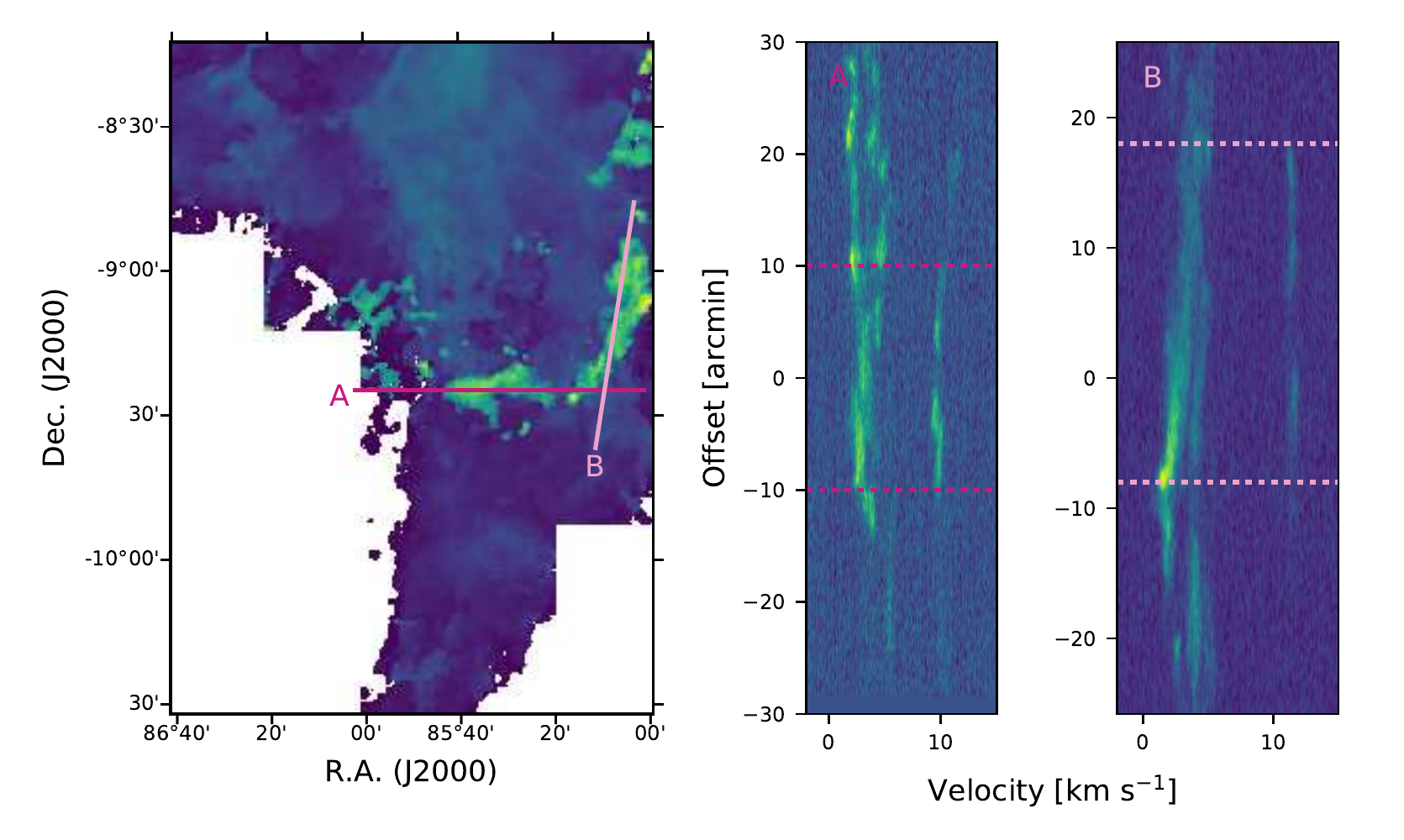}
\caption{Same as Figure \ref{fig_ISF_PV} except for the high-$\sigma_V$ regions in L1709. \label{fig_L1647_PV}}
\end{figure}

\begin{figure}
\epsscale{0.9}
\plotone{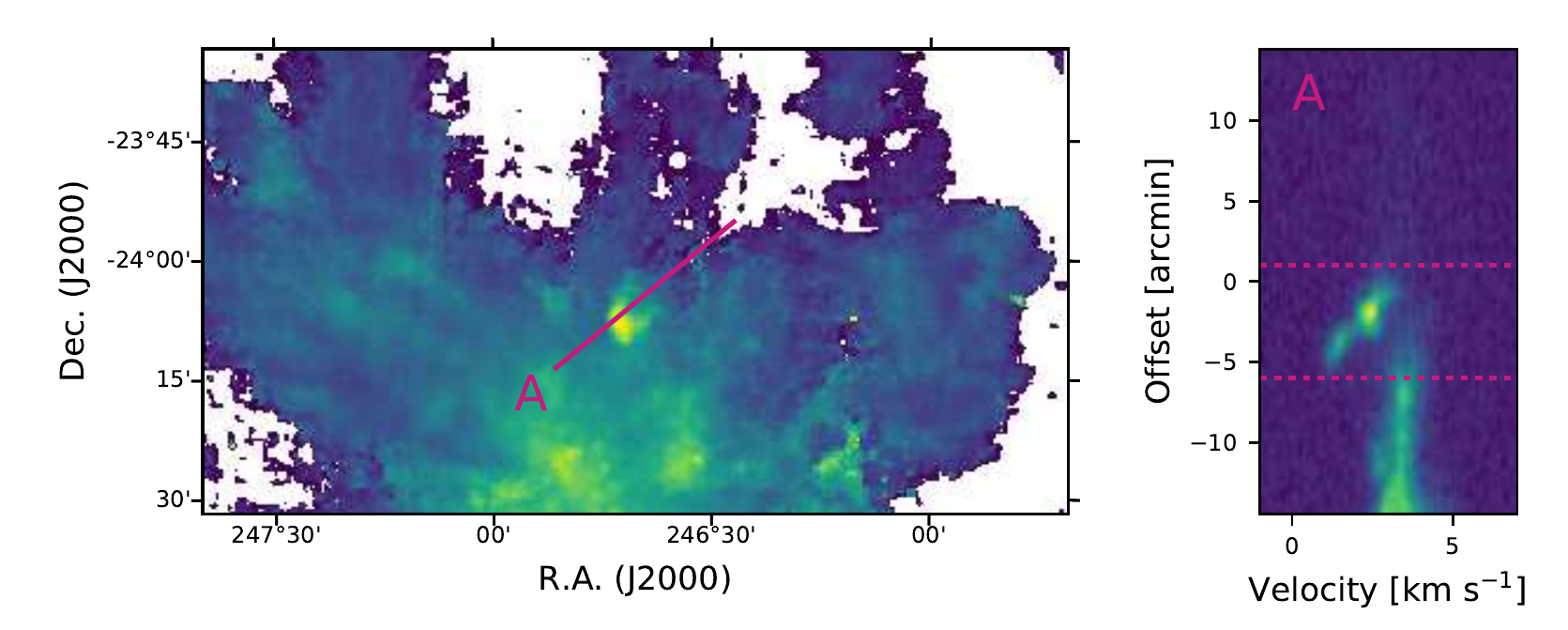}
\caption{Same as Figure \ref{fig_ISF_PV} except for the high-$\sigma_V$ region in L1688. \label{fig_L1688_PV}}
\end{figure}
\clearpage

\section{Random motion of the L1688 cloud} \label{App_random}
Because the observed $^{13}$CO line can be optically thick toward the dense part of L1688, the $V_\mathrm{lsr}$ map for $^{13}$CO (Figure \ref{fig_Oph_13CO_mmt1}) cannot exactly present the global motion of the cloud. Therefore, the optically thinner lines, such as the C$^{18}$O and CS lines, should be used to trace the global motion. However, the $V_\mathrm{lsr}$ maps of the C$^{18}$O and CS lines do not present any large-scale motions of the L1688 cloud (Figures \ref{fig_Oph_C18O_mmt1} and \ref{fig_Oph_HCOp_mmt1}). \citet{Lor89b} also suggested that the gas motion in the L1688 cloud is more complex than simple rotation. They measured $V_\mathrm{lsr}$ of the Ophiuchus cores using the $^{13}$CO and DCO$^+$ lines and found that the velocity variation for the DCO$^+$ lines is larger than that for the $^{13}$CO lines. Therefore, we assume that there is no global motion in L1688. 

\bibliographystyle{aasjournal.bst}
\bibliography{TRAO_obs_refs_0228.bib}

\end{document}